\documentclass[useAMS,usenatbib,fleqn]{mn2e}

\usepackage{amssymb}
\usepackage{times}
\usepackage{graphicx}
\usepackage{epstopdf}
\usepackage{subfigure}
\usepackage[T1]{fontenc}
\usepackage{aecompl} 
\usepackage{multirow}
\usepackage{pdflscape}
\usepackage{rotating}

\usepackage[fleqn]{amsmath}
\usepackage{mathrsfs}
\usepackage{xcolor}

\usepackage{anyfontsize}

\usepackage[breaklinks=true]{hyperref}

\newcommand{\beq}{\begin{equation}}
\newcommand{\eeq}{\end{equation}}

\newcommand{\bftheta}{\mbox{\boldmath $\theta$}}

\newcommand{\bfd}{\mathbfss{d}}

\newcommand{\bfP}{\mathbfss{P}}

\newcommand{\bfY}{\mathbfss{Y}}

\def\gs{\mathrel{\lower0.6ex\hbox{$\buildrel {\textstyle >}\over{\scriptstyle \sim}$}}}
\def\ls{\mathrel{\lower0.6ex\hbox{$\buildrel {\textstyle <}\over{\scriptstyle \sim}$}}}
\newcommand{\simgt}{\lower.5ex\hbox{$\; \buildrel > \over \sim \;$}}
\newcommand{\simlt}{\lower.5ex\hbox{$\; \buildrel < \over \sim \;$}}

\def\bm{\mbox{\boldmath $m$}}
\def\bs{\mbox{\boldmath $s$}}
\def\bc{\mbox{\boldmath $c$}}

\newcommand{\aap}{A\&A}

\newcommand{\apj}{ApJ}
\newcommand{\apjl}{ApJ}
\newcommand{\apjs}{ApJS}
\newcommand{\aj}{AJ}

\newcommand{\pasj}{PASJ}

\newcommand{\mnras}{MNRAS}
\newcommand{\physrep}{Phisycs Rep.}
\newcommand{\apss}{Astroph. Sp. Science}
\newcommand{\ssr}{Space Science Reviews}
\newcommand{\procspie}{SPIE Conference Series}

\defcitealias{se+et15_comalit_I}{CoMaLit-I}
\defcitealias{ser+al15_comalit_II}{CoMaLit-II}
\defcitealias{ser15_comalit_III}{CoMaLit-III}
\defcitealias{se+et15_comalit_IV}{CoMaLit-IV}
\defcitealias{xxl_I_pie+al15}{XXL-I}
\defcitealias{xxl_II_pac+al15}{XXL-II}
\defcitealias{xxl_III_gil+al15}{XXL-III}
\defcitealias{xxl_IV_lie+al15}{XXL-IV}
\defcitealias{xxl_XIII_eck+al15}{XXL-XIII}

\begin{document}

\title[CLUMP-3D. MACS1206]{CLUMP-3D. Three dimensional lensing and multi-probe analysis of MACS J1206.2-0847, a remarkably regular cluster}
\author[Sereno et al.]{
Mauro Sereno$^{1,2}$\thanks{E-mail: mauro.sereno@oabo.inaf.it (MS)}, Stefano Ettori$^{1,3}$, Massimo Meneghetti$^{1,3}$, Jack Sayers$^4$,  \newauthor  \ Keiichi Umetsu$^5$, Julian Merten$^6$, I-Non Chiu$^5$, Adi Zitrin$^7$\\
$^1$INAF - Osservatorio Astronomico di Bologna, via Ranzani 1, 40127 Bologna, Italy\\
$^2$Dipartimento di Fisica e Astronomia, Universit\`a di Bologna, viale Berti Pichat 6/2, 40127 Bologna, Italy\\
$^3$INFN, Sezione di Bologna, viale Berti Pichat 6/2, I-40127 Bologna, Italia\\
$^4$Division of Physics, Math, and Astronomy, California Institute of Technology, Pasadena, CA 91125\\
$^5$Institute of Astronomy and Astrophysics, Academia Sinica, P. O. Box 23-141, Taipei 10617, Taiwan\\
$^6$Department of Physics, University of Oxford, Keble Road, Oxford OX1 3RH, UK\\
$^7$Physics Department, Ben-Gurion University of the Negev, P.O. Box 653, Be'er-Sheva 84105, Israel
}


\maketitle

\begin{abstract}
Multi-wavelength techniques can probe the distribution and the physical properties of baryons and dark matter in galaxy clusters from the inner regions out to the peripheries. We present a full three-dimensional analysis combining strong and weak lensing, X-ray surface brightness and temperature, and the Sunyaev-Zel'dovich effect. The method is applied to MACS~J1206.2-0847, a remarkably regular, face-on, massive, $M_{200}=(1.1\pm0.2)\times10^{15}M_\odot/h$, cluster at $z=0.44$. The measured concentration, $c_{200}=6.3\pm1.2$, and the triaxial shape are common to halos formed in a $\Lambda$CDM scenario. The gas has settled in and follows the shape of the gravitational potential, which is evidence of pressure equilibrium via the shape theorem. There is no evidence for significant non-thermal pressure and the equilibrium is hydrostatic.
\end{abstract}


\section{Introduction}
\label{sec_intro}

Clusters of galaxies correspond to the densest regions to undergo gravitational relaxation in the hierarchical scenario of structure formation, where the universe is dominated early by cold dark matter and later by dark energy in the form of a cosmological constant. Tiny density fluctuations rise and grow in the early Universe under the influence of gravity to create the massive, dark matter dominated structures we observe today. Clusters are the most massive and latest objects to near viral equilibrium. 

This prominent role makes clusters de facto laboratories to test cosmology, astrophysics and fundamental physics \citep{voi05}. The accurate measurement of their mass and intrinsic properties is crucial to astrophysical investigation \citep{men+al10,pos+al12,ras+al12,lim+al13}. 

We want a complete picture of the cluster properties and, in the spirit of the scientific method, we want to compare observations to theory. However, we face the problem that theoretical models and numerical simulations of formation and evolution of cosmic structure are naturally expressed in terms of mass and three-dimensional properties, whereas we have access to only partial information. We can easily measure only projected quantities.

Clusters of galaxy can be observed at very different wave-lengths to provide independent data-sets, from X-ray surface brightness and spectral observations of the intra-cluster medium (ICM), to gravitational lensing (GL) observations of the total mass distribution to the Sunyaev-Zel'dovich effect (SZe) in the radio-band. 

Weak lensing (WL) analyses are in principle independent of the equilibrium state but can measure only the projected mass. To infer the true mass, we have to deproject the lensing maps assuming a cluster shape, which we actually ignore. The assumption of spherical symmetry can introduce biases in the measurement of mass and concentration  \citep{ogu+al05,se+zi12}.  An unbiased analysis has to take into account shape and orientation \citep{gav05,lim+al13}. 

X-ray observations open a window on the cluster thermodynamics, but hypotheses on the hydrostatic equilibrium (HE) are needed to further the analysis and measure the mass. This is not optimal. By assuming HE we can strongly bias the mass measurement \citep{ras+al12}. 

Simplifying assumptions can bias the results. This is critical also in the context of large present and future surveys \citep{eucl_lau_11,planck_2015_XXVII,xxl_I_pie+al16}, when reliable masses of selected clusters are used to calibrate large samples through scaling relations \citep{se+et16_comalit_V}.

Furthermore, when strong working hypotheses are enforced a priori, we cannot investigate anymore fundamental cluster properties. The cluster shape shows how matter aggregates from large-scale perturbations \citep{wes94,ji+su02}. Assessing the equilibrium status is crucial to determine evolution and mechanisms of interaction of baryons and dark matter \citep{le+su03,kaz+al04}.

In the era of precision astronomy, we have to use all the information and analyse it coherently. The distribution of three dimensional shapes of a population of astronomical objects can be obtained by inverting statistical samples of projected maps, see \citet{hub26}, \citet{noe79}, \citet{bin80}, \citet{bi+de81}, \citet{fa+vi91}, \cite{det+al95}, \citet{moh+al95}, \citet{bas+al00}, \citet{coo00}, \citet{th+ch01}, \citet{al+ry02}, \citet{ryd96}, \citet{pli+al04}, \citet{paz+al06}, \citet{kaw10}. 

Individual clusters can be investigated with multi-probe approaches, see \citet{zar+al98}, \citet{reb00}, \citet{dor+al01}, \citet{fo+pe02}, \citet{pu+ba06}, \citet{ma+ch11}, \citet{mor+al12}. Joint X-ray and SZ analyses can probe the gas physics in detail, e.g. the occurrence and mass distribution of infalling gas clumps \citep{eck+al16,tch+al16}.

Cluster maps can be deprojected combining different constraints of the cluster potential. However, despite the growing interest, this topic is still in its infancy. See \citet{lim+al13} for a recent review on the asphericity of galaxy clusters.

CLUster Multi-Probes in Three Dimensions (CLUMP-3D) is a project to get the unbiased intrinsic properties of galaxy clusters. By exploiting rich data-sets ranging from X-ray, to optical, to radio wavelengths, the mass and concentration can be determined together with the intrinsic shape and equilibrium status of the cluster as required by precision astronomy. The inversion problem is tackled with a Bayesian inference method.

This project builds on a series of methods developed by the same authors. \citet{def+al05} and \citet{ser+al06} first studied a sample of 25 clusters with X-ray and SZe data to find signs of a quite general triaxial morphology. The method was later generalised in a Bayesian framework in \citet{ser+al12a}. The triaxial analysis of strong and weak lensing was introduced in \citet{se+um11}, and a method combining lensing, X-ray and SZe was presented in \citet{ser+al13}. The latest application to Abell 1689 \citep{ume+al15a} showed that the cluster is elongated and not in equilibrium. The triaxial analysis reduces the apparent tension with theoretical predictions. 

As a test case, we consider MACS J1206.2-0847 (hereafter MACS1206) an X-ray luminous cluster at $z =0.439$ originally discovered in the  Massive Cluster Survey \citep[MACS,][]{ebe+al01,ebe+al09}. MACS1206 was included in the CLASH \citep[Cluster Lensing And Supernova survey with Hubble,][]{pos+al12} sample on the basis of being massive and relatively relaxed. \citet{zit+al12b} carried out a detailed strong-lensing analysis of the cluster exploiting CLASH {\it HST} (Hubble Space Telescope) imaging and Very Large Telescope (VLT)/VIMOS spectroscopic observations. Based on a strong lensing analysis, \citet{eic+al13} found evidence for tidally stripped halos of the cluster galaxies. 

\citet{ume+al12} performed an accurate mass reconstruction of the cluster from a combined weak-lensing distortion, magnification, and strong-lensing analysis of wide-field Subaru $BVR_cI_cz'$ and {\it HST} imaging. Alternative lensing analyses of the full CLASH sample were later presented in \citet{mer+al15,ume+al16}. 

\citet{biv+al13} exploited a rich data-set of $\sim 600$ spectroscopic redshifts, obtained as part of the VLT/VIMOS program, to constrain the mass, the velocity-anisotropy, and the pseudo-phase-space density profiles using the projected phase-space distribution. The overall agreement among different studies further suggests that the cluster is in a relaxed dynamical state.

The deep optical coverage is completed by X-ray data and ancillary measurements of the SZe collected with Bolocam, operating at 140 GHz at the Caltech Submillimeter Observatory \citep{cza+al15}. This makes MACS1206 an ideal target for a detailed lensing and multi-wavelength analysis.

The paper is as follows. Sections \ref{sec_tria_mat} and \ref{sec_tria_gas} are devoted to the triaxial parametric modelling of the matter and gas distribution, respectively. In Sec.~\ref{sec_obse}, we list the observational constraints. In Sec.~\ref{sec_data}, we present the data-sets used for the analysis. The Bayesian inference method is introduced in Sec.~\ref{sec_infe}. Results are presented and discussed in Sec.~\ref{sec_resu}. A check for systematics is performed in Sec.~\ref{sec_syst}. Section~\ref{sec_conc} is devoted to the conclusions. Appendix~\ref{sec_proj} summarises the basics of the projection of an ellipsoidal volume density. Appendix~\ref{sec_pote_shap} details how we approximated the shape of the gravitational potential.

\subsection{Notations and conventions}

Throughout the paper, the frame-work cosmological model is the concordance flat $\Lambda$CDM universe with matter density parameter $\Omega_\text{M}=1-\Omega_\Lambda=0.3$, Hubble constant $H_0=100h~\mathrm{km~s}^{-1}\mathrm{Mpc}^{-1}$ with $h=0.7$, and power spectrum amplitude $\sigma_8=0.82$. $H(z)$ is the redshift dependent Hubble parameter and $E_z\equiv H(z)/H_0$.  

The ellipsoid is our reference geometric shape for halos and distributions. We may further distinguish triaxial ellipsoids with three axes of different lengths, oblate or prolate spheroids, or spherical halos.

$O_{\Delta}$ denotes a global property of the cluster measured within a region which encloses a mean over-density of $\Delta$ times the critical density at the cluster redshift, $\rho_\text{cr}=3H(z)^2/(8\pi G)$. For an ellipsoidal halo, this region is the ellipsoid of semi-major axis $\zeta_\Delta$ and volume $(4\pi/3) q_{\text{mat},1}q_{\text{mat},2} \zeta_\Delta^3$, where $q_{\text{mat},1}$ and $q_{\text{mat},2}$ are the axis ratios. $O_{\text{sph},\Delta}$ indicates that the quantity is computed in a spherical region of radius $r_{\text{sph},\Delta}$.

The alternative subscript $\Delta\text{m}$ indicates that the overdensity region is computed with respect to the mean cosmological matter density.

Throughout the paper, `$\log$' is the logarithm to base 10 and `$\ln$' is the natural logarithm.

The parameters of our model are listed in Table~\ref{tab_parameters}. Coordinates and derived quantities are listed in Table~\ref{tab_symbols}. The ellipticity $\epsilon$ of the cluster refers to its projected two-dimensional shape in the plane of the sky; the line of sight elongation $e_\parallel$ quantifies the extent of the cluster along the line of sight, see App.~\ref{sec_proj}.

Typical values and dispersions of the parameter distributions are usually computed as bi-weighted estimators \citep{bee+al90} of the marginalised posterior distributions.

\section{Triaxial matter distribution}
\label{sec_tria_mat}

\begin{table*}
\caption{List of the parameters of the regression scheme. Units and description are in Col.~2 and 3, respectively. The default priors used in the regression scheme are listed in Col.~4. ${\cal U}$ is the uniform prior; $\delta$ is the Dirac delta function for parameters set to fixed values. In Col. 5, we refer to the section where the parameter is introduced and in Col.~6 we refer to the main equations involving the parameters. Units of $\zeta_\text{c}\text{kpc}/h$ mean that if e.g the truncation radius is $x$ times the core radius then $\zeta_\text{t}=x$.
}
\label{tab_parameters}
\begin{tabular}{ l l l l l l}     
\hline
Symbol & Units &\multicolumn{1}{c}{Description} & Default prior & Sec. & Eqs. \\
\hline
\multicolumn{6}{c}{\bf Total matter distribution} \\
\noalign{\smallskip}
$M_{200}$     		&	$10^{15}M_\odot/h$  & Total mass within the ellipsoid of mean density $200\rho_\text{cr}$ & ${\cal U}(10^{-1},10)$ & \ref{sec_nfw} & \ref{eq_nfw_1} \\
$c_{200}$     		&	& Concentration parameter & ${\cal U}(0.1,20)$ & \ref{sec_nfw} & \\ 
\noalign{\smallskip}
\multicolumn{6}{c}{\bf Matter shape and orientation} \\
\noalign{\smallskip}
$q_{\text{mat},1}$       		&	& minor to major axis ratio of the total matter distribution & ${\cal U}(0.1,1)$ &\ref{sec_tria_mat_shape} & \\
$q_{\text{mat},2}$       		&	& minor to major axis ratio of the total matter distribution & ${\cal U}(q_{\text{mat},1},1)$ &\ref{sec_tria_mat_shape} &\\
$\cos \vartheta$ 	&	& Cosine of the inclination angle of the ellipsoid major axis & ${\cal U}(0,1)$ &\ref{sec_tria_mat_shape} & \\
$\varphi$      		&	& Second Euler angle  & ${\cal U}(-\pi/2,\pi/2)$ &\ref{sec_tria_mat_shape} & \\
$\psi$      			&	& Third Euler angle  & ${\cal U}(-\pi/2,\pi/2)$ &\ref{sec_tria_mat_shape} & \\
\noalign{\smallskip}
\multicolumn{6}{c}{\bf Gas shape} \\
\noalign{\smallskip}
$q_\text{ICM,1}$    	&	& minor to major axis ratio of the ICM distribution & ${\cal U}(q_{\text{mat},1},1)$ &\ref{sec_tria_gas_shape} & \\
$q_\text{ICM,2}$  	&	& intermediate to major axis ratio of the ICM distribution & ${\cal U}(q_\text{ICM,1},1)$&\ref{sec_tria_gas_shape} & \\
\noalign{\smallskip}
\multicolumn{6}{c}{\bf Gas distribution} \\
\noalign{\smallskip}
$n_0$			&$\text{cm}^{-3}$	& central scale density of the distribution of electrons & ${\cal U}(10^{-6},10)$ & \ref{eq_gas_term} & \ref{eq_nprof_1}\\
$\zeta_\text{c}$  		& $\text{kpc}/h$	& Ellipsoidal core radius of the gas distribution & ${\cal U}(0,10^4)$  &\ref{eq_gas_term} & \ref{eq_nprof_1}\\
$\zeta_\text{t}$ 		&$\zeta_\text{c}\text{kpc}/h$	&	Ellipsoidal truncation radius of the gas distribution & ${\cal U}(0,10)$  &\ref{eq_gas_term} & \ref{eq_nprof_1}\\
$\beta$     		&	&      Slope of the gas distribution & ${\cal U}(0,3)$  &\ref{eq_gas_term} & \ref{eq_nprof_1}\\	
$\eta$      			&	&      Inner slope of the gas distribution & ${\cal U}(0,3)$  &\ref{eq_gas_term} & \ref{eq_nprof_1}\\
$\gamma_\text{ICM}$&	&      Outer slope of the gas distribution & ${\cal U}(0,3)$  &\ref{eq_gas_term} & \ref{eq_nprof_1} \\ 	
\noalign{\smallskip}
\multicolumn{6}{c}{\bf Gas temperature} \\
 \noalign{\smallskip}
$T_0$  			& keV	&      Typical temperature of the gas& ${\cal U}(10^{-2},10^{2})$  &\ref{eq_gas_term} & \ref{eq_Tprof_1}, \ref{eq_Tprof_3}\\ 		
$\zeta_{cT}$		&$\zeta_\text{c}\text{kpc}/h$	& Ellipsoidal truncation radius of the temperature profile & ${\cal U}(0,10)$ &\ref{eq_gas_term} & \ref{eq_Tprof_1}, \ref{eq_Tprof_3}\\
$a_T$       	&	& Intermediate slope of the temperature profile & $\delta(0)$ &\ref{eq_gas_term} & \ref{eq_Tprof_1}, \ref{eq_Tprof_3}\\
$b_T$       	&	& Steepness of the temperature profile & $\delta(2)$ &\ref{eq_gas_term} & \ref{eq_Tprof_1}, \ref{eq_Tprof_3}\\
$c_T$       	&	& Outer slope of the temperature profile & ${\cal U}(0,3)$ &\ref{eq_gas_term} & \ref{eq_Tprof_1}, \ref{eq_Tprof_3}\\	
$T_\text{cc}$     	& keV	&	 Temperature of the cool core & ${\cal U}(10^{-2},T_0)$ &\ref{eq_gas_term} & \ref{eq_Tprof_1}, \ref{eq_Tprof_2}\\
$\zeta_\text{cc}$		&$\zeta_\text{c}\text{kpc}/h$	& Ellipsoidal radius of the cool core & ${\cal U}(0,10)$ &\ref{eq_gas_term} & \ref{eq_Tprof_1}, \ref{eq_Tprof_2}\\
$\alpha_\text{cc}$		&	& Steepness of the cool core & $\delta(1.9)$ &\ref{eq_gas_term} & \ref{eq_Tprof_1}, \ref{eq_Tprof_2}\\
\hline	
\end{tabular}
\end{table*}

The spherical cow is a humorous metaphor but it is very far from an overly simplified model. If we cannot tell the head from the tail, if we do not know the ground, if we are not even sure the cow is a cow, make it spherical and we can still study the system without committing cancerous errors. But if we have some tools to tell the head from the side and we are sure that the cow is a cow, we can make it ellipsoidal to have a better insight.

For very irregular systems, the spherical approximation is still the better option. But if the galaxy cluster is well shaped, with the ellipsoidal model we can determine the properties of the system unbiased by shape and orientation.

The ellipsoidal model can be an improvement but it is not the final deal. Even in regular clusters, the eccentricity and the orientation of the matter distribution can change with the radius \citep{veg+al16,sut+al16}. The gas distribution changes with radius too. Axial ratios and orientation of the ellipsoidal distribution have to be meant as effective.

\subsection{Shape and orientation}
\label{sec_tria_mat_shape}

The main assumption of our triaxial modelling is that the total mass distribution of galaxy clusters is approximately ellipsoidal. This is the natural extension of the spherical modelling. The matter isodensities are approximated as a family of concentric, coaxial ellipsoids. We assume that the ellipsoids are self-similar, i.e., ellipsoids are concentric and share the same axis ratio and orientation. The halo shape is determined by the axis ratios, which we denote as $q_{\text{mat},1}$ (minor to major axis ratio) and $q_{\text{mat},2}$ (intermediate to major axis ratio). The eccentricity is $e_i=\sqrt{1-q_i^2}$. 

Cosmological simulations showed that self-similarity is not strict  \citep{kaz+al04}. At $z=0$, inner regions of simulated halos are less spherical than outer regions \citep{sut+al16,veg+al16}. The radial dependence gradually changes with time \citep{sut+al16}. At $z=1$, the axis ratios $q_{\text{mat},1}$ increases toward the inner regions. On the other hand, the axis ratio always steeply decreases in the outskirts due to filamentary structure around the halos.

The precise assessment of radial variations depends on how axis ratios are measured. Several methods have been proposed and some of them can be well-defined only in simulations. This may not be the case of actual observed clusters. Furthermore, radial variations can be below the accuracy reached by present-day observational campaigns.

The self-similar ellipsoidal distribution with fixed axis ratios can then provide a good description of galaxy clusters \citep{ji+su02,bon+al15}. Measured axis ratios have to be intended as effective radially weighted averages.

\subsubsection{Flat distribution}
As reference prior distribution, we considered a nearly flat distribution covering the range $q_\text{min}\le q_{\text{mat},1} \le 1$ and $q_{\text{mat},1} \le q_{\text{mat},2} \le 1$ \citep{se+um11}. We assume that the marginalised probability $p(q_{\text{mat},1})$ and the conditional probability $p(q_{\text{mat},2}|q_{\text{mat},1})$ are constant. In formulae,
\beq
\label{flat1}
p(q_{\text{mat},1}) =1/(1-q_\text{min})
\eeq 
for $q_\text{min}<q_{\text{mat},1} \le 1$ and zero otherwise, and
\beq
\label{flat2}
p(q_{\text{mat},2}|q_{\text{mat},1}) = 1/(1-q_{\text{mat},1})
\eeq 
for $q_{\text{mat},1} \le q_{\text{mat},2} \le 1$ and zero otherwise. We fixed $q_\text{min}=0.1$. The flat distribution is compatible with very triaxial clusters ($q_{\text{mat},1} \la q_{\text{mat},2} \ll1$), which are preferentially excluded by $N$-body simulations. Here and in the following, $p$ denotes the probability density, which can be larger than 1.

\subsubsection{$N$-body prior}

A population of ellipsoidal, coaligned, triaxial clusters fits well the relaxed clusters of galaxies produced in $N$-body simulations \citep{ji+su02}. The distributions of the minor to major and intermediate to major axis ratios can be well described by simple functional forms. Most of the dependences on mass and redshift can be expressed in terms of the peak height, $\nu$, from the spherical collapse theory \citep{bon+al15}. 

As a shape prior based on $N$-body simulations, we consider the results of \citet{bon+al15}, who analysed relaxed halos from the Millennium XXL simulation and provided statistically significant predictions in the mass range above $3\times10^{14}M_\odot/h$ at two redshifts ($z=0$ and $z=1$). Unrelaxed clusters were removed by selecting only haloes for which the offset between the most bound particle and the centre of mass of the particles enclosed by the ellipsoid was less than 5 per cent of their virial radius.

\citet{bon+al15} found that the minor-to-major axis ratio, after rescaling in terms of the peak height, $\tilde{q}_1 = q_1 \nu^{0.255}$, follows a log-normal distribution,
\beq
p (\ln\tilde{q}_1)\sim {\cal N}(\mu=-0.49,\sigma=0.20),
\eeq
where ${\cal N}$ is the Gaussian distribution; the conditional probability for the rescaled ratio $\tilde{q}_2=(q_2-q_1)/(1-q_1)$ can be written as a beta distribution,
\beq
P(\tilde{q}_2|q_1) \sim {\cal B}(\alpha,\beta) ,
\eeq
where
\begin{eqnarray}
\beta   & =& 1.389 q_1^{-1.685} \\
\alpha & = & \beta/\left[1/(0.633 q_1 - 0.007) - 1\right],
\end{eqnarray}

Alternatively, we also consider the priors based on \citet{ji+su02}, where the distribution of $q_1$ is approximated as,
\beq
\label{nbod3}
p(q_1) \sim {\cal N}(q_\mu/r_{q_1}, \sigma_\mathrm{s}=0.113),
\eeq
with $q_\mu=0.54$, and
\beq
r_{q_1} = (M_\mathrm{vir}/M_*)^{0.07 \Omega_\mathrm{M}(z)^{0.7}},
\eeq
with $M_*$ the characteristic nonlinear mass at redshift $z$ and $M_\mathrm{vir}$ the virial mass. The conditional probability of $q_2$ can be expressed as  
\beq
\label{nbod4}
p(q_1/q_2|q_1)=\frac{3}{2(1-r_\mathrm{min})}\left[ 1-\frac{2q_1/q_2-1-r_\mathrm{min}}{1-r_\mathrm{min}}\right],
\eeq
for $q_1/q_2 \geq r_\mathrm{min} \equiv \max[q_1,0.5]$, whereas is null otherwise.

\subsubsection{Random orientation}

The orientation of the halo is established by three Euler's angles, $\vartheta, \varphi$ and $\psi$, with $\vartheta$ quantifying the inclination of the major axis with respect to the line of sight. 

A priori, we considered a population of randomly oriented clusters with
\beq
\label{flat3}
p(\cos \vartheta) = 1
\eeq 
for $0 \le \cos \vartheta \le 1$,
\beq
p(\varphi)=\frac{1}{\pi}
\eeq
for $-\pi/2 \le \varphi \le \pi/2$, and
\beq
p(\psi)=\frac{1}{\pi}
\eeq
for $-\pi/2 \le \psi \le \pi/2$.

\subsection{Density profile}
\label{sec_nfw}

The Navarro-Frenk-White (NFW) density profile embodies the most relevant features of matter halos  \citep{nfw96,nav+al97},
\begin{equation}
\label{eq_nfw_1}
	\rho_\text{NFW}=\frac{\rho_\text{s}}{(\zeta/\zeta_\text{s})(1+\zeta/\zeta_\text{s})^2},
\end{equation}
where $\zeta$ is the ellipsoidal radius and $\zeta_\text{s}$ is the scale radius. In the coordinate frame oriented along the principal axes of the ellipse
\beq
\label{eq_nfw_1a}
\zeta^2 = \frac{x_1^2}{q_1^2}+\frac{x_2^2}{q_2^2}+x_3^2
\eeq
The NFW density profile can be described by two parameters. $M_{200}$ is the mass within the ellipsoid, 
\beq
\label{eq_nfw_2}
M_{200}\equiv(800\pi/3) \rho_\text{cr}\ q_{\text{mat},1}q_{\text{mat},2} \zeta_{200}^3.
\eeq 
The concentration is $c_{200} \equiv \zeta_{200}/ \zeta_\text{s}$. Ellipsoidal mass and concentration follow the same relations found in numerical $N$-body simulations for spherically averaged halos \citep{cor+al09}.


\begin{table}
\caption{List of coordinates and derived quantities. Symbols and descriptions are in Cols.~1 and 2, respectively. Quantities may refer to the total matter ($\text{mat}$), to the gas ($\text{ICM}$) and the gravitational potential ($\Phi$). In the manuscript, the component will be indicated with a subscript, e.g. $q_1$ is the generic axis ratio whereas $q_{\text{mat},1}$, $q_{\text{ICM},1}$ and $q_{\Phi,1}$ refer to the total matter, the gas, and the gravitational potential.  In Col. 3, we refer to the section where the parameter is introduced and in Col.~4 we refer to the main equations involving the parameters.}
\label{tab_symbols}
\resizebox{\hsize}{!} {
\begin{tabular}{ l l ll }     
\hline
Symbol & \multicolumn{1}{c}{Description} & Sec. & Eq. \\
\hline
\multicolumn{4}{c}{\bf 3D coordinates} \\
\noalign{\smallskip}
$\zeta$     	&	Ellipsoidal radial coordinate & \ref{sec_tria_mat_shape} & \ref{eq_nfw_1a}\\
$r$     		&	spherical radial coordinate & \ref{sec_tria_mat_shape} &\\
\noalign{\smallskip}
\multicolumn{4}{c}{\bf 2D coordinates} \\
\noalign{\smallskip}
$\xi$     		&	elliptical radial coordinate & \ref{sec_tria_mat_shape} &  \\
$\theta_\xi$     		&	elliptical angular coordinate & \ref{sec_tria_mat_shape} &  \\
$R$     		&	circular radial coordinate & \ref{sec_tria_mat_shape} &  \\
$\theta_R$     		&	circular angular coordinate & \ref{sec_tria_mat_shape} &  \\
\noalign{\smallskip}
\multicolumn{4}{c}{\bf 3D shape parameters} \\
\noalign{\smallskip}
$\mathcal{T}$       		&	triaxial parameter & \ref{sec_tria_gas_shape} & \ref{eq_shape_1} \\
$e_i$       			&	eccentricity of the $i$-th axis & \ref{sec_tria_mat_shape}  &  \\
\noalign{\smallskip}
\multicolumn{4}{c}{\bf 3D radii} \\
\noalign{\smallskip}
$\zeta_\Delta$     	&	Semi-major axis of the ellipsoid of average density & \ref{sec_intro} &  \\
 			     	&	equal to $\Delta$ times the critical density $\rho_\text{cr}$ &  &  \\
$\zeta_{\Delta\text{m}}$     	&	Semi-major axis of the ellipsoid of average density  & \ref{sec_intro} &  \\
     						&	equal to $\Delta$ times the cosmological matter density &  &  \\
$r_{\text{sph},\Delta}$     		&	Radius of the sphere of average density equal to & \ref{sec_intro} &  \\
    						&	$\Delta$ times the cosmological critical density &  &  \\
\noalign{\smallskip}
\multicolumn{4}{c}{\bf 2D projected parameters} \\
\noalign{\smallskip}
$q_\perp$    	&	axis ratio of the projected ellipse & \ref{sec_proj}  &  \ref{eq_tri1} \\
$\epsilon$    	&	projected ellipticity & \ref{sec_proj}  & \ref{eq_tri1a} \\
$\theta_\epsilon$    	&	orientation angle of the projected ellipse & \ref{sec_proj}  & \ref{eq_tri5}  \\
\noalign{\smallskip}
\multicolumn{4}{c}{\bf 2D NFW parameters} \\
\noalign{\smallskip}
$\kappa_\text{s}$    	& scale-convergence	 & \ref{sec_obse_lens}  &  \ref{eq_nfw_3} \\
$\xi_\perp$    	&	projected elliptical scale radius & \ref{sec_obse_lens}  &  \ref{eq_nfw_3} \\
\noalign{\smallskip}
\multicolumn{4}{c}{\bf Line-of-sight parameters} \\
\noalign{\smallskip}
$f$    	&	line of sight projection & \ref{sec_proj}  &  \ref{eq_tri8}  \\
$e_\parallel$    	&	elongation & \ref{sec_proj}  &  \ref{eq_tri9}  \\
\noalign{\smallskip}
\hline	
\end{tabular}
}
\end{table}

\section{Triaxial gas distribution}
\label{sec_tria_gas}

\subsection{Shape}
\label{sec_tria_gas_shape}

In our modelling, the gas distribution is ellipsoidal and co-aligned with the matter. This is supported by the observed morphologies of galaxy clusters in the X-ray band, which are nearly elliptical \citep{kaw10}.

In the ideal case of an halo in perfect hydrostatic equilibrium, the assumptions of ellipsoidal and fixed matter distribution and of ellipsoidal and fixed gas distribution are mutually exclusive. If the matter is ellipsoidal, the potential that originates from it cannot be ellipsoidal. The gas distribution traces the gravitational potential, and turns rounder in the outer regions. On the other hand, given an ellipsoidal gas density, the gravitational potential is ellipsoidal too and can turn unphysical for extreme axis ratios, with negative density regions or unlikely configurations.

However, the variation of eccentricity of the potential of an ellipsoidal mass distribution in the radial range covered by observations is small \citep{le+su03,le+su04} and the ellipsoidal approximation for the gas is suitable in the inner regions or when small eccentricities are considered \citep{bu+hu12}. 

In App.~\ref{sec_pote_shap}, we detail how the shape of the gravitational potential of an ellipsoidal halo can be approximated. The ratio of eccentricities of potential ($e_\Phi$) and matter ($e_\text{mat}$) is nearly constant up to the length scale, with $e_{\Phi,i}/e_{\text{mat},i} \simeq 0.7$ for $i=1,2$. Furthermore, the variation in projected ellipticity of the potential  is usually smaller than the observational error on the measured ellipticity of the X-ray surface brightness map.

Physical processes perturb the only hypothetical perfect equilibrium. Radiative cooling, turbulence or feedback mechanisms strongly affect the gas shape, which can show a distinctly oblate configuration towards the central regions compared to the underlying dark matter potential shape \citep{lau+al11}. Outside the core, radiative processes can make the ICM distribution rounder. The effect of filamentary accretion or merging events can have dramatic effects in the outskirts.

These mechanisms can be effective. In a conservative scheme, the overall triaxiality of the gas cannot be strictly related to the underlying shape of the dark matter potential. Total matter distribution and gas have to be modelled independently. We then explore different scenarios  to relate gas and total matter distributions. 


In the less informative one ($q_\text{ICM}\ge q_\text{mat}$), we assume that total matter and gas are shaped as coaligned ellipsoids with fixed, but different eccentricity. The angles $\vartheta, \varphi$ and $\psi$ set the orientation of both distributions. In this scheme, the gas is rounder than the total matter, i.e., $q_\text{ICM,1}\ge q_{\text{mat},1}$ and $q_\text{ICM,2}\ge q_{\text{mat},2}$. The prior on $q_{\text{ICM},1}$ is then similar to that of $q_{\text{mat},2}$ in the case of the flat distribution for the matter axis ratios.

In the second scheme ($\mathcal{T}_\text{mat}=\mathcal{T}_\text{ICM}$), the matter and gas distributions share the same triaxiality parameter, 
\beq
\label{eq_shape_1}
\mathcal{T}=(1-q_2^2)/(1-q_1^2).
\eeq
If two distributions have the same triaxiality, the misalignment angle between the orientations in the plane of the sky is zero \citep{ro+ko98}. The axis ratios of the gas distribution, $q_{\text{ICM},i}$, can be expressed in terms of the corresponding axis ratios of the matter distributions as
\beq
q_{\text{ICM},i}=\sqrt{1-(e_\text{ICM}/e_\text{mat})^2(1-q_{\text{mat},i}^2)}.
\eeq
Being $\mathcal{T}_\text{mat}=\mathcal{T}_\text{ICM}$, then $e_{\text{ICM},1}/e_{\text{mat},1}=e_{\text{ICM},2}/e_{\text{mat},2}=e_\text{ICM}/e_\text{mat}$. The above assumptions limit the number of free axis ratios to three: $q_{\text{mat},1}$ and $q_{\text{mat},2}$ for the matter and $q_{\text{ICM},1}$ for the gas. $q_{\text{ICM},2}$ is determined by $\mathcal{T}_\text{mat}$ and $q_{\text{ICM},1}$,
\beq
q_{\text{ICM},2}=\sqrt{1-\frac{1-q_{\text{ICM},1}^2}{\mathcal{T}_\text{mat}^2}}.
\eeq 
Under these hypotheses, total matter and gas have different projected ellipticities and elongations but share the same orientation in the plane of the sky, $\theta_\epsilon$. This is in agreement with what observed in MACS1206, where the centroid and the orientation of the surface brightness map coincide with those of the projected mass distribution  inferred from lensing, see Sec.~\ref{sec_res_tri}. 

In the third scheme ($q_\text{ICM}=q_\Phi$), we assume that the gas follows the potential. This scheme formally conflicts with our assumption that both total matter and gas are distributed as ellipsoids with fixed eccentricity. In fact, the potential of an ellipsoidal matter distribution is not ellipsoidal. However, in most cases the isopotential surfaces can be still well approximated as ellipsoids whose axial ratios vary slightly with the radius, see App.~\ref{sec_pote_shap}. We can then consider the axial ratios of the potential at a typical radius and assume that the gas distribution is ellipsoidal with those axial ratios at each radius.

In the following, we measure the effective axis ratios of the potential at $\zeta = \zeta_{200}/3\sim\zeta_{2500}$, the radius better probed by X-ray observations. Alternatively, we can also consider $\zeta = 2\zeta_{200}/3\sim\zeta_{500}$.

\subsection{Termodymanics}
\label{eq_gas_term}

We describe the thermodynamical properties of the intracluster medium in term of the distribution of the gas and its temperature.

The electronic density was modelled with the parametric profile \citep{vik+al06,ett+al09},
\beq
\label{eq_nprof_1}
n_\mathrm{e}=n_0  \left( \frac{\zeta}{\zeta_\text{c}} \right)^{-\eta}\left[ 1+\left( \frac{\zeta}{\zeta_\text{c}} \right)^2 \right]^{-3\beta/2+\eta/2} \left[ 1+\left( \frac{\zeta}{\zeta_\text{t}} \right)^3 \right]^{-\frac{\gamma_\text{ICM}}{3}},
\eeq
where $n_\mathrm{0}$ is the central electron density, $\zeta_\text{c}$ is the core elliptical radius, $\zeta_\text{t}(>\zeta_\text{c})$ is the tidal radius, $\beta$ is the slope in the intermediate regions, and $\eta$ and $\gamma_\text{ICM}$ are the inner and outer slope, respectively.  

For the intrinsic temperature, we used \citep{vik+al06,bal+al12}
\beq
\label{eq_Tprof_1}
T_\text{3D}(\zeta)=T_0 t_\text{cc}(\zeta)t_\text{out}(\zeta),
\eeq
where $T_0$ sets the temperature scale, $t_\text{cc}(r)$ describes the temperature decline in the cluster cool core,
\beq
\label{eq_Tprof_2}
t_\text{cc}(\zeta)= \frac{x+T_\text{cc}/T_0}{1+ x}, \ x=(\zeta/\zeta_\text{cc})^{\alpha_\text{cc}},
\eeq
and $t_\text{out}(r)$ parameterises the temperature profile outside the central cool region,
\beq
\label{eq_Tprof_3}
t_\text{out}= \frac{(\zeta/\zeta_{cT})^{-a_T} }{[1+ (\zeta/\zeta_{cT})^{b_T}]^{{c_T}/{b_T}}  };
\eeq
the radius $\zeta_{cT}$ is the transition radius. Some of the parameters describing the profile at the truncation are degenerate. We fixed $b_T=2$.

\section{Observational constraints}
\label{sec_obse}

The intrinsic properties of the cluster can be obtained by deprojection of the observed maps.

\subsection{Lensing}
\label{sec_obse_lens}

Lensing analysis can provide the projected mass density of the cluster,
\beq
\Sigma_\text{mat} = \int _\parallel \rho_\text{mat}\ dl ,
\eeq
where the subscript $\parallel$ denotes integration along the line of sight.

Ellipsoidal 3D halos project as elliptical 2D profiles, see App.~\ref{sec_proj}. For the NFW halo, the convergence $\kappa$, i.e., the surface mass density in units of the critical density for lensing, $\Sigma_\text{cr}=c^2D_\text{s}/(4\pi G\,D_\mathrm{d}\,D_\mathrm{ds})$, where $D_\text{s}$, $D_\mathrm{d}$ and $D_\mathrm{ds}$ are the source, the lens and the lens-source angular diameter distances respectively, can be written as
\beq
\label{eq_nfw_3}
\kappa_\text{NFW}(x)=\frac{2 \kappa_\text{s}}{1-x^2}\left[ \frac{1}{\sqrt{1-x^2}} \mathrm{arccosh}\left(\frac{1}{x}\right) -1\right],
\eeq
where $x$ is the dimensionless elliptical radius, $x \equiv \xi /\xi_\perp$. The elliptical isodensities are characterised by the ellipticity $\epsilon_\text{mat}$ and the direction angle $\theta_{\epsilon,\text{mat}}$. The central strength $\kappa_\text{s}$ and the projected scale radius are related to mass and concentration and depend on shape and orientation parameters too. 

Following App.~\ref{sec_proj}, explicit formulae between the intrinsic parameters, $M_{200}$ and $c_{200}$, shape and orientation and measurable projected parameters, $\kappa_\text{s}$ and $\xi_\perp$, can be written as, see also \citet{ser+al10b},
\beq
\delta_c= \frac{1}{e_\parallel} \frac{\kappa_\text{s}}{\xi_\perp}\frac{\Sigma_\text{cr}}{\rho_\text{cr}},
\eeq
where, as usual,
\beq
\delta_c= \frac{200}{3}\frac{c_{200}}{\ln(1+c_{200})-c_{200}/(1+c_{200})}, 
\eeq
and
\beq
M_{200} =\frac{4\pi}{3}200\rho_\text{cr}c_{200}^3 (e_\parallel \sqrt{f} \xi_\perp)^3 .
\eeq

The convergence map can be derived from lensing analyses \citep{mer+al15,ume+al16}. The corresponding $\chi^2$ function can be expressed as \citep{ogu+al05}
\begin{equation}
\label{eq:chi2wl}
 \chi_\text{GL}^2 =\sum_{m,n=1}^{N_\text{GL}}
\left[\kappa(\bftheta_m)-\hat{\kappa}(\bftheta_m)\right]
\left( C_\text{GL}^{-1} \right)_{mn}
\left[\kappa(\bftheta_n)-\hat{\kappa}(\bftheta_n)\right],
\end{equation}
where $\mbox{\boldmath $\kappa$}=\{\kappa(\bftheta_m)\}_{m=1}^{N_\text{GL}}$ is the convergence map from the lensing analysis, $C_\text{GL}^{-1}$ is the inverse of the error covariance matrix, and the hat symbol denotes a modelled quantity. Here, $\kappa(\bftheta_m)$ can be seen as either the convergence at a particular position, e.g. the convergence in a pixel, or the mean convergence in a large area, e.g. the mean convergence in an annular bin. $N_\text{GL}$ is the number of convergence measurements.

\subsection{X-ray Surface Brightness}

The observed X-ray surface brightness (SB) in an energy band due to bremsstrahlung and line radiation resulting from electron-ion collisions in the high temperature plasma can be written as
\begin{equation}
\text{SB} = \frac{1}{4 \pi (1+z )^3} \int _\parallel n_\mathrm{e}^2 \Lambda_\mathrm{eff}(T_\mathrm{e}, {\cal{Z}}) dl ,
\label{eq_SB1}
\end{equation}
where $T_\mathrm{e}$ is the intrinsic temperature, ${\cal{Z}}$ the metallicity and $\Lambda_\mathrm{eff}$ is the effective cooling function of the ICM in the cluster rest frame, which depends on the energy-dependent area of the instrument \citep{ree+al10}.  

In the present paper, we performed a full 2D analysis of the X-ray map and we determined ellipticity and orientation at once with the gas distribution. Photon number counts follow the Poisson distribution. The $\chi^2$ analog is the Cash statistic \citep{cas79},
\beq
\label{eq_SB2}
{\cal C}_\text{SB,2D} =2 \sum_{i=1}^{N_{\text{SB,2D}}} \left[ \hat{N}_i -N_i \ln \hat{N}_i \right],
\eeq
where $N_{\text{SB,2D}}$ is the number of pixels, $N_i$ is the number of observed photons in the $i$-th pixel and $\hat{N}_i(=\hat{S}_i+\hat{B}_i$) is the sum of source $\hat{S}_i$ and background  $\hat{B}_i$ model amplitudes,
\beq
\label{eq_SB3}
\hat{S}_i = t_\text{exp}A_i\text{SB}_i,
\eeq
where $t_\text{exp}$ is the exposure time and $A_i$ the pixel collecting area. The formalism in Eqs.~(\ref{eq_SB2} and \ref{eq_SB3}) assumes that the background is known and measured in regions well outside the cluster emission. This approximation holds as far as the background is smaller than the source amplitude, otherwise we should model the background emission with a second independent Poisson process.

In the outer regions where the signal is comparable to the background, it can be convenient to measure the mean signal in annular regions. Due to the large total number counts, the statistics is approximately Gaussian and the $\chi^2$ function for the averaged surface brightness can be can be written as
\begin{equation}
\label{eq_SB4}
\chi^2_\text{SB,1D}  = \sum_{i=1}^{N_{\text{SB,1D}}}  \left(\frac{S_{\text{X},i}-\hat{S}_{\text{X},i}}{\delta_{S,i}}\right)^2 ,
\end{equation}
where $\hat{S}_{\text{X},i}$ is the model prediction for the X-ray surface brightness in the $i$-th annulus, $\delta_{S,i}$ is the measurements uncertainty and $N_{\text{SB,1D}}$ is the number of annular bins.

\subsection{Observed temperature}

The temperature $T_\mathrm{e}$ in Eq.~(\ref{eq_SB1}) is the intrinsic temperature. The spectroscopic temperature $T_\mathrm{sp}$ measured by space observatories is well approximated by \citep{maz+al04}
\beq
\label{eq_proj7}
T_\mathrm{sp}= \frac{\int W T_\mathrm{e} dV}{\int W dV}, \, \, W=\frac{n^2}{T_\mathrm{e}^{3/4}} .
\eeq

The $\chi^2$ function for the temperature can be written as \citep{ser+al12a}
\begin{equation}
  \chi^2_{T}  = \sum_{i=1}^{N_T} \left(\frac{T_{\text{sp},i}-\hat{T}_{\text{sp},i}}{\delta_{T,i}}\right)^2,
\end{equation}
where $\hat{T}_{\text{sp},i}$ is the model prediction for the corresponding observed spectroscopic temperatures $T_{\text{sp},i}$ in the $i$-th angular bin and $\delta_{T_i}$ is the measurement uncertainty.

\subsection{The Sunyaev-Zel'dovich effect}

The Sunyaev-Zeldovich effect (SZe) is the distortion of the cosmic microwave background (CMB) spectrum due to inverse Compton scattering by the hot ICM energetic electrons \citep{su+ze70,bir99}. The amplitude of the signal can be expressed in terms of the Compton-$y$ parameter, which is proportional to the integral of the electron pressure along the line of sight,
\beq
\label{eq:sze1}
y \equiv \frac{ \sigma_{\rm T} k_{\rm B} }{m_\mathrm{e} c^2} \int_\parallel n_\mathrm{e}  T_\mathrm{e} dl ,
\eeq
where $k_{\rm B}$ is the Boltzmann constant, $\sigma_{\rm T}$ is  the Thompson cross section, $m_\mathrm{e}$ the electron mass, $c$ the speed of light in vacuum.

The measured temperature decrement $\Delta T_\text{SZ}$ of the CMB for an isothermal plasma is given by
\begin{equation}
\label{eq:sze2}
\Delta T_\text{SZ} = f_\text{SZ}(\nu, T) T_\mathrm{CMB} y
\end{equation}
where $T_\mathrm{CMB}$ is the temperature of the CMB and $f_\text{SZ}(\nu, T)$ accounts for relativistic corrections at frequency $\nu$.

As overall measure of the thermal energy content in a cluster we consider $Y$, i.e. the Compton-$y$ parameter integrated over a cluster region,
\beq
\label{eq:sze3}
Y_\Omega = \int_{\Omega}y(\theta) d \theta,
\eeq
where $\Omega$ is the angular area. $Y$ is nearly independent of the model of gas distribution used for the analysis and it is a robust quantity for observational tests \citep{ben+al04}. In addition, integrating the Compton-$y$ diminishes (though does not completely remove) effects resulting from the presence of strong entropy features in the central regions of clusters \citep{mca+al03}.

The $\chi^2$ function can be written as
\begin{equation}
 \chi^2_\text{SZ}=\sum_{m,n=1}^{N_\text{SZ}}
\left[Y(\Omega_m)-\hat{Y}(\Omega_m)\right]
\left( C_\text{SZ}^{-1} \right)_{mn}
\left[Y(\Omega_n)-\hat{Y}(\Omega_n)\right],
\end{equation}
where $\bfY=\{ Y(\Omega_m)\}_{m=1}^{N_\text{SZ}}$ is the set of integrated Compton parameters in the circular annuli, $C_\text{SZ}^{-1}$ is the inverse of the uncertainty covariance matrix, and the hat symbol denotes a modelled quantity.

\begin{figure*}
       \includegraphics[width=15cm]{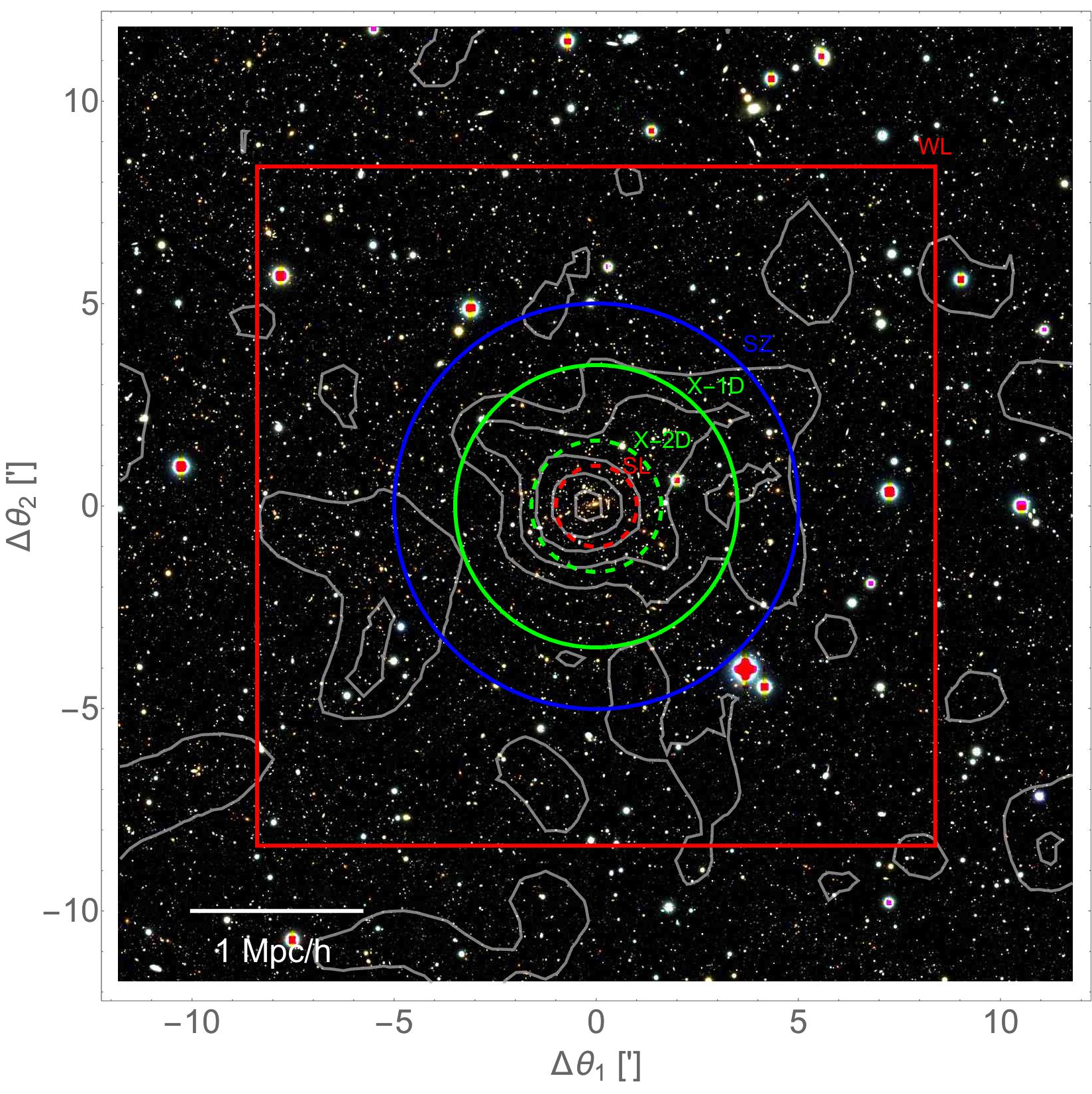}
       \caption{Subaru $BVR_\text{c}$ composite colour images of the galaxy cluster MACS1206. 
       The geometric forms centred in the optical centre enclose the regions exploited for inference by each probe: the dashed red circle at $1\arcmin$ and the red square of semi-size $2\text{Mpc}/h\sim8.39\arcmin$ for SL and WL; the dashed and full green circles at $\theta_{80\%}=1.61\arcmin$ and $3.49\arcmin$ for the 2D- and 1D-X-ray analysis, respectively; the blue circle at 5\arcmin for the SZe. The smoothed mass contours from the WL analysis of the Subaru observations are overlaid in white. The convergence levels (for a reference source redshift of $z_\text{s}=20000$) go from $\kappa= 0.1$ to 0.5, with increments of 0.1. The image size is $24\arcmin\times24\arcmin$. The horizontal bar represents $1\text{Mpc}h^{-1}$ at the cluster redshift. North is top and east is left.}
	\label{fig_M1206_rgb}
\end{figure*}

\section{Data analysis}
\label{sec_data}

In this section, we present the data-sets used for the analysis. The different data-sets cover a large radial range, from the cluster core up to $\ga2\text{Mpc}/h$, see Fig.~\ref{fig_M1206_rgb}.

\subsection{Chandra}

\begin{figure}
     \resizebox{\hsize}{!}{\includegraphics{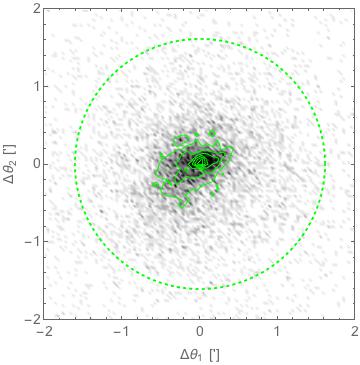}}
       \caption{{\it Chandra} exposure-corrected image in the [0.7--2]~keV band used for the spatial analysis. The dashed green circle of radius $\theta_{80\%}=1.61\arcmin$ encloses 80 per cent of the source light. Green contours are X-ray surface brightness contours at arbitrary levels. The image is centred on the optical centre. North is top and east is left.}
	\label{fig_M1206_SB_2D}
\end{figure}

\begin{figure}
       \resizebox{\hsize}{!}{\includegraphics{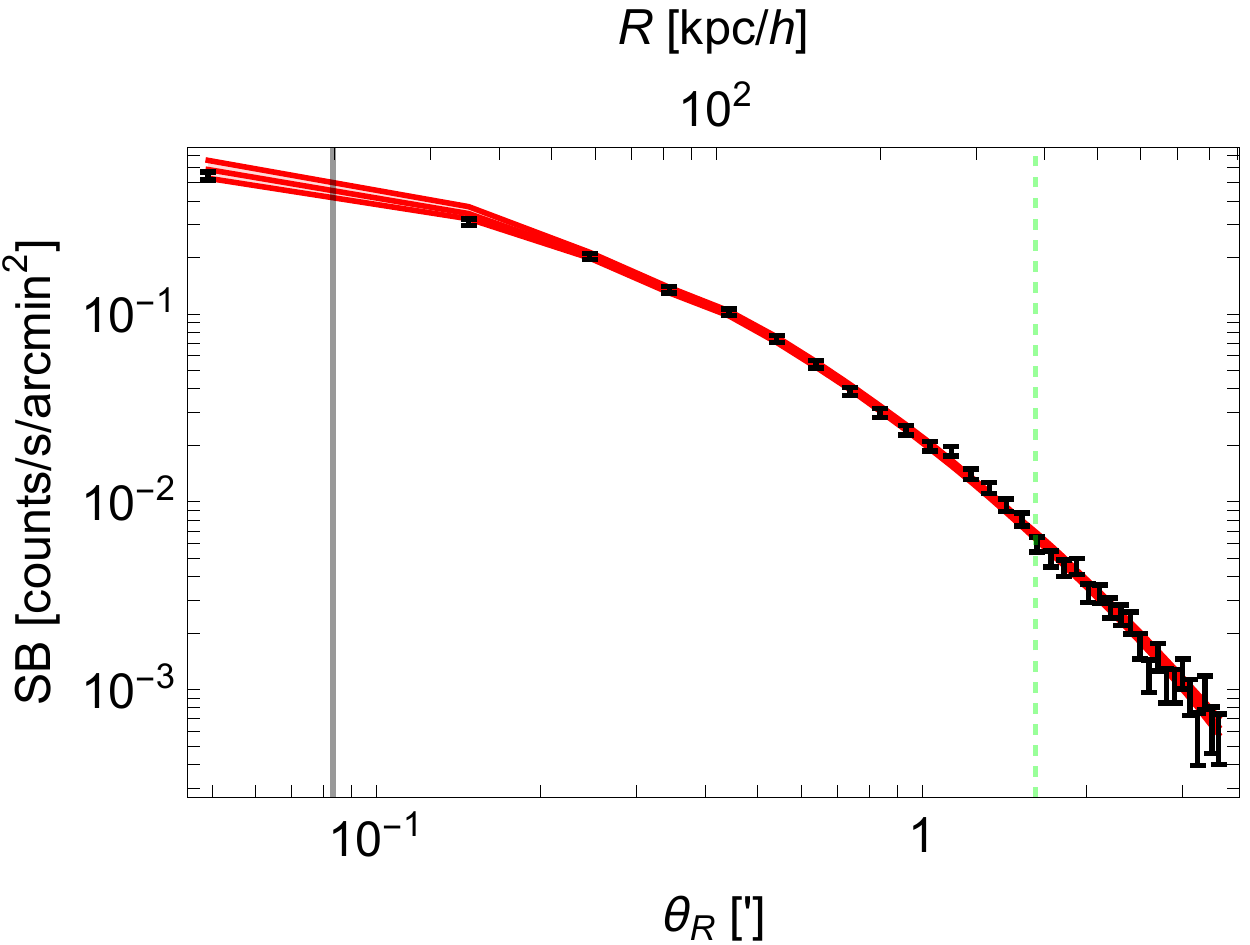}}
       \caption{Radial profile along the projected spherical axis of the [0.7--2]~keV surface brightness measured by {\it Chandra} (black points with error bars) in circular annuli. The innermost vertical (black) line is at $5\arcsec$, i.e., the minimum radius of the fitting region. The outermost (dashed green) line is at $R_{80\%}$, the radius of the circle including 80\% of the total cluster emission. The red line and the shadowed region plot the predicted median profile and the 68.3 per cent region obtained assuming a flat prior for the axis ratio $q_{\text{mat},1}$ of the matter distribution and $q_\text{ICM}\ge q$. The red line is not a fit to the plotted points, but it stems from the combined fit to all probes.}
	\label{fig_M1206_SB_profile}
\end{figure}

\begin{figure}
       \resizebox{\hsize}{!}{\includegraphics{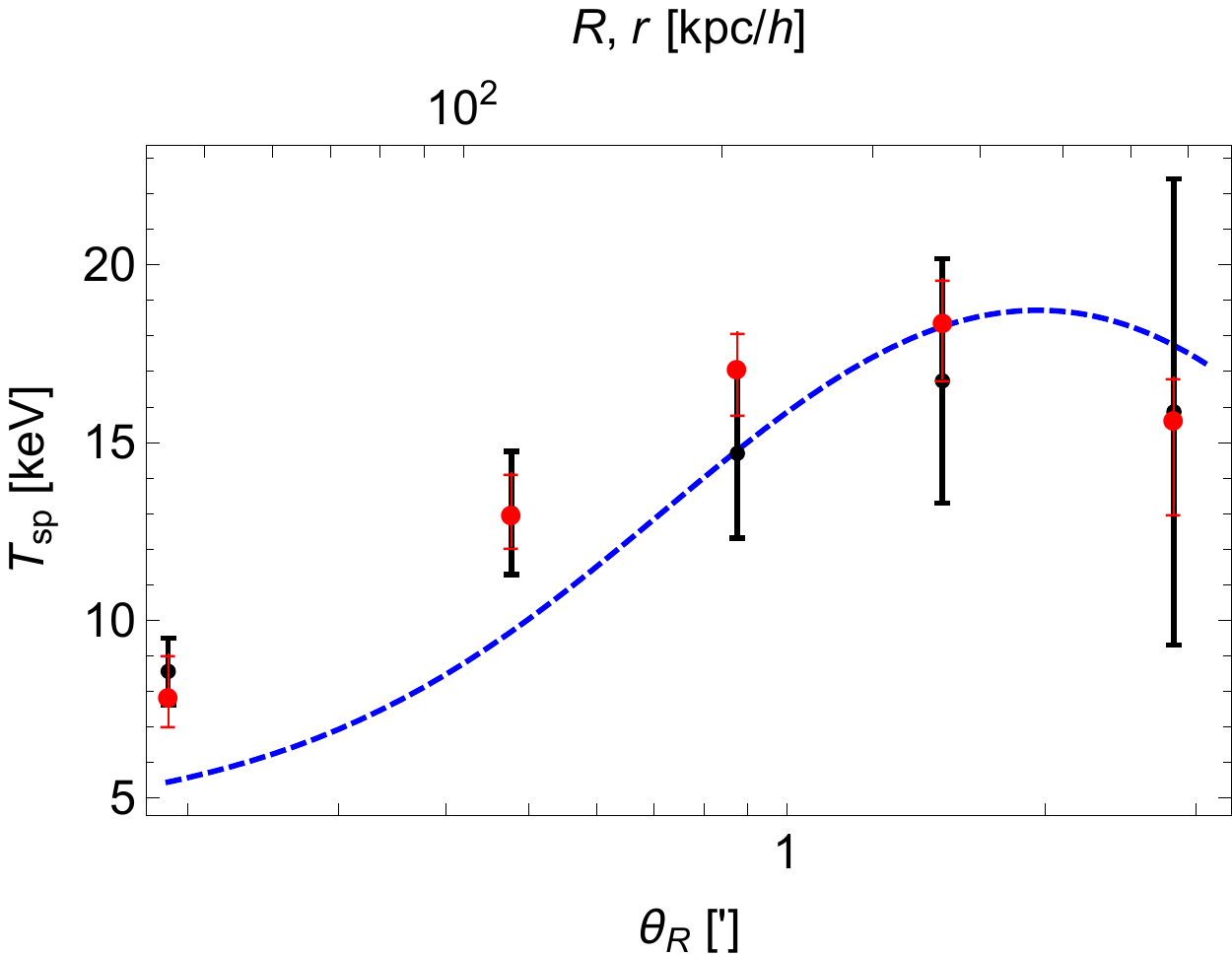}}
       \caption{Radial profile of the projected temperature measured by {\it Chandra} (points with error bars) in circular annuli. Red points and bars denote the median spectroscopic-like temperatures and 68.3 per cent uncertainty obtained assuming a flat prior for the axis ratio $q_{\text{mat},1}$ of the matter distribution and $q_\text{ICM}\ge q$. The projected spectroscopic-like temperature is a function of the spherical radius $r$. The dashed blue line represents the temperature of the gas distribution $T_\mathrm{e}$ as measured along the ellipsoidal radius $\zeta$. The plotted lines stems from the combined fit to all probes.}
	\label{fig_M1206_Te_profile}
\end{figure}

In the X-ray band, we make use of the archived {\it Chandra} exposure of MACS1206 obtained in AO3 in ACIS-I configuration, see Fig.~\ref{fig_M1206_SB_2D}. Using CIAO 4.8 software \citep{fru+al06} and the calibration database CALDB 4.7.1, we prepared a cleaned (by grade, status, bad pixels, and time intervals affected from flares in the background count rate) events file for a total filtered exposure time of 22.9~ksec. The background has been extracted locally over three circular regions with radius of $2\arcmin$ located at $\sim 6$-$7\arcmin$ from the cluster X-ray peak. Exposure-corrected images in the [0.7-2] keV band are produced.

The point-sources identified with the tool {\tt wavedetct} were masked and their regions filled with values of counts from surrounding background areas through the CIAO tool {\tt dmfilth}. 

We performed the 2D analysis of the circular region enclosing 80 per cent of the total source emission, with radius $\theta_{80\%}=1.61\arcmin\sim385\text{kpc}/h$. Pixels were binned four by four, with a final resolution of $1.968\arcsec$. After the excision of the inner region of radius $5\arcsec$, we ended up with 7589 binned pixels. Outside the 80 per cent region, we extracted the surface brightness profiles in 20 circular annuli up to $\sim 3.5\arcmin$, see Fig.~\ref{fig_M1206_SB_profile}.

We combined the 2D and the 1D analysis of the surface brightness by summing the Cash, see Eq.~(\ref{eq_SB2}), and the $\chi^2$, see Eq.~(\ref{eq_SB4}), statistics\footnote{The left-hand side of Eq.~\ref{eq_SB5} is not properly a $\chi^2$.},
\beq
\label{eq_SB5}
\chi^2_\text{SB}= {\cal C}_\text{SB,2D}+\chi^2_\text{SB,1D}.
\eeq

Spectra were accumulated in 5 circular annuli up $\la 2.8\arcmin$ and fitted in XSPEC software package \citep[v.12.9,][]{arn96} with an absorbed thermal model represented by the components {\tt tbabs}, with a fixed Galactic absorption $n_H = 4.35 \times 10^{20}$ cm$^{-2}$ extrapolated from HI radio maps in \citet{kal+al05} and {\tt apec}, with redshift fixed to the value of 0.439 and leaving free 3 parameters: normalisation, temperature and metallicity. 

The same emission model with metallicity fixed to the median value, ${\cal Z}=0.29$, was used for regression. The temperature profile is shown in Fig.~\ref{fig_M1206_Te_profile}. 


\subsection{Bolocam}

\begin{figure}
       \resizebox{\hsize}{!}{\includegraphics{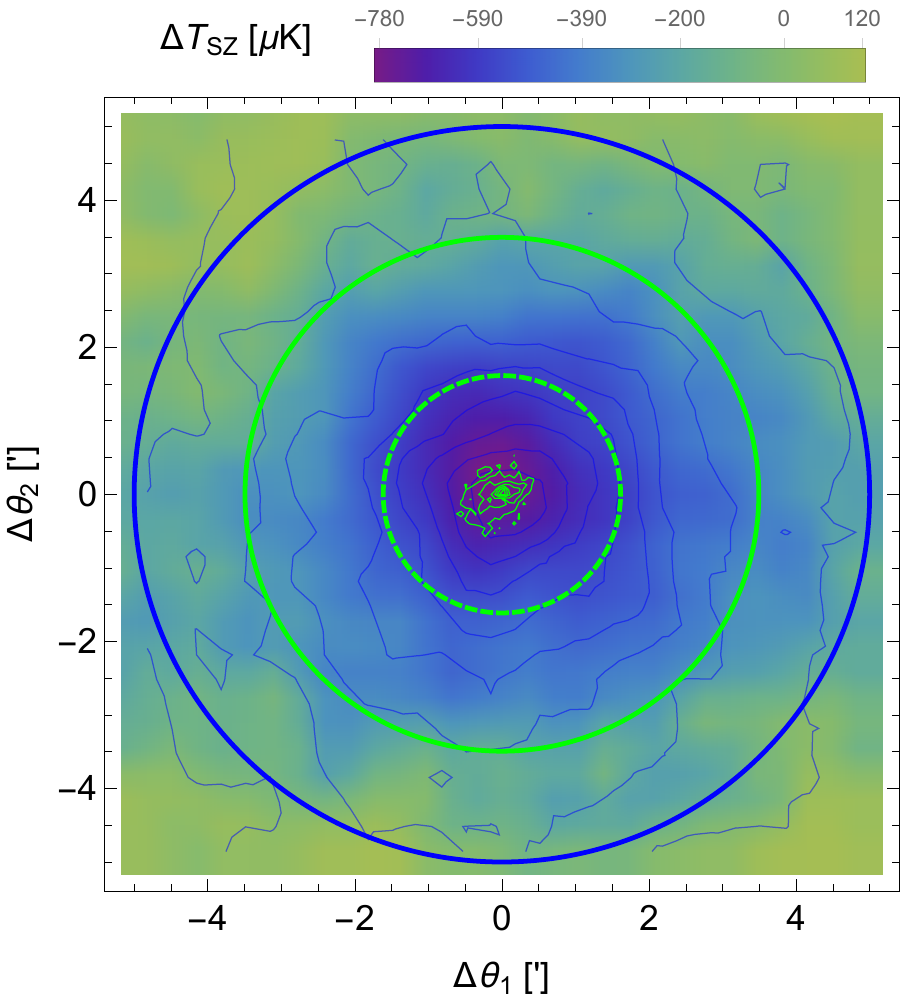}}
       \caption{Bolocam/{\it Planck} deconvolved SZe decrement image of MACS1206. The image is $10\arcmin\times10\arcmin$ in size and centred on the optical cluster centre. For visualisation purposes, the image was smoothed with a Gaussian kernel of standard deviation $1.0\arcmin$. The colour bar indicates the temperature decrement. The blue contours go from $\Delta T_\text{CMB}= -900$ to $0~\mu\text{K}$, with increments of $100~\mu\text{K}$. The blue circle at $5\arcmin$ encloses the region considered for inference. The dashed green circle of radius $\theta_{80\%}=1.61\arcmin$ encloses 80 per cent of the X-ray emission. Green contours are X-ray surface brightness contours at arbitrary levels. North is top and east is left.}
	\label{fig_M1206_DeltaTSZ_2D}
\end{figure}

\begin{figure}
       \resizebox{\hsize}{!}{\includegraphics{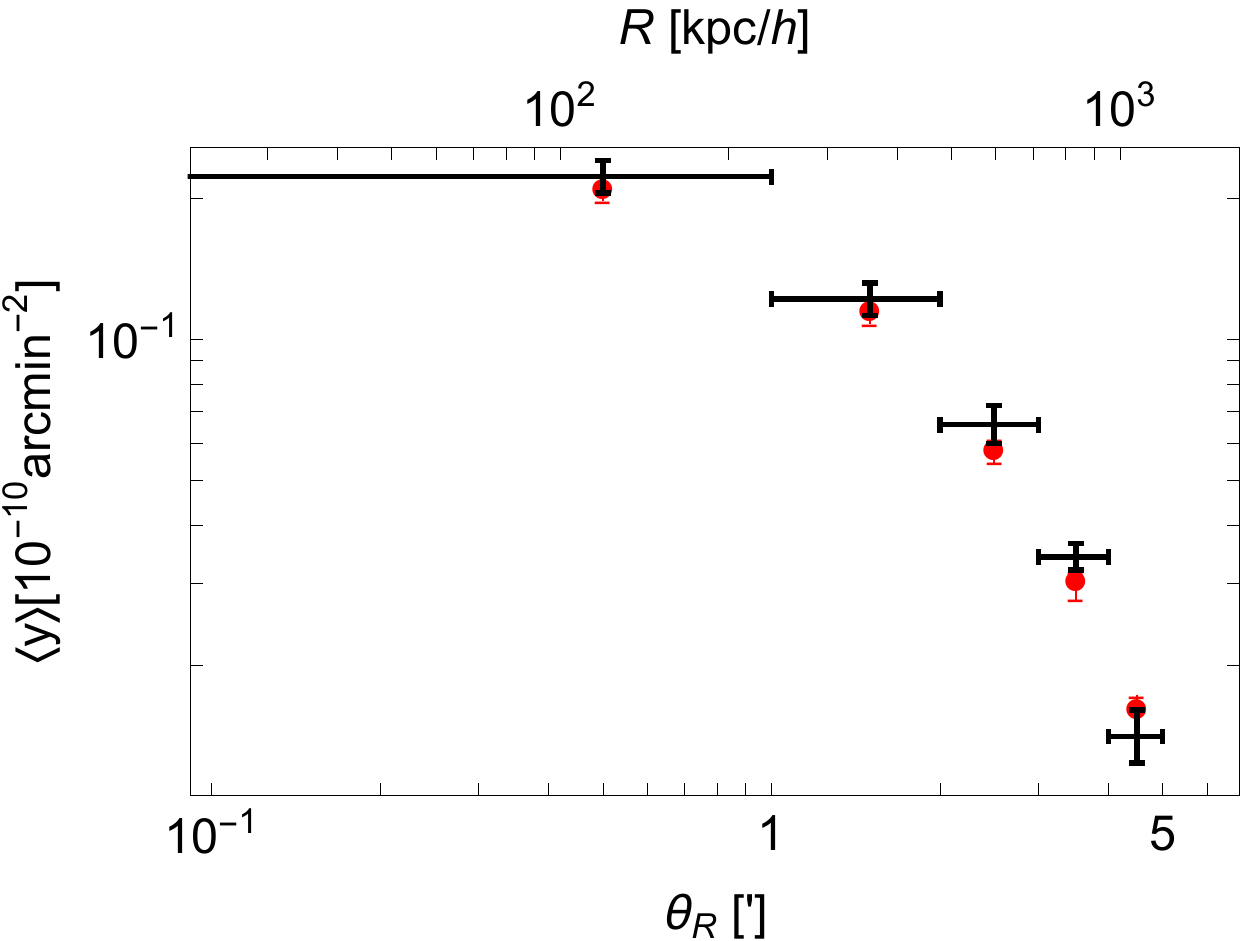}}
       \caption{Mean Compton parameter in circular annuli. Black points denotes the combined Bolocam/{\it Planck} data, the horizontal black bar cover the radial range of the annulus. Red points and bars denote the median fitted values obtained assuming a flat prior for the axis ratio $q_{\text{mat},1}$ of the matter distribution and $q_\text{ICM}\ge q$. The plotted lines stems from the combined fit to all probes.}
	\label{fig_M1206_ySZ_profile}
\end{figure}

MACS1206 belongs to the Bolocam X-ray SZ (BoXSZ) sample, a set of 47 clusters with publicly available data from Bolocam\footnote{\url{http://irsa.ipac.caltech.edu/data/Planck/release\_2/ancillary-data/bolocam/}} and {\it Chandra}. 

Due to the filtering applied to the Bolocam data, the mean signal level of the maps is unconstrained by the Bolocam alone. Because the maps are not in general large enough to include areas outside the cluster with negligible SZ emission, there is no way to directly determine this mean level. As a result, the mean signal level in the publicly available images was constrained based on a parametric model fit to the data, which effectively provided an extrapolation beyond the edges of the map to regions where the SZ signal is approximately zero, see \citet{cza+al15} for more details. In general, this method results in statistical uncertainties on the mean signal level that are significant when computing aperture fluxes from the maps. Furthermore, because the parametric model must be extrapolated beyond the edge of the data, there is the potential for un-quantified systematic errors if the cluster is not well described by the model in the extrapolation region. 

For this analysis, external SZ measurements from the {\it Planck} all-sky survey were employed to obtain more precise mean signal estimates. Specifically, the parametric model used to set the mean signal level was jointly constrained by the Bolocam data and the publicly available {\it Planck} $y$-maps\footnote{\url{http://irsa.ipac.caltech.edu/data/Planck/release\_2/docs/}} according to the procedures described in \citet{say+al16}. Furthermore, the same F-test procedure described in \citet{cza+al15} was used to determine the minimal parametric model required by the joint data, which was a spherical model with a floating scale radius (the Bolocam data alone required a spherical model with a fixed scale radius). Because the {\it Planck} $y$-maps extend well beyond the edge of the cluster, this joint model can be constrained without extrapolation. Compared to the publicly available Bolocam data, the updated mean signal level is shifted upward by approximately 30 $\mu$K$_\text{CMB}$, resulting in a 5\% reduction in the absolute value of the peak SZ signal (which is negative at the Bolocam wavelength).

For our analysis we use this updated unfiltered Bolocam map, which contains the SZ image after deconvolving the effects of the filtering due to the Bolocam data processing. As a result, this image provides a representation of the SZ signal from the cluster suitable for aperture photometry. The details of the reduction are given in \citet{say+al11,cza+al15}. Since the deconvolution results in significant noise on large angular scales, the image is truncated to a size of $10\arcmin\times10\arcmin$, see Fig.~\ref{fig_M1206_DeltaTSZ_2D}.

The aperture flux was computed in circular annuli based on all of the pixels with centres falling inside each given annulus. Due to the relatively coarse pixelisation of the map, a geometric correction factor is then applied to the flux values. Specifically, the deficit or excess of area at the inner or the outer border with respect to the smooth circular area is computed and $Y$ is corrected by attributing to the area difference the median $y$ value measured at the respective border. In this analysis, we measured the integrated Compton parameter in 5 equally spaced annular bins up to a maximum radius of $5\arcmin$, see Fig.~\ref{fig_M1206_ySZ_profile}. The width of the annuli is $1\arcmin$, comparable to the PSF FWHM, in order to limit correlation effects.

If the integration radius is not significantly larger than the Bolocam PSF, some of the SZe emission within the aperture appears outside due to beam smearing and some of the emission from the outside is mapped inside (mostly at the inner border). As a result, the estimates of $Y$ obtained from directly integrating the images can be biased. To estimate the boosting factor, we followed \citet{cza+al15}. The Compton flux  is computed both before and after deconvolution. The convolution Gaussian kernel accounts for both the Bolocam PSF (58\arcsec FWHM) and the pointing accuracy (5\arcsec). The Bolocam measured value is then corrected by the ratio of the $Y$ values determined from the un-smoothed and beam-smoothed maps.

For precise error estimates, we computed the aperture flux using identical apertures for the 1000 noise realisations of the Bolocam maps. These realisations fully encapsulate all of the noise statistics of the data, including instrumental, atmospheric, and astronomical sources of noise, and all pixel-pixel correlations due to these noise sources. Furthermore, these realisations also include fluctuations based on the mean signal level, which have been updated according to the joint Bolocam/{\it Planck} fitting procedure described above. The distribution of these 1000 values was then used to measure the uncertainty covariance matrix for the aperture fluxes within the annuli.


%


\subsection{Lensing}

\begin{figure}
       \resizebox{\hsize}{!}{\includegraphics{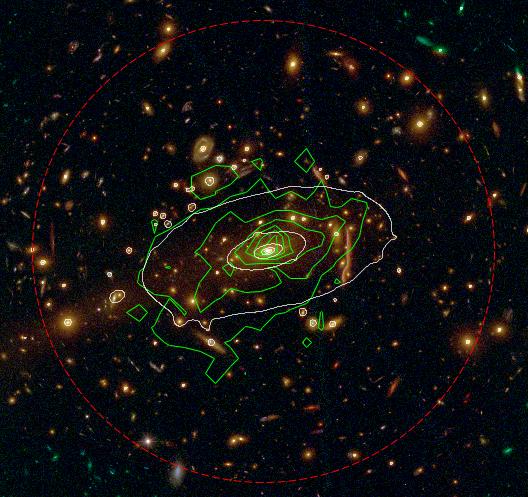}}
       \caption{{\it HST} composite colour images centred of the core of the galaxy cluster MACS1206. The dashed red circle at $1\arcmin$  encloses the region considered for the strong lensing analysis. The mass contours from the analysis of {\it HST} observations from \citet{zit+al15} based on the NFW parameterisation are overlaid in white. The mass contours follows the convergence and go from $\kappa= 0.1$ to 1.0, with increments of 0.1 (for a reference source redshift of $z_\text{s}=20000$). The green contours are the smoothed arbitrary levels of the X-ray surface brightness. North is top and east is left.}
	\label{fig_M1206_hst}
\end{figure}

\begin{figure}
       \resizebox{\hsize}{!}{\includegraphics{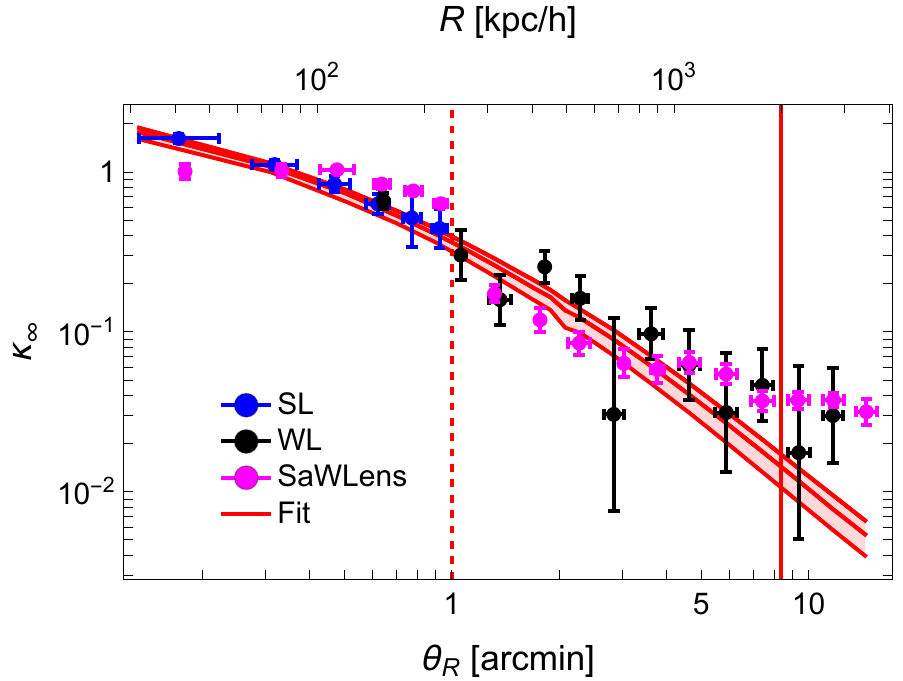}}
       \caption{Radial profile of the convergence map in circular annuli. Convergence is rescaled to $z=20000$. The black and blue points are obtained from the reference weak and strong lensing analysis, respectively. The magenta points were obtained with SaWLens. The vertical lines denote the fitting regimes. The innermost is at $1\arcmin$, i.e., the maximum radius for SL. The outermost is at $2~\text{Mpc}/h$, which include the considered WL region. The red profile and the shadowed region plot the median profile and the 68.3 per cent region around the median obtained assuming a flat prior for the axis ratio $q_{\text{mat},1}$ of the matter distribution and $q_\text{ICM}\ge q$. The red line is not a fit to the plotted points, but it stems from the combined fit to all probes.}
	\label{fig_M1206_kappa_profile}
\end{figure}

The lensing features of MACS1206 were observed and studied in the context of the CLASH collaboration. We refer to \citet{ume+al12,mer+al15,ume+al16} for a full presentation of the data and of the methods.

\subsubsection{Strong lensing}

For our analysis, we considered the joint analysis of the inner regions of MACS1206 based on {\it HST} data presented by \citet{zit+al15}\footnote{The mass models are available through the Hubble Space Telescope Archive at \url{https://archive.stsci.edu/prepds/clash/}.}. Constraints in the inner core come mostly from strong-lens modelling of multiple image systems identified with deep {\it HST} imaging and VLT/VIMOS spectroscopy. MACS1206  hosts a well known giant arc system at $z_\text{s} = 1.03$ \citep{ebe+al09}. \citet{zit+al12} identified 12 more candidate multiple image systems of distant sources, bringing the total known for this cluster to 50 multiply lensed images of 13 sources. The images cover fairly evenly the central region, $3\arcsec \la \theta \la 1\arcmin$, and span a wide redshift range of $1\la z_\text{s}\la5.5$. 

Additional constraints comes from weak lensing in the {\it HST} field, which exploits the high density of lensed background galaxies, $\sim 50~\text{gal/arcmin}^2$. 

The analysis was performed in two distinct parameterisations: the first one adopts light-traces-mass for both galaxies and dark matter \citep{zit+al09b} while the other assumes an analytical, elliptical NFW form for the dark matter halo components \citep{zit+al13}. As a reference, we considered the NFW parametrisation, which conforms better to our modelling. It does not employ any external shear and it lets the DM halos to be elliptical.

The very inner region of the cluster core is dominated by the BCG and by massive ellipticals, see Fig.~\ref{fig_M1206_hst}. Whereas the average mass distribution and radial slope follow the general profile, local substructures can strongly impact the inferred local projected shape and orientation of the core. On the geometrical side, the analysis of the cluster core can deviate from the overall picture. 

We did not consider the full fine-resolution 2D map but we applied a conservative azimuthal binning scheme. We computed the mean convergence in 6 equally spaced angular annuli centred in the optical centre. The innermost and the outermost radii were set to an angular scale of 5\arcsec and 1\arcmin, respectively. 

This conservative approach for the SL analysis is also justified by the number of noise realisations is not big enough to properly compute the uncertainty covariance matrix of the 2D-SL map. 

The formal statistical errors of the models are under-estimated \citep{zit+al15}. The actual (and much larger) uncertainties that account for model-dependent systematics can be inferred by comparing the convergence profiles derived under the two distinct modellings from \citet{zit+al15}. The projected mass enclosed within the critical curves of the CLASH clusters agrees typically within $\sim15\%$, which gives an empirical assessment of the true underlying errors.

We computed the standard deviation of the differences of the convergences of the pixels in each annulus and added this error in quadrature to the formal statistical error. For this computation, original maps were rebinned with a final formal resolution of $0.65\arcsec$.

Since the uncertainty budget is dominated by this second term, we considered only the diagonal part of the covariance matrix.

The convergence profile is plotted in Fig.~\ref{fig_M1206_kappa_profile}. The SL- and WL-based convergences are in remarkable agreement in the overlapping region.

\subsection{Subaru Weak-lensing Analysis}

\begin{figure}
       \resizebox{\hsize}{!}{\includegraphics{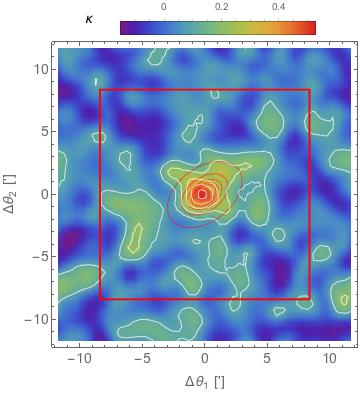}}
       \caption{Two-dimensional weak-lensing mass map of MACS1206 reconstructed using wide-field multi-colour imaging observations. The image is $24\arcmin\times24\arcmin$ in size and centred on the optical cluster centre. For visualisation purposes, the image was smoothed with a Gaussian kernel of standard deviation 1.5\arcmin. The colour bar indicates the lensing convergence $\kappa$ for a reference source at $z_\text{s}=20000$. The convergence isolevels go from $\kappa= 0.1$ to 1.0, with increments of 0.1. White and red contours follow the smoothed map and the predicted median NFW profile obtained assuming a flat prior for the axis ratio $q_{\text{mat},1}$ of the matter distribution and $q_\text{ICM}\ge q$, respectively. The red square of semi-size $2\text{Mpc}/h\sim8.39\arcmin$ encloses the WL region used for inference. North is top and east is left.}
	\label{fig_M1206_kappa_2D_WL}
\end{figure}

For our Subaru weak-lensing mapmaking, we used the nonparametric method of \citet{ume+al15a}, a two-dimensional extension of the Cluster Lensing Mass Inversion (\texttt{CLUMI}) code developed by \citet{ume+al11a}. We reconstructed the projected mass distribution around MACS1206 from a joint analysis of two-dimensional shear and azimuthally averaged magnification. The combined constraints allows us to break the mass-sheet degeneracy. Full details will be given in a forthcoming paper that will present a systematic two-dimensional weak-lensing analysis of 20 CLASH clusters (K. Umetsu et al.\, in preparation). Here, we briefly outline the methods and data used in our analysis.

\subsubsection{Data}

The weak lensing analysis of the wide field exploits Subaru data and photometry. MACS1206 was observed in deep $BVR_\text{c}I_\text{c}z'$ with the wide-field camera Suprime-Cam at the prime focus of the 8.3 m Subaru telescope. The mosaicked images cover a field of approximately $36\arcmin\times34\arcmin$ \citep{ume+al12}. 

In the present study, we used the weak-lensing shear and magnification data as obtained by \citet{ume+al14}. We refer the reader to \citet[section 4]{ume+al14} and \citet[sections 3.2--3.4] {ume+al16} for a summary of our Subaru data and data analysis. We performed a new two-dimensional shear analysis using  the shear catalog presented in \citet{ume+al14}. The magnification bias measurements used in this study are presented in \citet[figure 2]{ume+al14}.
The systematic uncertainty in the absolute mass calibration using the CLASH weak-lensing measurements was estimated to be $\sim 8\%$ \citep{ume+al14}. 

\subsubsection{Estimators and covariance matrix}

For mapmaking, the lensing field was pixelised into a regular Cartesian grid of $N_\text{pix}=48^2 = 2304$ independent pixels with $\Delta\theta=0.5\arcmin$ spacing, covering the central $24\arcmin\times 24\arcmin$ region. The magnification bias analysis in this region was performed by \citet{ume+al14}. 

The lensing signal is described by the vector  $\bs=\{\Sigma_m\}_{m=1}^{N_\text{pix}}$ whose elements contain the cell-averaged values of the two-dimensional surface mass density field $\Sigma(\btheta)$, sampled on the grid. Both the convergence, $\kappa(\btheta)$, and the shear, $\gamma(\btheta)$, fields can be written as linear combinations of $\bs$ \citep{ume+al15a}. In our approach, we combine the observed spatial distortion pattern $g(\btheta)$ with the azimuthally averaged magnification measurements $n(\theta)$, which impose a set of azimuthally integrated constraints on the mass distributions.

The magnification is locally related to $\kappa$ and the magnification constraints provide the otherwise unconstrained normalisation of $\Sigma(\btheta)$ over a set of concentric annuli where count measurements are available, thus effectively breaking the mass-sheet degeneracy. No assumption is made of azimuthal symmetry or isotropy of the underlying mass distribution.


The (complex) reduced shear field can be reconstructed from shape measurements of background galaxies. The weighted average $g_m\equiv g(\btheta_m)$ on the grid ($m=1,2,..., N_\text{pix}$) can be written as
\begin{equation}
\label{eq:bin_shear} 
g_m = \left[
\displaystyle\sum_k
S(\btheta_{(k)},\btheta_m)
w_{(k)}g_{(k)}
\right]
\left[
\displaystyle\sum_{k} 
S(\btheta_{(k)},\btheta_m)w_{(k)}
\right]^{-1} ,
\end{equation}
where $g_{(k)}$ is an estimate of $g(\btheta)$ for the $k$-th galaxy at $\btheta_{(k)}$. The statistical weight $w_{(k)}$ can be written as $w_{(k)} = 1/(\sigma^2_{g(k)}+\alpha^2_g)$, with $\sigma^2_{g(k)}$ the error variance of $g_{(k)}$, and $\alpha_g^2(=0.16)$ the softening variance \citep{ume+al14}, as typical of the mean variance $\overline{\sigma_g^2}$ found in Subaru observations \citep{ume+al09,oka+al10}. $S(\btheta_{(k)},\btheta_m)$ is the window function. We adopted a top-hat window of filtering radius $\theta_\mathrm{f}=0.4\arcmin$ \citep{ume+al14}.

The error covariance matrix for the weighted average $g_m$ is 
\begin{equation}
  \mathrm{Cov}(g_{\alpha,m},g_{\beta,n}) 
 \equiv 
  \delta_{\alpha\beta}\left(C_g\right)_{mn}
 = 
  \frac{\delta_{\alpha\beta}}{2} \sigma_{g,m} \sigma_{g,n}
  \xi_{H}(|\btheta_m-\btheta_n|),
\end{equation}
where the indexes $\alpha$, $\beta \in \{1,2\}$, $\sigma_{g,m}=\sigma_g(\btheta_m)$ denotes the error dispersion for $g_m=g_{1,m}+ig_{2,m}$ ($m=1,2,...,N_\text{pix}$), and $\xi_H(x; \theta_\mathrm{f})$ is the auto correlation of a pillbox of radius $\theta_\mathrm{f}$ normalised as $\xi_H(0;\theta_\mathrm{f})=1$ \citep[equation 11]{ume+al15a}. 

We interpreted the observed weak-lensing signal, Eq.~\ref{eq:bin_shear}, accounting for the nonlinear effect on the source-averaged reduced shear. The signal can be written as
\citep{se+sc97},
\begin{equation}
\label{eq:g_ave}
\hat{g}(\btheta_m) = \frac{\gamma(\btheta_m)}{1-f_{W,g} \,\kappa(\btheta_m)},
\end{equation}
where $f_{W,g}=\langle \beta^2\rangle_g/\langle \beta\rangle_g^2$ is a correction factor of the order unity estimated from the source-averaged
lensing depths
$\langle\beta\rangle_g=\langle D_\mathrm{ls}/D_\mathrm{os}\rangle_g$
and
$\langle\beta^2\rangle_g=\langle (D_\mathrm{ls}/D_\mathrm{os})^2\rangle_g$ for the shear weak-lensing analysis. We used $\langle\beta\rangle_g=0.54\pm 0.03$ and $f_{W,g}=1.06$ \citep[table 3]{ume+al14}.

We excluded from our analysis the pixels within the Einstein radius $\theta_\text{Ein}(z_\text{s}=2)=26.8\arcsec$ \citep[table 1]{ume+al16}, where $\Sigma(\btheta)$ can be close to or greater than the critical value $\Sigma_\mathrm{crit}$, as well as those containing no background galaxies with usable shape measurements. For distortion measurements, $g_1(\btheta) and g_2(\btheta)$, a total of 2293 measurement pixels is usable, corresponding to 4586 constraints.


For magnification measurements, we used the azimuthally averaged source number counts $\{n_{\mu,i}\}_{i=1}^{N_\text{bin}}$ and their total errors $\{\sigma_{\mu,i}\}_{i=1}^{N_\text{bin}}$ as obtained by \citet{ume+al14} using their flux-limited sample of $BR_\mathrm{C}z'$-selected red background galaxies. For additional analysis details, we refer to \citet[section 3.4]{ume+al16}. The magnification constraints were measured in $N_\text{bin}=10$ log-spaced circular annuli centred on the cluster, spanning the range $[\theta_\mathrm{min},\theta_\mathrm{max}]=[0.9\arcmin,16\arcmin]$ with a constant logarithmic spacing, $\Delta\ln\theta\simeq 0.29$.

The theoretical azimuthally averaged source number counts is \citep{ume+al15a,ume+al16}
\begin{equation}
 \label{eq:nb_th}
\hat{n}_{\mu,i} =
\overline{n}_\mu
\sum_{m} {\cal P}_{im} \mu^{2.5s-1}(\btheta_m),
\end{equation}
where $\mu=[(1-\kappa)^2-|\gamma|^2]^{-1}$ is the lensing magnification in the subcritical regime, $\overline{n}_\mu$ is the (unlensed) mean background surface number density of source galaxies, and $s$ is the logarithmic count slope evaluated at the fainter magnitude limit $m_\mathrm{lim}$, $s=[d\log_{10}N(<m)/dm]_{m=m_\mathrm{lim}}$; ${\cal P}_{im}=(\sum_m A_{mi})^{-1}A_{mi}$ is the projection matrix normalised in each annulus as $\sum_m {\cal P}_{im}=1$. Here, $A_{mi}$ represents the fraction of the area of the $m$-th cell lying within the $i$-th annular bin  ($0\le A_{mi} \le 1$).

The background sample used for lensing magnification differs from the galaxy population used for shape measurements. The mean lensing depth of the background sample is $\langle\beta\rangle_\mu=0.51\pm 0.03$  with $z_\text{eff}=1.04$ \citep[table 4]{ume+al14}. The count normalisation and slope parameters are $\overline{n}_\mu=(11.4\pm 0.4)$\,arcmin$^{-2}$ and $s=0.13\pm 0.05$ at the fainter magnitude limit $z'=24.6$\,ABmag \citep[table 4]{ume+al14} using the source counts in the outskirts at $[10\arcmin,\theta_\text{max}]$  \citep{ume+al14}.

\subsubsection{Joint reconstruction of the mass map}

We reconstructed the mass distribution of MACS1206 from a joint likelihood analysis of the shear and magnification measurements, $\{g_{1,m},g_{2,m}\}_{m=1}^{N_\text{pix}}$  and $\{n_{\mu,i}\}_{i=1}^{N_\text{bin}}$. The model $\bm$ is specified by $\bs$ with $N_\text{pix}=2304$ signal parameters and a set of calibration parameters to marginalise over, $\bc=\{\langle\beta\rangle_g, f_{W,g}, \langle\beta\rangle_\mu,\overline{n}_\mu, s\}$. We have a total of $N_\mathrm{data}=4586 + 10 = 4596$ constraints, yielding $N_\text{data}-N_\text{pix}=N_\mathrm{dof}=2292$ degrees of freedom.

The posterior probability of a model $\bm$ is proportional to the product of the likelihood ${\cal L}(\bm)$ and the prior probability. We used Gaussian priors on the calibration nuisance parameters $\bc$, given in terms of quadratic penalty terms with mean values and errors estimated from data \citep{ume+al14} as stated above.

The joint likelihood function ${\cal L}(\bm)={\cal L}_g(\bm){\cal L}_\mu(\bm)$ for combined weak-lensing data is given as the product of the likelihood functions for shear, ${\cal L}_g$, and magnification, ${\cal L}_\mu$. 
The shear log-likelihood function $l_g\equiv -2{\cal L}_g$ can be written as
\begin{equation}
\begin{aligned}
l_g  = & \frac{1}{2}
 \sum_{m,n=1}^{N_\text{pix}}
 \sum_{\alpha=1}^{2}
[g_{\alpha,m}-\hat{g}_{\alpha,m}(\bm)]
\left({\cal W}_g\right)_{mn}%
 [g_{\alpha,n}-\hat{g}_{\alpha,n}(\bm)] \\
 & + \text{const.},
\end{aligned}
\end{equation}
where $\hat{g}_{\alpha,m}(\bm)$ is the theoretical expectation for $g_{\alpha,m}=g_{\alpha}(\btheta_m)$; ${\cal W}_g$ is the shear weight matrix,
\begin{equation}
\left({\cal W}_g\right)_{mn} = M_m M_n \left(C_g^{-1}\right)_{mn},
\end{equation}
where $M_m$ is the mask weight, defined such that $M_m=0$ if the $m$-th cell is masked out and $M_m=1$ otherwise.
 
The log-likelihood function for magnification-bias data, $l_\mu\equiv -2\ln{\cal L}_\mu$, is given by
\begin{equation}
l_\mu= \frac{1}{2}\sum_{i=1}^{N_\text{bin}}
\frac{[n_{\mu,i}-\hat{n}_{\mu,i}(\bm)]^2}{\sigma_{\mu,i}^2} + \text{const.},
\end{equation}
where $\hat{n}_{\mu,i}(\bm)$ is the theoretical prediction for the observed counts $n_{\mu,i}$, see Eq.~(\ref{eq:nb_th}). Following \citet{ume+al15a}, we used Monte Carlo integration to calculate the projection matrix ${\cal P}_{im}$ of size $N_\text{bin}\times N_\text{pix}$, which is needed to predict $\{n_{\mu,i}\}_{i=1}^{N_\text{bin}}$ for a given $\bm=(\bs,\bc)$.

The negative log-posterior function $F(\bm)$ is expressed as the linear sum of the log-likelihood functions ($l_g,l_\mu$) and the Gaussian prior terms. The {\em best-fit} solution, $\hat{\bm}$ corresponds the global maximum of the joint posterior distribution. The resulting mass map is plotted in Fig.~\ref{fig_M1206_kappa_2D_WL}.

Uncertainties are estimated by evaluating the Fisher matrix at $\bm=\hat{\bm}$ as \citep{ume+al15a}
\begin{equation}
{\cal F}_{pp'} = 
\left\langle \frac{\partial^2 F(\bm)}{\partial m_p \partial m_{p'}}
\right\rangle\Big|_{\bm=\hat{\bm}}
\end{equation}
where the angular brackets represent an ensemble average over all possible (an infinite number of) noise realizations, and the indices $(p,p')$ run over all model parameters $\bm=(\bs,\bc)$.

The posterior covariance matrix is then obtained as
\begin{equation}
\mathrm{Cov}(m_p,m_{p'})\equiv \left(C^\mathrm{stat}\right)_{pp'} = \left({\cal F}^{-1}\right)_{pp'}.
\end{equation}

Additionally, we accounted for the uncorrelated large scale structures projected along the line of sight. The cosmic noise covariance matrix $(C^\mathrm{lss})_{mn}$ for the pixelised surface mass density distribution $\bs=\{\Sigma_m\}_{m=1}^{N_\mathrm{pix}}$ was computed by projecting the nonlinear matter power spectrum of \citet{smi+al03} for the {\em Wilkinson Microwave Anisotropy Probe} ({\em WMAP}) seven-year cosmology \citep{kom+al11}. 

The total covariance matrix for the pixelised surface mass density distribution is $(C)_{mn}=(C^\mathrm{stat})_{mn}+(C^\mathrm{lss})_{mn}$.

\subsubsection{Parametric analysis of the mass map}

We finally fitted the convergence map to constrain the parametric mass distribution. Following \citet{mer+al15}, we set the fitting area to a square with physical scale of $2~\text{Mpc}/h$ centred in the optical centre. 

We summed the $\chi^2$ contributions from the fine resolution grid dominated by the strong lensing observations $\chi^2_\text{SL}$ and the term from the weak lensing observations by the Subaru telescope, $\chi^2_\text{WL}$,
\beq
\chi^2_\text{GL}=\chi^2_\text{SL}+\chi^2_\text{WL}.
\eeq

\subsection{Cluster centre}

Misidentification of the cluster centre is a potential source of systematic errors for joint, multi-wavelength analyses. We fixed the cluster centre at the sky position of the brightest cluster galaxy (BCG) of R.A.=12:06:12.15, DEC=--08:48:03.4 (J2000). Miscentring effects are small in MACS1206. The BCG position agrees with the peak of X-ray emission (R.A.=12:06:12.08, DEC=--08:48:02.6) within 1.3\arcsec or a projected offset distance of $\sim5~\text{kpc}/h$ at the cluster redshift, see Fig.~\ref{fig_M1206_hst}. Furthermore the BCG and the peak of the total mass distribution are off-set by just 1\arcsec, well within the uncertainties \citep{ume+al12}, see Figs.~\ref{fig_M1206_rgb} and \ref{fig_M1206_hst}. 

In the present work, we conservatively limited our analysis to radii greater than 5\arcsec ($\sim20~\text{kpc}/h$), which slightly exceeds the location of the innermost strong-lensing constraint and is sufficiently large to avoid the BCG contribution. Our inner radial limit corresponds roughly to 4 times the offset between the BCG centre and the X-ray peak, beyond which smoothing from the cluster miscentring effects on the convergence profile are sufficiently negligible \citep{joh+al07,ume+al12}.


\section{Inference}
\label{sec_infe}

The problem of finding the volume density distribution of halos whose projected isocontours are similar ellipses has no unique solution \citep{sta77}.

In the generic terms, we have to constrain the five unknown geometrical intrinsic properties (two axis ratios and three orientation angles) of the ellipsoidal halo. However, we can only measure three observable quantities, i.e. the ellipticity $\epsilon$, the orientation $\theta_\epsilon$ and the elongation $e_\parallel$, see App.~\ref{sec_proj}. Ellipticity and orientation can be inferred from a single map whereas the elongation can be derived by combining data sets with a different dependence on density. For example, we can combine X-ray and SZ to directly infer the elongation of the gas distribution. For an isothermal plasma \citep{def+al05},
\beq
\label{eq_infe_1}
e_\parallel = \frac{1}{D_\text{d}}\frac{\Delta T^2_\text{SZ,0}}{\text{SB}_0}\frac{\Lambda}{T^2} .
\eeq
No assumption is needed about hydrostatic equilibrium but clumpiness and contamination from structures along the line of sight can bias the result.

The problem is then under-constrained even with an ideal multi-probe data-set without noise \citep{ser07}.

However, as far as the effect on observations is considered, some intrinsic geometrical parameters are more equal than others \citep{orw49,se+um11,ser+al12a}. In a triaxial analysis, ellipticity and elongation strongly depend on the minor to major axis ratio, $q_{\text{mat},1}$, and on the inclination angle, $\vartheta$. The projected orientation angle is just $\psi$ plus an arbitrary constant. That makes three main parameters for three observables.

Bayesian inference is suitable to cluster deprojection. Whereas some main parameters are essentially derived from the data, others can be more subject to priors. For example, we obtain similar results for the minor-to-major axial ratio and the halo orientation either using $q_{\text{mat},1}$, and $q_\text{ICM,1}$ as free parameters and $q_{\text{mat},2}$, and $q_\text{ICM,2}$ as functions of the other axis ratios or considering all four axis ratios as free.

To assess realistic probability distributions for the parameters we exploited the Bayes' theorem, which states that
\beq
\label{eq_infe_2}
p(\bfP | \bfd) \propto {\cal L}( \bfP|\bfd) p(\bfP),
\eeq
where $p(\bfP | \bfd)$ is the posterior probability of the parameters $\bfP$ given the data $\bfd$, ${\cal L}( \bfP|\bfd)$ is the likelihood of the data given the model parameters and $p(\bfP)$ is the prior probability distribution for the model parameters. 

For our multi-wavelength analysis, the likelihood can be written as \citep{ser+al13},
\beq
\label{eq_infe_3}
{\cal L} \propto \exp\left[-\left(\chi^2_\text{GL}+\chi^2_\text{SB}+\chi^2_\mathrm{T}+\chi^2_\text{SZ}\right)/2\right]. 
\eeq
Our method relies on a minimum number of assumption, e.g. we do not require equilibrium. Under this general setting, the matter part (GL) and the ICM part (X+SZ) communicate only through the orientation, which is shared by gas and matter, and the shape parameters, whose relations is defined through priors.

Under the strong assumption of spherical symmetry, the mass and the concentration, which only appear in the $\chi^2_\text{GL}$ are determined by lensing alone. Under the triaxial assumption, the inference of mass and concentration is affected by the X+SZ part too through the shape and orientation parameters. In fact, the estimate of the concentration is strongly correlated with shape and orientation \citep{ogu+al05,ser+al10a}.

The present analysis presents some major developments with respect to the methodology used in \citet{ser+al13}. The priors cover a larger range of scenarios than those used in \citet{ser+al13} and are updated to latest results from numerical simulations. Priors for shape and orientation have been introduced in Secs.~\ref{sec_tria_mat} and \ref{sec_tria_gas}. 

Furthermore, we now fit the 2D map of the X-ray surface brightness rather than the averaged 1D profile only.

The last major development with respect to \citet{ser+al13,ume+al15a} is in our treatment of the likelihood. \citet{ser+al13} used a step procedure, where the data were first fitted in terms of projected parameters ($\kappa_\text{s}$, $\epsilon_\text{mat}$, $e_{\text{ICM},\parallel}$, ...) and then the inferred probability of the projected parameters (approximated with either a smooth kernel distribution or a multi-variate Gaussian distribution) was used as likelihood to infer the intrinsic parameters ($M_{200}$, $q_{\text{mat},1}$, $\cos \vartheta$, ...). Now, we fit the data in terms of the intrinsic parameters in a single step, see Eq.~(\ref{eq_infe_3}).

\section{Results}
\label{sec_resu}

\begin{table*}
\caption{Results of the regression under different priors for the mass and gas shape. For the mass, we considered either spherical, flat or $N$-body priors. For the gas, we considered $q_\text{ICM}\ge q$ and, additionally, either $\mathcal{T}_\text{mat}=\mathcal{T}_\text{ICM}$ or $q_\text{ICM}=q_\Phi$. The reference case, i.e., $q$-flat and $q_\text{ICM}\ge q$, is listed in Col.~2. Parameters in brackets are held fixed. Units are as in Table~\ref{tab_parameters}. Typical values and dispersions are computed as bi-weighted estimators of the marginalised posterior distributions.}
\label{tab_results}
\begin{tabular}{ l r@{$\,\pm\,$}l r@{$\,\pm\,$}l r@{$\,\pm\,$}l r@{$\,\pm\,$}l r@{$\,\pm\,$}l}     
\hline
			&		\multicolumn{2}{c}{$q$-flat}	&  \multicolumn{2}{c}{$q$-spherical}&	\multicolumn{2}{c}{$q$-flat}	&	\multicolumn{2}{c}{$q$-flat}	&	\multicolumn{2}{c}{$q$-$N$body}	\\
	&	\multicolumn{2}{c}{$q_\text{ICM}\ge q$}	&  \multicolumn{2}{c}{}&	\multicolumn{2}{c}{$\mathcal{T}_\text{mat}=\mathcal{T}_\text{ICM}$}	&	\multicolumn{2}{c}{$q_\text{ICM}=q_\Phi$}	&	\multicolumn{2}{c}{$q_\text{ICM}\ge q$}	\\ 
	\hline
$M_{200}$                       	&	1.137  	&	0.229 	&	1.057	&	0.157	&	1.027  	&	0.362	&	0.951  	&	0.242	&	1.069  	&	0.238	\\
$c_{200}$                       	&	6.277  	&	1.188 	&	5.047	&	0.730	&	6.682  	&	1.823	&	6.469  	&	1.537	&	6.051  	&	1.126	\\
$q_{\text{mat},1}$              	&	0.466  	&	0.119 	&	\multicolumn{2}{c}{[1]}	&	0.406  	&	0.136	&	0.409  	&	0.111	&	0.439  	&	0.077	\\
$q_{\text{mat},2}$              	&	0.735  	&	0.176 	&	\multicolumn{2}{c}{[1]}	&	0.759  	&	0.143	&	0.544  	&	0.096	&	0.586  	&	0.098	\\
$\cos \vartheta$                	&	0.297  	&	0.204 	&	\multicolumn{2}{c}{...}	&	0.319  	&	0.217	&	0.220  	&	0.184	&	0.272  	&	0.194	\\
$\varphi$                       	&	-0.438 	&	1.609 	&	\multicolumn{2}{c}{...}	&	-0.075 	&	1.257	&	0.147  	&	1.058	&	0.154  	&	1.474	\\
$\psi$                          	&	1.027  	&	0.123 	&	\multicolumn{2}{c}{...}	&	0.972  	&	0.278	&	0.937  	&	0.061	&	0.931  	&	0.152	\\
$q_\text{ICM,1}$                	&	0.587  	&	0.109 	&	\multicolumn{2}{c}{...}	&	0.498  	&	0.119	&	0.734  	&	0.055	&	0.501  	&	0.071	\\
$q_\text{ICM,2}$                	&	0.779  	&	0.057 	&	\multicolumn{2}{c}{...}	&	0.783  	&	0.113	&	0.791  	&	0.052	&	0.784  	&	0.070	\\
$n_0$                           	&	0.010  	&	0.001 	&	\multicolumn{2}{c}{...}	&	0.011  	&	0.002	&	0.010  	&	0.001	&	0.011  	&	0.002	\\
$\zeta_\text{c}$                	&	169.780	&	12.02	&	\multicolumn{2}{c}{...}	&	167.300	&	8.520	&	163.900	&	3.466	&	166.430	&	6.748	\\
$\zeta_\text{t}/\zeta_\text{c}$ 	&	8.250  	&	1.635 	&	\multicolumn{2}{c}{...}	&	8.125  	&	1.358	&	7.576  	&	1.563	&	7.921  	&	1.663	\\
$\beta$                         	&	0.600  	&	0.025 	&	\multicolumn{2}{c}{...}	&	0.591  	&	0.031	&	0.578  	&	0.030	&	0.592  	&	0.025	\\
$\eta$                          	&	0.627  	&	0.047 	&	\multicolumn{2}{c}{...}	&	0.628  	&	0.060	&	0.638  	&	0.069	&	0.631  	&	0.049	\\
$\gamma_\text{ICM}$             	&	1.810  	&	0.632 	&	\multicolumn{2}{c}{...}	&	2.322  	&	0.390	&	2.383  	&	0.506	&	2.335  	&	0.449	\\
$T_0$                           	&	24.245 	&	2.646 	&	\multicolumn{2}{c}{...}	&	21.607 	&	2.197	&	20.319 	&	3.337	&	21.727 	&	2.193	\\
$\zeta_{cT}/\zeta_\text{c}$	&	9.069  	&	1.034 	&	\multicolumn{2}{c}{...}	&	8.670  	&	0.817	&	8.559  	&	1.137	&	8.667  	&	0.738	\\
$a_T$                    	&	\multicolumn{2}{c}{[0.]} 	&	\multicolumn{2}{c}{[0.]}	&	\multicolumn{2}{c}{[0.]}	&	\multicolumn{2}{c}{[0.]}	&	\multicolumn{2}{c}{[0.]}	\\
$b_T$                    	&	\multicolumn{2}{c}{[2.]} 	&	\multicolumn{2}{c}{[2.]}	&	\multicolumn{2}{c}{[2.]}	&	\multicolumn{2}{c}{[2.]}	&	\multicolumn{2}{c}{[2.]}	\\
$c_T$                    	&	2.576  	&	0.313 	&	\multicolumn{2}{c}{...}	&	2.779  	&	0.169	&	2.770  	&	0.220	&	2.782  	&	0.167	\\
$T_\text{cc}$                   	&	4.221  	&	1.328 	&	\multicolumn{2}{c}{...}	&	4.046  	&	1.159	&	3.398  	&	1.076	&	3.871  	&	1.053	\\
$\zeta_\text{cc}/\zeta_\text{c}$	&	1.140  	&	0.211 	&	\multicolumn{2}{c}{...}	&	0.998  	&	0.147	&	0.967  	&	0.193	&	0.969  	&	0.144	\\
$\alpha_\text{cc}$                    	&	\multicolumn{2}{c}{[1.9]} 	&	\multicolumn{2}{c}{[1.9]}	&	\multicolumn{2}{c}{[1.9]}	&	\multicolumn{2}{c}{[1.9]}	&	\multicolumn{2}{c}{[1.9]}	\\
\hline	
\end{tabular}
\end{table*}

\begin{figure*} 
     \resizebox{\hsize}{!}{\includegraphics{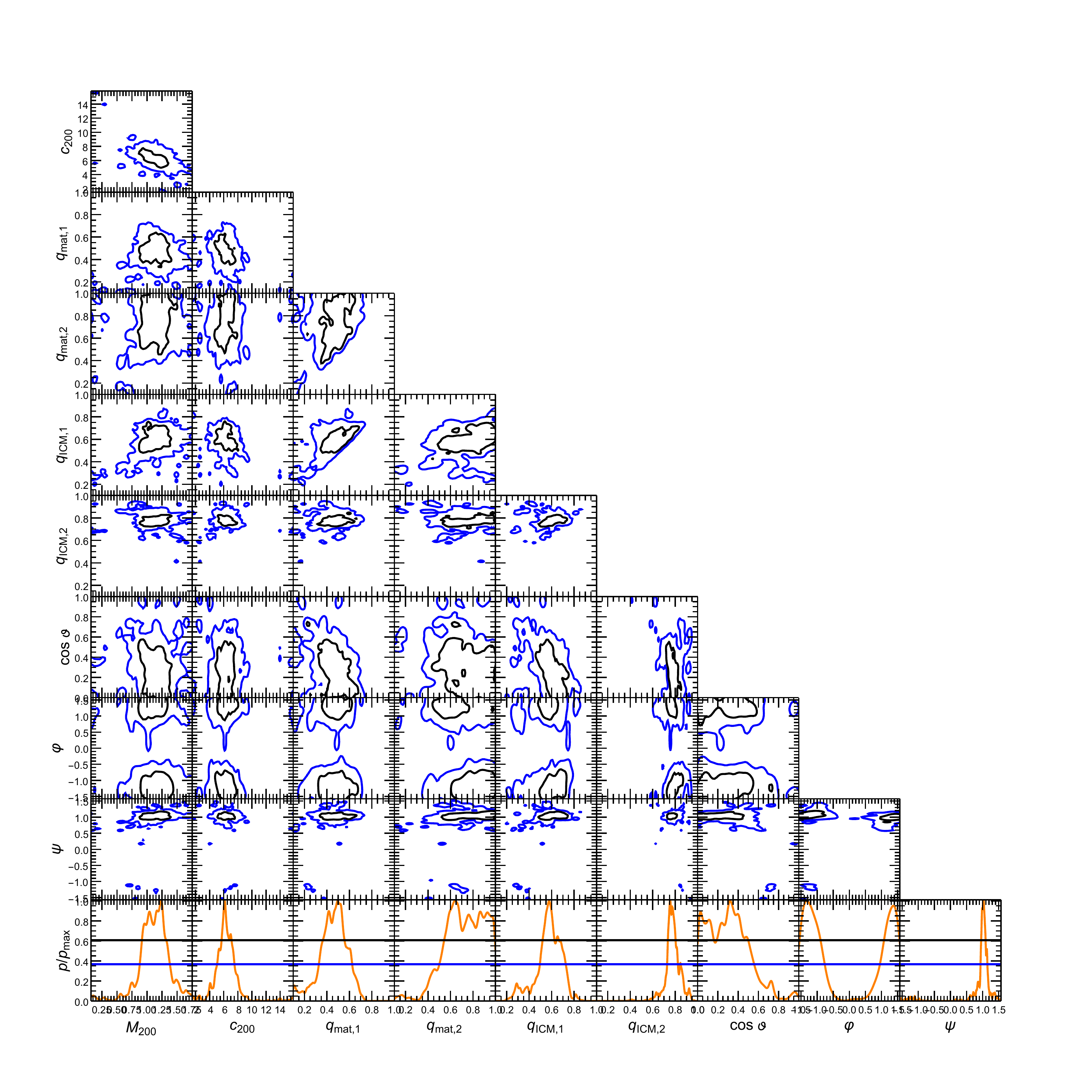}}
     \caption{Probability distributions of the parameters of mass distribution, shape and orientation as derived from the regression assuming a flat prior for the matter axis ratios and $q_\text{ICM}\ge q$. The black and blue contours include the 1-, 2-$\sigma$ confidence regions in two dimensions, here defined as the regions within which the probability is larger than $\exp[-2.3/2]$, or $\exp[-6.17/2]$ of the maximum, respectively. The bottom row plots the marginalised 1D distributions, renormalised to the maximum probability. The blue and black levels denote the confidence limits in one dimension, i.e. $\exp[-1/2]$ and $\exp[-4/2]$ of the maximum.}
	\label{fig_M1206_PDF_Mat}
\end{figure*}

The Bayesian regression scheme was applied to infer the properties of total matter and gas. We used different priors to perform the regression under five schemes:
\begin{itemize}
\item Flat prior for the minor-to-major axial ratio of the matter distribution ($q$-flat) and gas rounder than total matter ($q_\text{ICM}\ge q$). This is our reference setting.
\item $q$-flat and shared triaxiality ($\mathcal{T}_\text{mat}=\mathcal{T}_\text{ICM}$).
\item $q$-flat and gas following the potential ($q_\text{ICM}=q_\Phi$).
\item $N$-body prior for the matter shape ($q$-$N$body) and $q_\text{ICM}\ge q$.
\item Spherical symmetry.
\end{itemize}

Apart from the spherical case, we always take the gas to be rounder than the matter distribution, either through $q_\text{ICM}\ge q$, $\mathcal{T}_\text{mat}=\mathcal{T}_\text{ICM}$ or $q_\text{ICM}=q_\Phi$, and we assume a priori randomly oriented clusters. When not stated otherwise, priors for other parameters are set to their default distributions listed in Table~\ref{tab_parameters}.

Results are presented in Table~\ref{tab_results}. The marginalised 2D and 1D inferred probability distributions are illustrated in Fig.~\ref{fig_M1206_PDF_Mat} for the reference case.

With no regard to the priors, estimates of mass and concentration are consistent. Only the results based on the spherical assumption deviate significantly. Stronger priors, i.e $q$-$N$body or $q_\text{ICM}=q_\Phi$, enforce smaller values of the axis ratios, but they still are compatible with other less informative schemes, i.e. $q_\text{ICM}\ge q$ or $\mathcal{T}_\text{mat}=\mathcal{T}_\text{ICM}$.

Large inclination angles ($\cos \vartheta \la 0.3$) are always favoured. 

The parameters of the gas distributions are remarkably stable through the different setting of priors, which mostly affect shape. The estimate of the central gas density is strongly anti-correlated with the elongation parameter $e_{\text{ICM},\parallel}$ \citep{ser+al12a}. The shape and orientation parameters vary but still conjure to keep $e_{\text{ICM},\parallel}$ stable, and the gas density is stable too. 

The estimate of the intrinsic ellipsoidal core radius $\zeta_\text{c}$ is prior-dependent too. Assuming an ellipsoidal geometry, if the projected radius is vey well constrained and the shape parameters are stable, the intrinsic radius can be deprojected through a geometrical factor, see App.~\ref{sec_proj}. Any variation in the geometrical factor, which can be induced by biased priors, should be compensated by variations in the core radius to keep the projected one nearly fixed. However, our estimates of $\zeta_\text{c}$ are remarkably stable, which is a consequence of the well determined shape.

The parametric model offers an excellent fit to the data, see Figs.~\ref{fig_M1206_SB_profile}, \ref{fig_M1206_Te_profile}, \ref{fig_M1206_ySZ_profile} and \ref{fig_M1206_kappa_profile}. Whereas the surface brightness and the temperature constraints are uncorrelated, the SZ aperture photometry is not. If a point is above the fit, the following leans to do the same. 

In Figs.~\ref{fig_M1206_SB_profile}, \ref{fig_M1206_Te_profile}, \ref{fig_M1206_ySZ_profile} and \ref{fig_M1206_kappa_profile}, we plot the predicted profiles as computed for the typical parameter values, i.e. the bi-weighted estimators of the marginalised posterior distributions, see Table~\ref{tab_results}. These are not the `best-fit' parameters found with a maximum likelihood analysis for the respective plots, seen in isolation, still they provide an excellent fitting to the plotted profiles.

We fitted the 2D WL map, so the result for the averaged profile in circular annuli showed in Fig.~\ref{fig_M1206_kappa_profile} is given for illustration purposes. Similarly, in Fig.~\ref{fig_M1206_SB_profile} we plotted the surface-brightness averaged in circular annuli in the inner regions too, even though we fitted the 2D map.

\subsection{Mass and concentration}

\begin{table*}
\caption{Published mass and concentration measurements of MACS1206 from gravitational lensing. The used data-sets are in Col.~3, where $\gamma$WL, $\mu$WL,  denote shear and magnification weak lensing data, respectively. Values in square brackets were assumed as fixed. The masses are in units of $10^{15}M_\odot h^{-1}$.}
\label{tab_comparison}
\begin{tabular}{ l l l l r@{$\,\pm\,$}l r@{$\,\pm\,$}l }     
\hline
Author			&     Geometry&	Data-set&	 GL observatory&	\multicolumn{2}{c}{$M_{200}$}	&  \multicolumn{2}{c}{$c_{200}$}\\
\hline
\citet{foe+al12}     	&	Spherical&	$\gamma$WL				& CFHT&		\multicolumn{2}{c}{$1.06^{+0.20}_{-0.13}$}	&	\multicolumn{2}{c}{$[4]$}		\\
\citet{mer+al15}      	&	Spherical&	$\gamma$WL, SL			&Subaru, {\it HST} &		0.86	&	0.11	&	4.3	&	1.5		\\
\citet{ume+al16}      	&	Spherical&	$\gamma$WL, $\mu$WL, SL	&Subaru, {\it HST} &		1.28	&	0.29	&	3.7	&	1.1		\\
This	work			&	Spherical&	$\gamma$WL, $\mu$WL, SL			&Subaru, {\it HST} &		1.06	&	0.16	&	5.0	&	0.7		\\
This	work			&	Triaxial&		$\gamma$WL, $\mu$WL, SL, X, SZ		& Subaru, {\it HST}	&		1.14	&	0.23	&	6.3	&	1.2		\\
\hline	
\end{tabular}
\end{table*}

\begin{table}
\caption{Ellipsoidal and spherically enclosed mass estimates, as derived from the regression assuming a flat prior for the axis ratio $q_{\text{mat},1}$ of the matter distribution and $q_\text{ICM}\ge q$. The ellipsoidal mass $M_\Delta$ is computed within the ellipsoid of semi-major axis $\zeta_\Delta$. The overdensity radii are given in units of $\text{Mpc}~h^{-1}$. The enclosed masses are in units of $10^{15}M_\odot h^{-1}$. Typical values and dispersions are computed as bi-weighted estimators.}
\label{tab_masses}
\resizebox{\hsize}{!} {
\begin{tabular}{ l r@{$\,\pm\,$}l r@{$\,\pm\,$}l r@{$\,\pm\,$}l r@{$\,\pm\,$}l}     
\hline
Overdensity			&		\multicolumn{4}{c}{Ellipsoidal}	&  \multicolumn{4}{c}{Spherically enclosed}	\\
$\Delta$	& \multicolumn{2}{c}{$\zeta_\Delta$}   &	 \multicolumn{2}{c}{$M_\Delta$}	&  \multicolumn{2}{c}{$r_\Delta$}&	 \multicolumn{2}{c}{$M_\text{sph}(<r_\Delta)$} \\
\hline
$2500\text{c}$     	&	0.65	&	0.09	&	0.42	&	0.07	&	0.44	&	0.02	&	0.39	&	0.06	\\
$500\text{c}$      	&	1.40	&	0.20	&	0.85	&	0.16	&	0.95	&	0.05	&	0.79	&	0.14	\\
$200\text{c}$      	&	2.10	&	0.31	&	1.14	&	0.23	&	1.43	&	0.09	&	1.08	&	0.20	\\
$\Delta_\text{vir}$	&	2.49	&	0.37	&	1.27	&	0.27	&	1.69	&	0.11	&	1.21	&	0.24	\\
200\text{m}        	&	2.68	&	0.40	&	1.33	&	0.29	&	1.83	&	0.12	&	1.27	&	0.26	\\
\hline	
\end{tabular}
}
\end{table}

\begin{figure}
       \resizebox{\hsize}{!}{\includegraphics{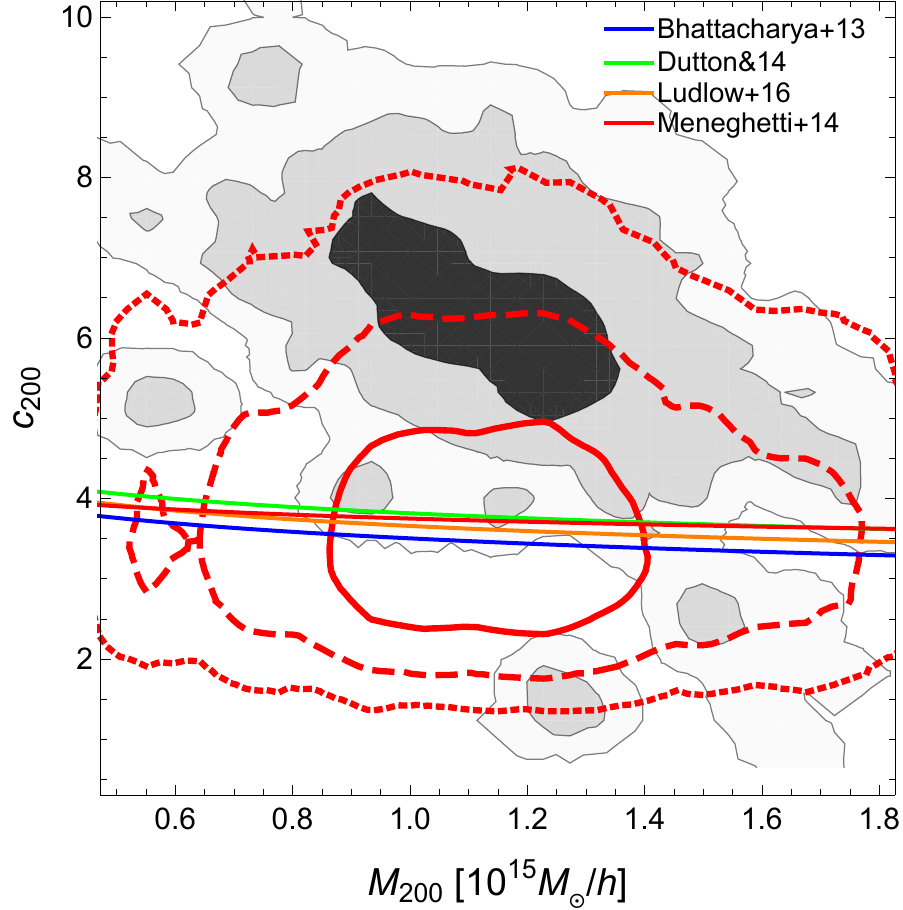}}
       \caption{Marginalised probability distribution of mass and concentration. The grey shadowed regions include the 1-, 2-, 3-$\sigma$ confidence region in two dimensions, here defined as the regions within which the probability density is larger than $\exp[-2.3/2]$, $\exp[-6.17/2]$, and  $\exp[-11.8/2]$ of the maximum, respectively. The regression assumed a flat prior for the axis ratio $q_{\text{mat},1}$ of the matter distribution ($q$-flat) and $q_\text{ICM}\ge q$. The blue, green, orange and red lines plot the mass-concentration relations of \citet{bha+al13}, \citet{du+ma14}, \citet{lud+al16}, and \citet{men+al14}, respectively. The red contours trace the predicted concentration from \citet{men+al14} given the observed mass distribution and the predicted scatter of the theoretical mass-concentration relation. If needed, published relations were rescaled to our reference cosmology.}
	\label{fig_M1206_M200_c200}
\end{figure}

\begin{figure}
\begin{center}
$
\begin{tabular}{c}
\noalign{\smallskip}
\resizebox{\hsize}{!}{\includegraphics{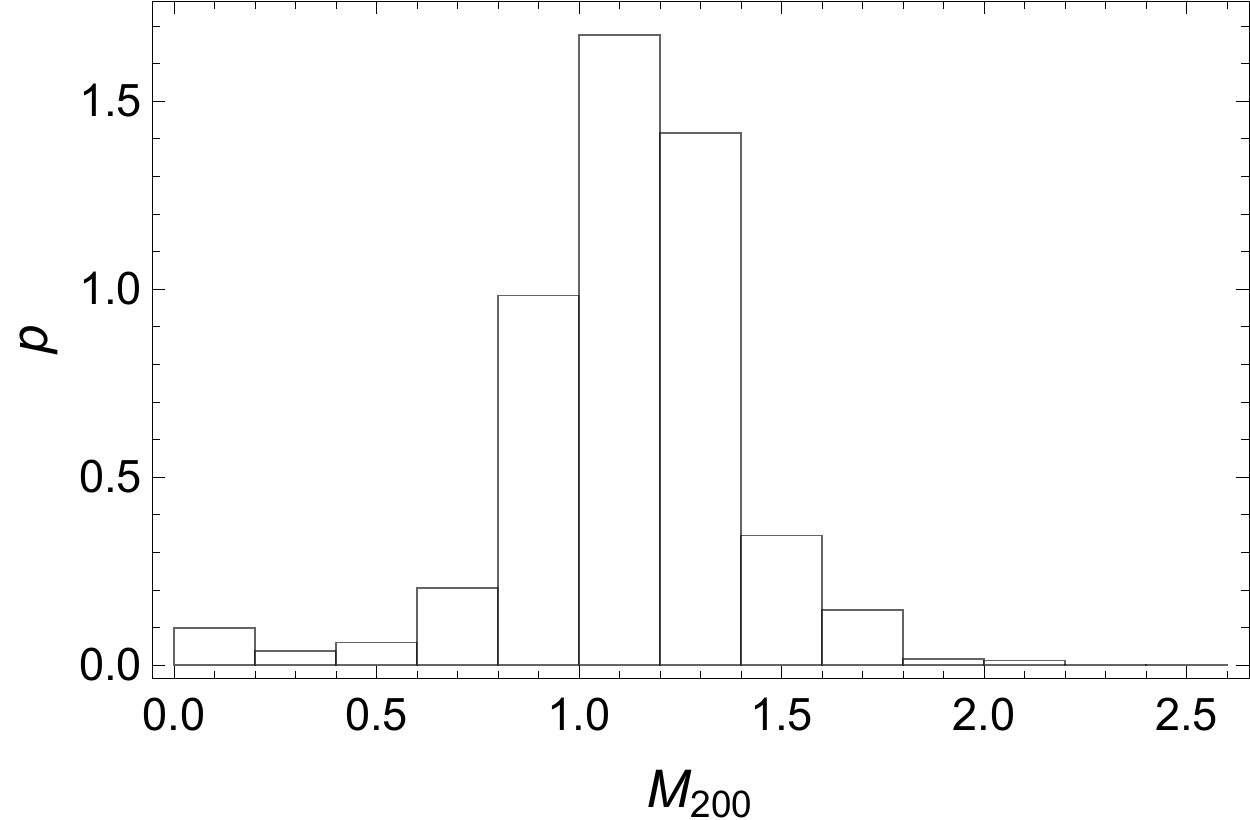}} \\  
 \resizebox{\hsize}{!}{\includegraphics{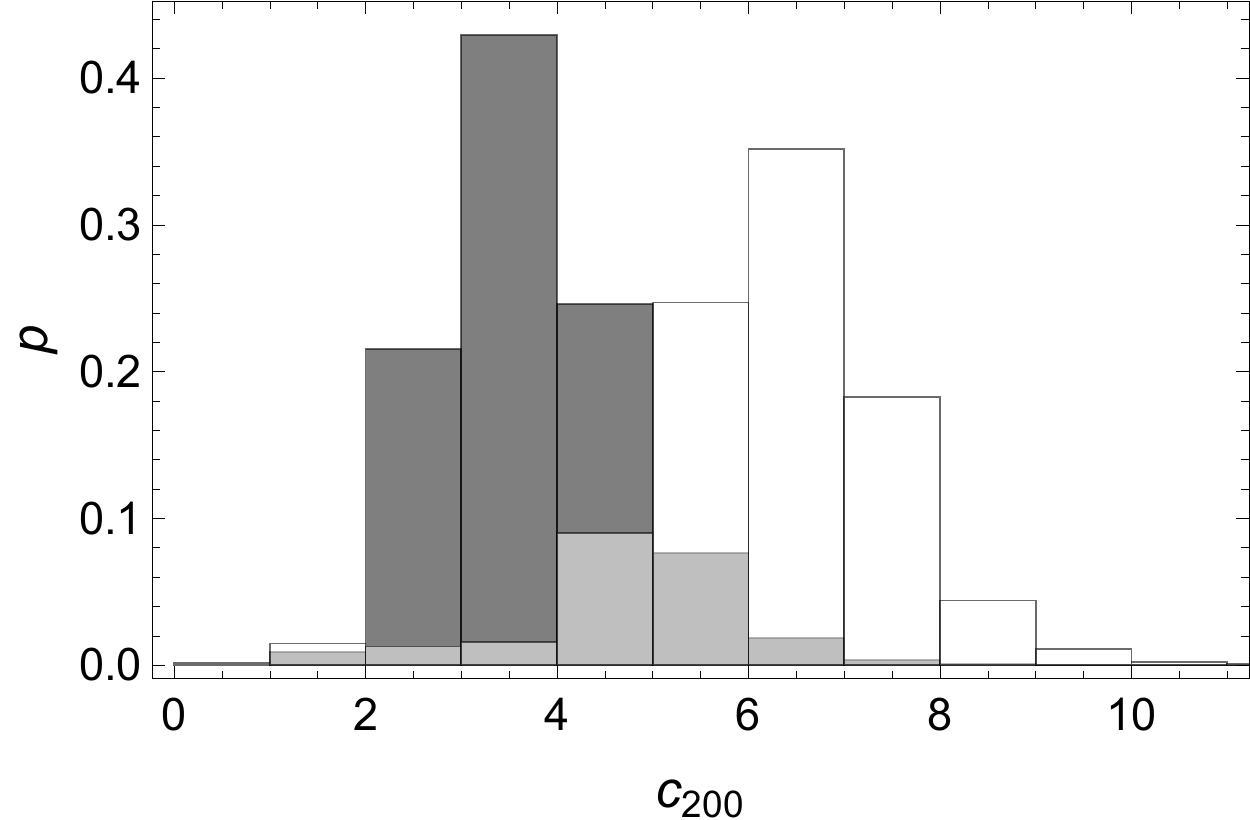}} \\
\end{tabular}
$
\end{center}
\caption{Marginalised PDFs (Probability Density Functions), plotted as white histograms, of mass (top panel) and concentration (bottom panel). The regression assumed a flat prior for the axis ratio $q_{\text{mat},1}$ of the matter distribution ($q$-flat) and $q_\text{ICM}\ge q$. The grey histogram in the bottom panel shows the theoretical prediction based on the inferred mass distribution and the scattered mass-concentration relation from \citet{men+al14}.
}
\label{fig_M1206_M_c_histo}
\end{figure}

MACS1206 is massive, $M_{200}=(1.1\pm0.2)\times 10^{15}M_\odot/h$. The measured concentration, $c_{200}=6.3 \pm 1.2$, as estimated in the reference regression, is slightly higher than but still consistent with predictions. Recent theoretical estimates based on $N$-body simulations of dark matter halos \citep{bha+al13,du+ma14,lud+al16} graze the 68.3 per cent confidence region, see Fig.~\ref{fig_M1206_M200_c200}. The most sensible comparison is to \citet{men+al14}, which studied a large set of nearly 1400 cluster-sized halos simulated at high spatial and mass resolution from the MUSIC-2 $N$-body/hydrodynamical runs. The sample was originally constructed by selecting all halos in the simulation box that were more massive than $10^{15}M_\odot/h$ at redshift $z = 0$. The evolved halos are distributed over the redshift range $0.25\le z\le 0.67$ and are suitable to make predictions about several properties of the clusters included in the CLASH sample

\citet{men+al14} defined clusters as regular if they showed unperturbed X-ray surface brightness distributions. These halos have small centroid shift, ellipticity, and power ratios and they have large surface-brightness concentrations. \citet{men+al14} used the term regular with reference to the X-ray appearance, so that regular clusters may be dynamically unrelaxed. 

From the comparison to the sample of simulated clusters above the completeness mass limit, MACS1206 is the only CLASH cluster less regular than the mean of the simulations. The comparison shows that the regularity of the CLASH clusters is not extreme, in the sense that the simulated sample has an extended tail of very regular clusters.

\citet{men+al14} measured the theoretical mass-concentration relation under different selection criteria and considered either projected or 3D concentrations and masses. The most sensible comparison for our analysis of MACS1206 is with the NFW fitting in 3D of the extended sample of simulated clusters, where no selection was applied except that based on the relaxation state.

The distribution of concentrations expected using the mass-concentration relation given the measured mass distribution is compared to the observed concentration in Fig.~\ref{fig_M1206_M200_c200} and in the bottom panel of Fig.~\ref{fig_M1206_M_c_histo} for the marginalised distribution. Agreement is substantial.

Triaxial analyses favour agreement of measured concentrations of massive lensing clusters with theoretical predictions \citep{ogu+al05,se+zi12}. In fact, they do not suffer by the orientation bias affecting lensing selected clusters preferentially elongated along the line of sight. This configuration makes the concentration estimated under the spherical hypothesis biased high. The opposite holds for clusters elongated orthogonally to the line-of-sight.

MACS1206 is X-ray selected and it is elongated in the plane of the sky, see Sec.~\ref{sec_orie}. Mass and concentration derived under the spherical assumption are then biased low, see Table~\ref{tab_results}. Our unbiased result is in good agreement with theoretical predictions.

Our estimates under the spherical hypothesis are consistent with previous lensing-based results from literature, see Table~\ref{tab_comparison}. Differences with the triaxial estimate show that the geometrical bias is significant when compared to the statistical errors.



We list the values of ellipsoidal and spherical overdensity mass in Table~\ref{tab_masses}. Ellipsoidal or spherically enclosed mass estimates are similar at a given overdensity.

\subsection{Matter shape}

\begin{figure}
       \resizebox{\hsize}{!}{\includegraphics{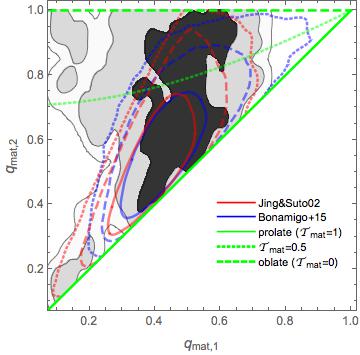}}
       \caption{Probability distribution of the matter axis ratios. The regression assumed a flat prior for the axis ratio $q_{\text{mat},1}$ of the matter distribution ($q$-flat) and $q_\text{ICM}\ge q$. The grey shadowed regions include the 1-, 2-, 3-$\sigma$ confidence region in two dimensions, here defined as the region within which the probability density is larger than $\exp[-2.3/2]$, $\exp[-6.17/2]$, and  $\exp[-11.8/2]$ of the maximum, respectively. The blue and red contours plot the theoretical predictions from \citet{bon+al15} and  \citet{ji+su02}, respectively, smoothed for the inferred mass distribution. Contours are drawn at 1- (full), 2-(dotted), 3-(dashed) $\sigma$. The green full, dashed and long dashed lines denotes the loci of points corresponding to prolate ($T_\text{mat}=1$), triaxial  ($T_\text{mat}=0.5$) and oblate  ($T_\text{mat}=0$) haloes.}
	\label{fig_M1206_q1_q2}
\end{figure}

\begin{figure}
\begin{center}
$
\begin{tabular}{c}
\noalign{\smallskip}
  \resizebox{\hsize}{!}{\includegraphics{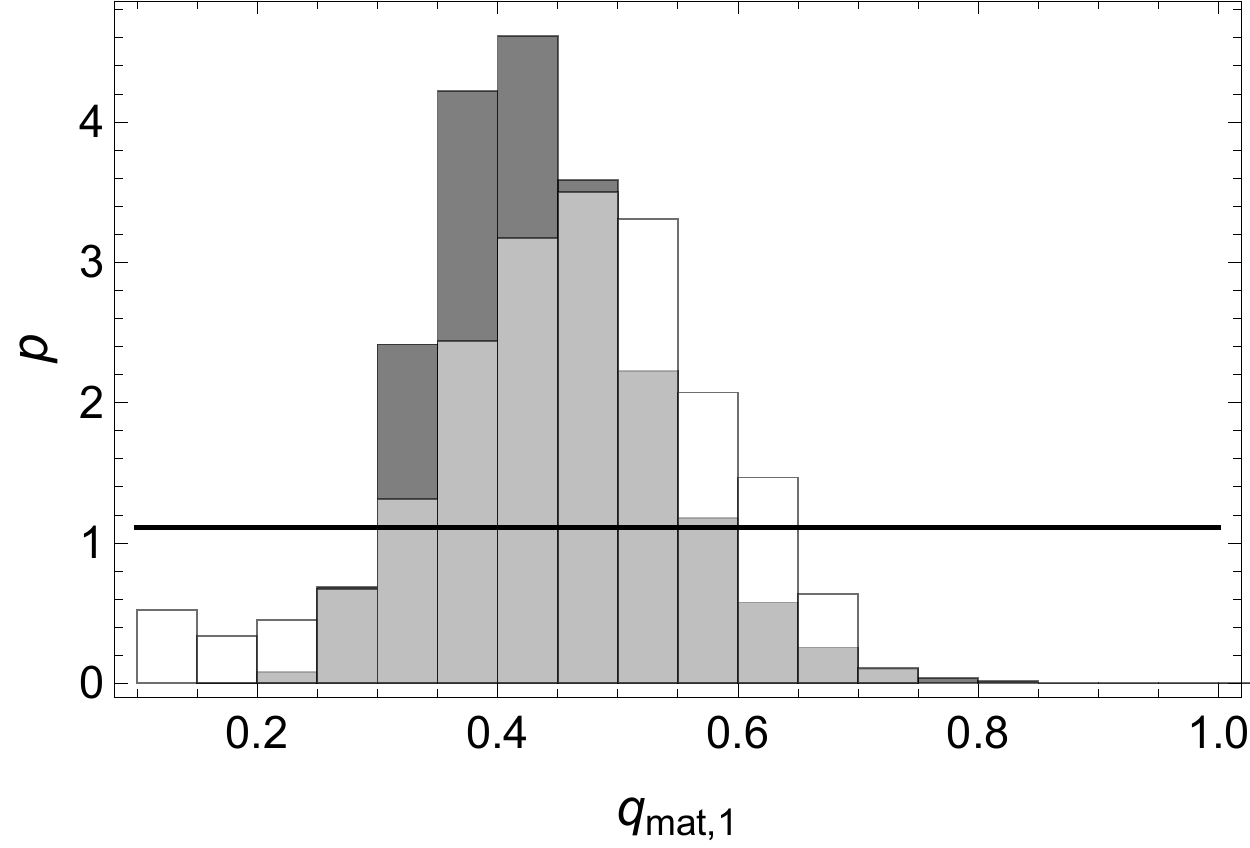}} \\
 \resizebox{\hsize}{!}{\includegraphics{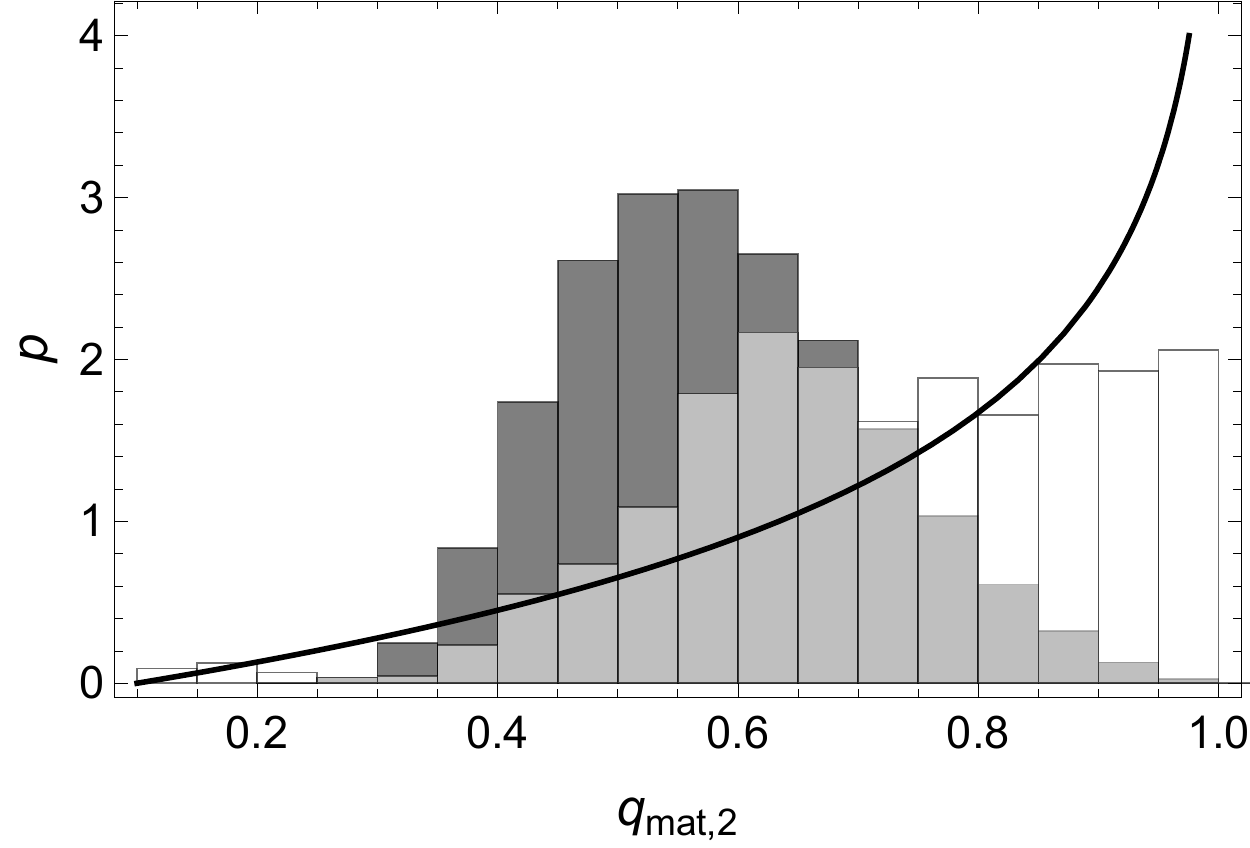}} \\
\end{tabular}
$
\end{center}
\caption{Marginalised PDFs (plotted as white histograms) of the minor-to-major (top panel) and intermediate-to-major (bottom) axis ratios of the matter distribution. The regression assumed a flat prior for the axis ratio $q_{\text{mat},1}$ of the matter distribution ($q$-flat) and $q_\text{ICM}\ge q$. Grey histograms show the theoretical predictions based on the inferred mass distribution and the predicted probability distribution for axis ratios from \citet{bon+al15}. The black lines shows the marginalised a-priori distributions under the $q$-flat prior.
}
\label{fig_M1206_q_histo}
\end{figure}

MACS1206 shows a triaxial shape, see Figs.~\ref{fig_M1206_q1_q2} and \ref{fig_M1206_q_histo}. The axis ratios given the cluster mass and redshift are compatible with the theoretical predictions, even though the mean intermediate to major axis ratio slightly exceeds expectations.

In the more conservative reference analysis, when we only require the gas to be rounder than the matter, we cannot preferentially distinguish the prolate from the oblate configuration, see Fig.~\ref{fig_M1206_q1_q2}. However, with the slightly more informative prior assuming the same triaxial parameter for both matter and gas (${\cal T}_\text{mat}={\cal T}_\text{ICM}$), prolate configurations are preferred.

\subsection{Halo orientation}
\label{sec_orie}

\begin{figure}
       \resizebox{\hsize}{!}{\includegraphics{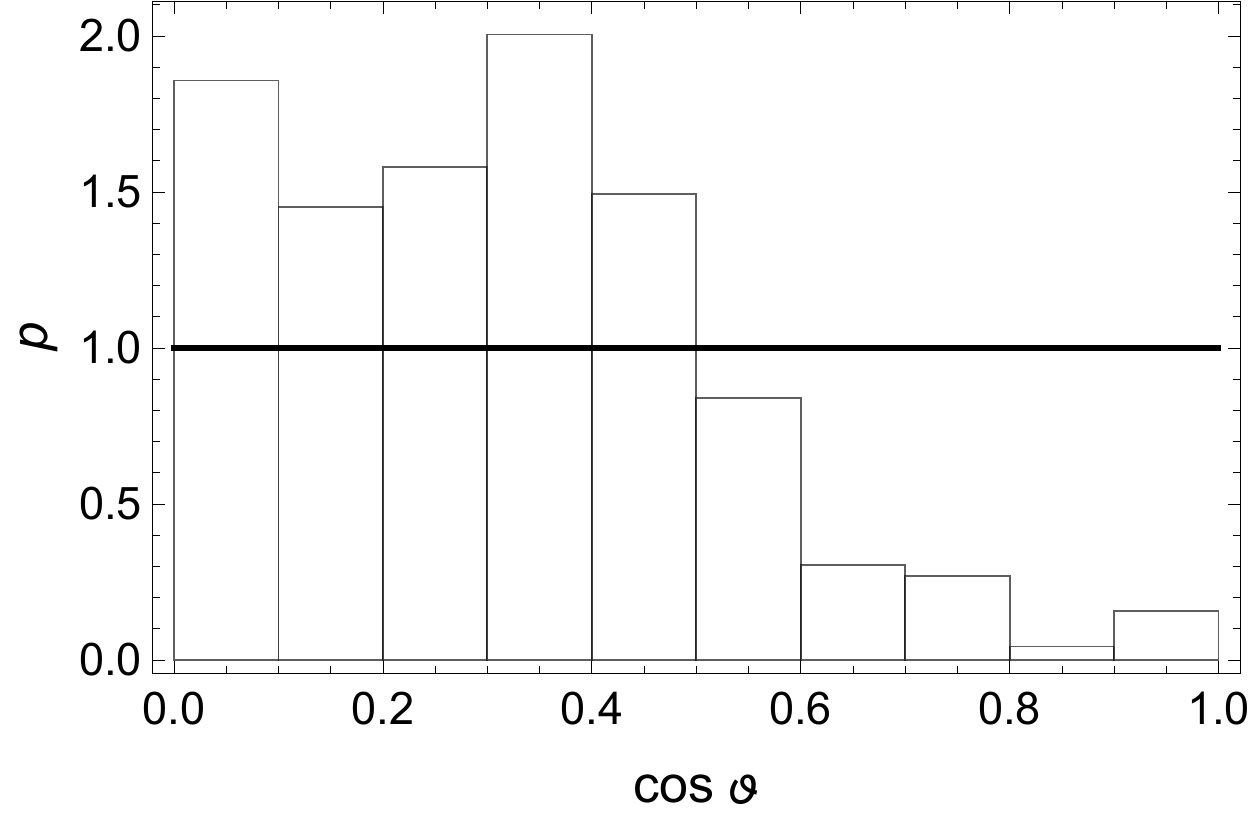}}
       \caption{Marginalised PDFs (plotted as white histogram) of the cosine of the inclination angle, $\cos \vartheta$. The regression assumed a flat prior for the axis ratio $q_{\text{mat},1}$ of the matter distribution ($q$-flat) and $q_\text{ICM}\ge q$. The black line denotes random orientation.}
	\label{fig_M1206_costheta}
\end{figure}

\begin{figure}
       \resizebox{\hsize}{!}{\includegraphics{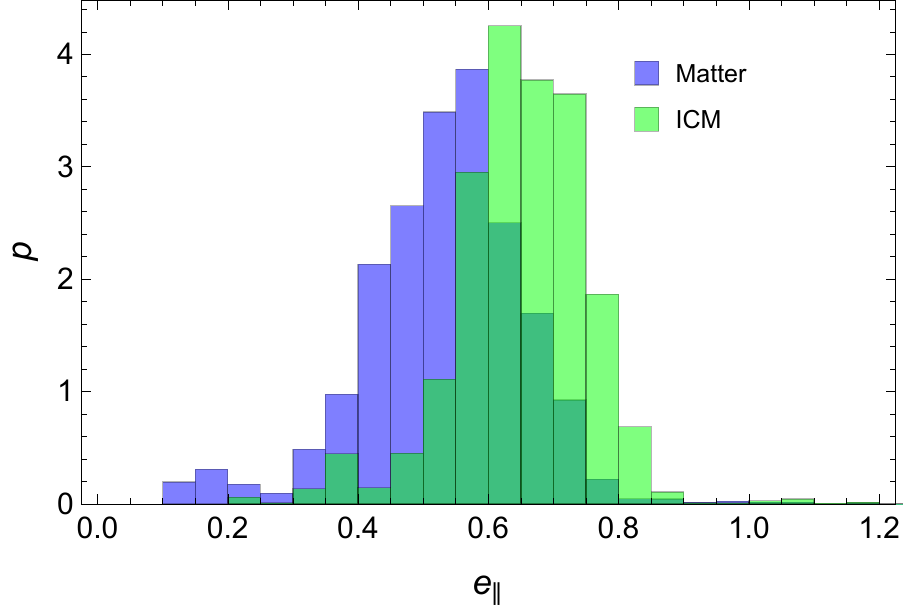}}
       \caption{Marginalised PDFs (plotted as white histogram) of the line-of-sight elongation parameter $e_\parallel$ of the total matter (blue) and of the gas distribution (green). The regression assumed a flat prior for the axis ratio $q_{\text{mat},1}$ of the matter distribution and $q_\text{ICM}\ge q$.}
	\label{fig_M1206_eParallel}
\end{figure}

We observe MACS1206 face-on, i.e. the main axis is near the plane of the sky. Large inclination angles, nearly orthogonal to the line of sight, are preferred, see Fig.~\ref{fig_M1206_costheta}. This is also confirmed by the analysis of the elongation parameter, which shows that the width in the plane of the sky is larger than the size along the line of sight, $e_\parallel < 1$, see Fig.~\ref{fig_M1206_eParallel}. This effect is more pronounced for the matter distribution $e_{\text{mat},\parallel} \la e_{\text{ICM},\parallel}$, but this is expected given the priors.

The second Euler angle $\varphi$ is poorly constrained with a bimodal distribution, whereas the precision on $\psi$ is reflective of the accuracy on the measured orientation angles of matter (from lensing) and gas (from X-ray surface brightness).

Based on the highly elliptical mass distribution in projection as inferred at large cluster radii from the Subaru weak lensing analysis, \citet{ume+al12} argued that that major axis of MACS1206 is not far from the sky plane. We could unambiguously prove this based on our multi-probe 3D analysis. This conclusion could not be obtained based on SZ data alone. \citet{rom+al16} noted that the Bolocam contours of MACS1206 do not exhibit much ellipticity, and by comparison with X-ray data analysed under the hypothesis of circular symmetry in the plane of the sky, found a major-to-minor axis ratio of $1.24\pm0.29$, where the major axis is along the line of sight.

\subsection{Gas distribution}

The inferred gas distribution is quite common to massive clusters. The slope $\beta$ is $\sim 0.6$, and there is some evidence for an inner spike ($\eta \sim 0.6$). There is no clear evidence for a truncation of the density profile, which would occur at $\zeta_\text{t}\ga 1.5~\text{Mpc}/h$, beyond the observational range. This makes the constraints on the outer slope $\gamma_\text{ICM}$ of poor significance.

The temperature distribution shows a cool core with a quite large cool core radius, $\zeta_\text{cc} \sim \zeta_\text{c}$. However, the temperature profile is sampled in just one point in the inner $100~\text{kpc}/h$, the presumptive size of the cool core, making this estimate a likely artefact of poor sampling and extrapolation.

\subsection{Gas shape}

\begin{table}
\caption{Comparison of the axis ratios and derived geometrical quantities of the distributions of gas (col.~1) and gravitational potential (cols.~2 and 3). The distributions share the same orientation, see Table~\ref{tab_results}. The regression assumed a flat prior for the axis ratio $q_{\text{mat},1}$ of the matter distribution ($q$-flat) and $q_\text{ICM}\ge q$. The shape parameters of the potential are computed considering the effective axis ratios at $r_{200}/3$ (Col.~2) and at $2\ r_{200}/3$ (col.~3). Projected orientation angles are measured in degrees North-over-East.}
\label{tab_ICM_shape}
\begin{center}
\begin{tabular}{ l  r@{$\,\pm\,$}l r@{$\,\pm\,$}l r@{$\,\pm\,$}l}     
\hline
				&    \multicolumn{2}{c}{ICM}	&	\multicolumn{2}{c}{$\Phi(r_{200}/3)$}	& \multicolumn{2}{c}{$\Phi(2\ r_{200}/3)$}\\
\hline
$q_1$                 	&	0.59  	&	0.11	&	0.77  	&	0.06	&	0.80  	&	0.05	\\
$q_2$                 	&	0.78  	&	0.06	&	0.88  	&	0.08	&	0.90  	&	0.07	\\
$\epsilon$            	&	0.22  	&	0.02	&	0.12  	&	0.07	&	0.10  	&	0.06	\\
$\theta_\epsilon$     	&	-58.52	&	3.53	&	-56.50	&	8.07	&	-56.51	&	8.06	\\
$e_\parallel$         	&	0.66  	&	0.09	&	0.83  	&	0.05	&	0.85  	&	0.05	\\
$e_1/e_{\text{mat},1}$	&	0.93  	&	0.06	&	0.72  	&	0.04	&	0.67  	&	0.04	\\
${\cal T}$            	&	0.63  	&	0.16	&	0.54  	&	0.31	&	0.54  	&	0.31	\\
\hline	
\end{tabular}
\end{center}
\end{table}

\begin{figure}
\begin{center}
$
\begin{tabular}{c}
  \resizebox{\hsize}{!}{\includegraphics{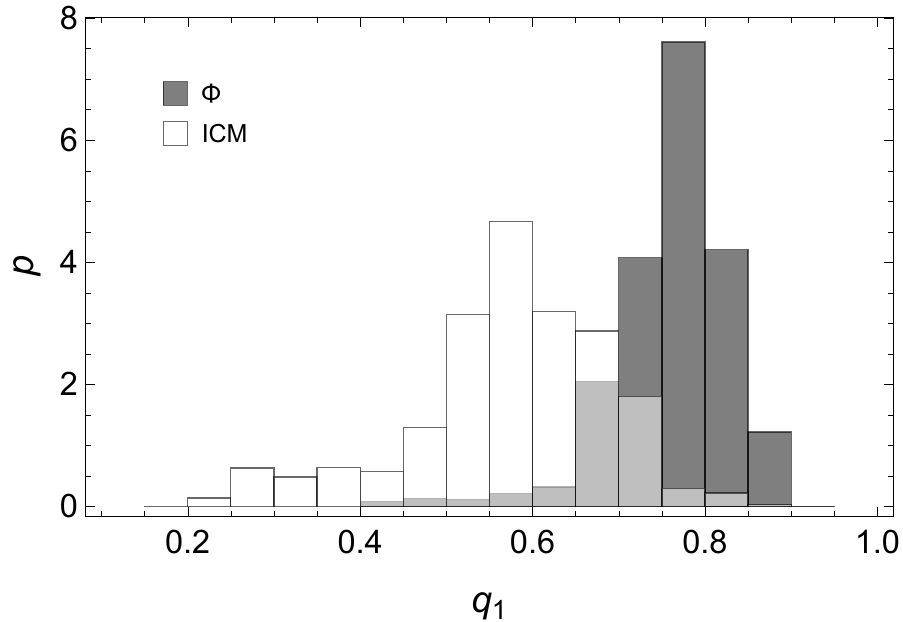}} \\
   \resizebox{\hsize}{!}{\includegraphics{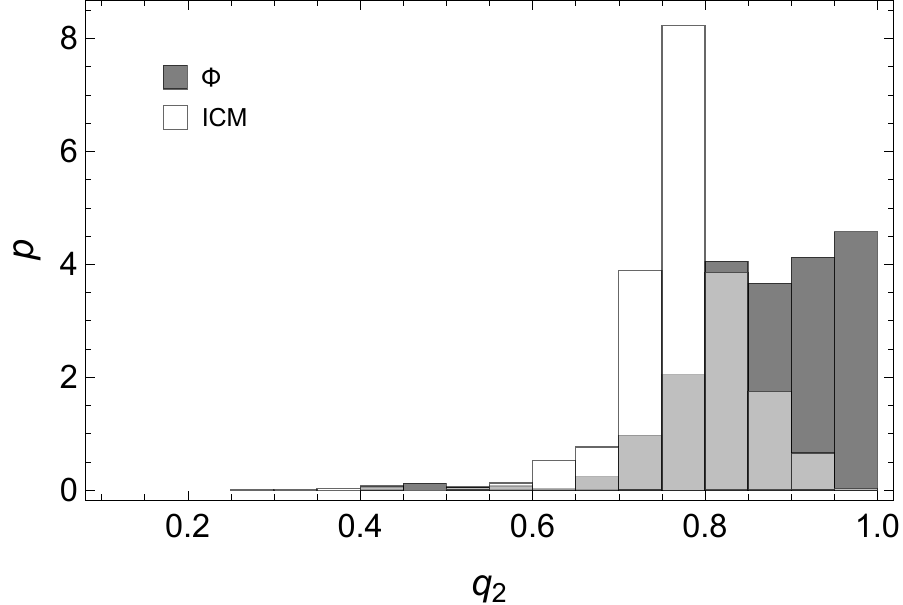}} \\
    \resizebox{\hsize}{!}{\includegraphics{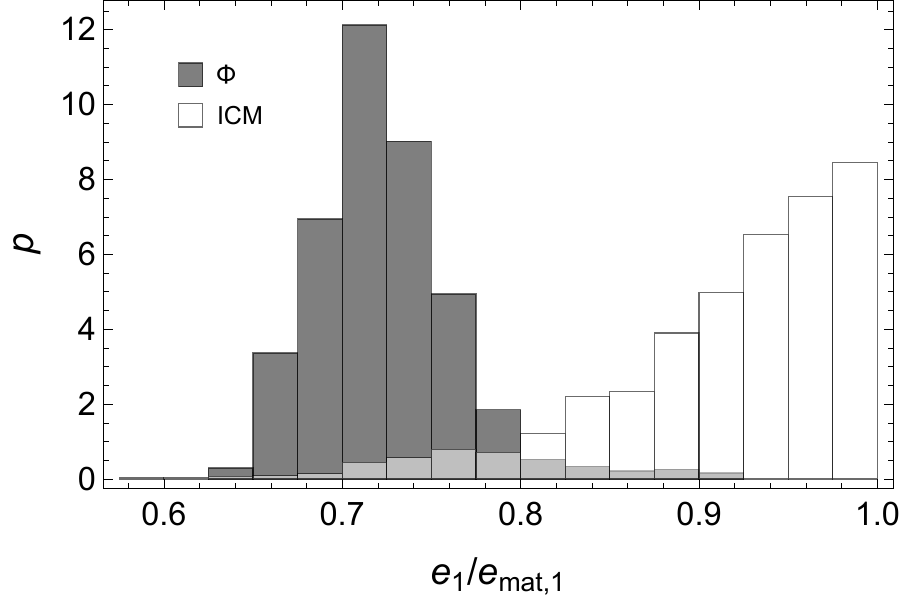}} \\
\end{tabular}
$
\end{center}
\caption{Marginalised PDFs of the axis ratios and the eccentricity of the gas distribution (plotted as white histograms) and of the gravitational potential (grey histograms). From top to bottom: minor-to-major axis ratio, intermediate-to-major axis ratio, and eccentricity. The regression assumed a flat prior for the axis ratio $q_{\text{mat},1}$ of the matter distribution ($q$-flat) and $q_\text{ICM}\ge q$.
}
\label{fig_M1206_qICM}
\end{figure}

\begin{figure}
       \resizebox{\hsize}{!}{\includegraphics{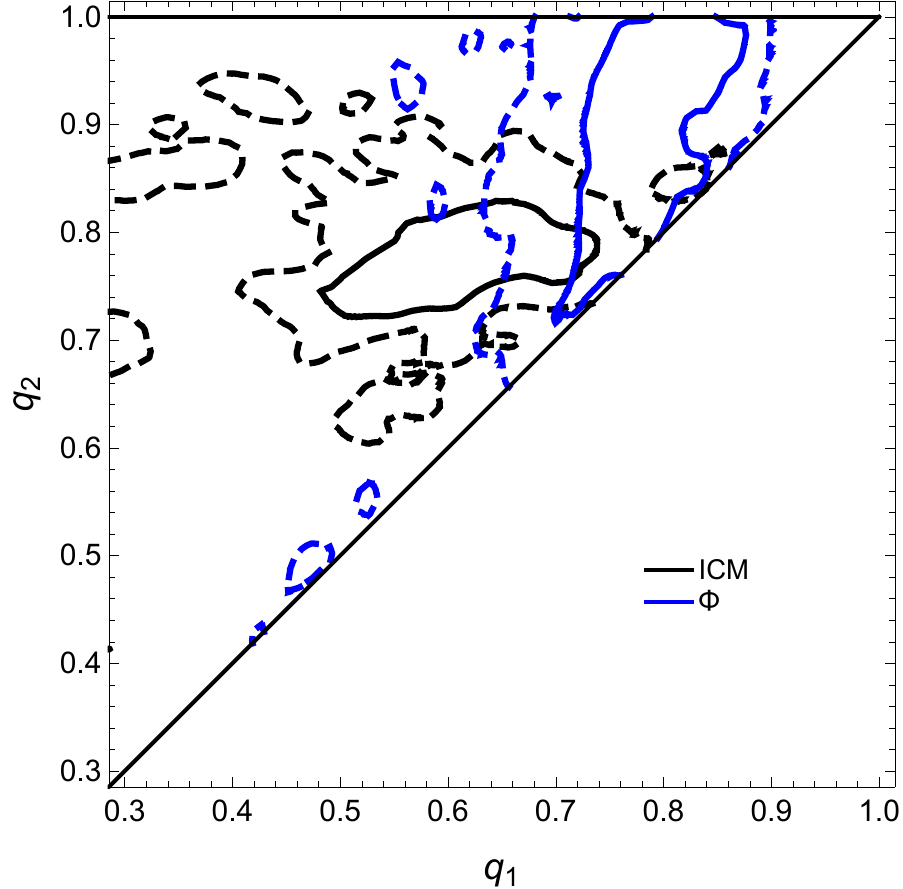}}
       \caption{Probability distribution of gas (black) and potential (blue) axis ratios. The regression assumed a flat prior for the axis ratio $q_{\text{mat},1}$ of the matter distribution ($q$-flat) and $q_\text{ICM}\ge q$. Contours are drawn at 1- (full), and 2-(dashed) $\sigma$ confidence region in two dimensions, here defined as the region within which the probability density is larger than $\exp[-2.3/2]$ or $\exp[-6.17/2]$ of the maximum, respectively.}
	\label{fig_M1206_qICM1_qICM2}
\end{figure}

A consequence of the pressure equilibrium is the {\em X-ray shape theorem} \citep{bu+ca94,bu+ca96,bu+ca98}, i.e. the gas in strict hydrostatic equilibrium follows the iso-potential surfaces of the underlying matter distribution. If the gas pressure can be written as a function of gas density and temperature and the gas is adequately described by a single phase, the hydrostatic equation demands that the potential, the gas density, the gas pressure, the gas temperature and the X-ray volume emissivity all share the same constant surfaces in three dimensions. This geometric test for dark matter is robust and independent of the temperature profile, which can be poorly constrained. 

The first application of the shape test compared the 2D ellipticities of the X-ray surface brightness with the gravitational potential after projection in the hypothesis of spheroidal symmetry for the emitting system \citep{bu+ca94,bu+ca96,bu+ca98}. Here, we can directly compare the axis ratios in three dimensions without restricting assumptions on the cluster shape.

Our method can determine the intrinsic structure, shape, and orientation of the gas without a priori assuming equilibrium. The direct comparison of gas shape, as determined by the regression, to the potential shape, as computed from the inferred mass distribution, is then a test of equilibrium.

The regular morphology of MACS1206 makes it an excellent candidate to the application of the shape theorem. The projected gravitational potential of MACS1206 as inferred from galaxy kinematics under the hypothesis of spherical symmetry agrees well with the reconstruction of the gravitational potential from weak and strong gravitational lensing or X-ray measurements \citep{sto+al15}.

We computed the effective shape of the potential as detailed in App.~\ref{sec_pote_shap}. The inferred gas shape is broadly compatible with the potential, see Table~\ref{tab_ICM_shape} and Figs.~\ref{fig_M1206_qICM} and \ref{fig_M1206_qICM1_qICM2}, even though rounder potential shapes are compatible with the data. In particular, the ratio of the measured eccentricities slightly exceeds the expected value of $\sim 0.7$, see bottom panel of Fig.~\ref{fig_M1206_qICM} , and the projected isocontours are more elongated, see Table~\ref{tab_ICM_shape}. The compatibility of gas and potential shapes is best evidenced by the marginalised distributions of the axis ratios, see Fig.~\ref{fig_M1206_qICM1_qICM2}.

The shape theorem is expected to be violated to some degree. The degree of hydrostatic equilibrium was investigated by \citet{bif+al16} on a sample of 29 massive clusters extracted from cosmological hydrodynamical simulations including several physical processes, i.e. stellar and AGN feedback. The radial balance between the gravitational and hydrodynamical forces can be assessed via comparison of the gas accelerations generated. They found an average deviation from equilibrium of 10-20 per cent out to the virial radius. However, the result is strongly dependent on the properties of the selected clusters.

\subsection{Triaxiality}
\label{sec_res_tri}

\begin{figure}
       \resizebox{\hsize}{!}{\includegraphics{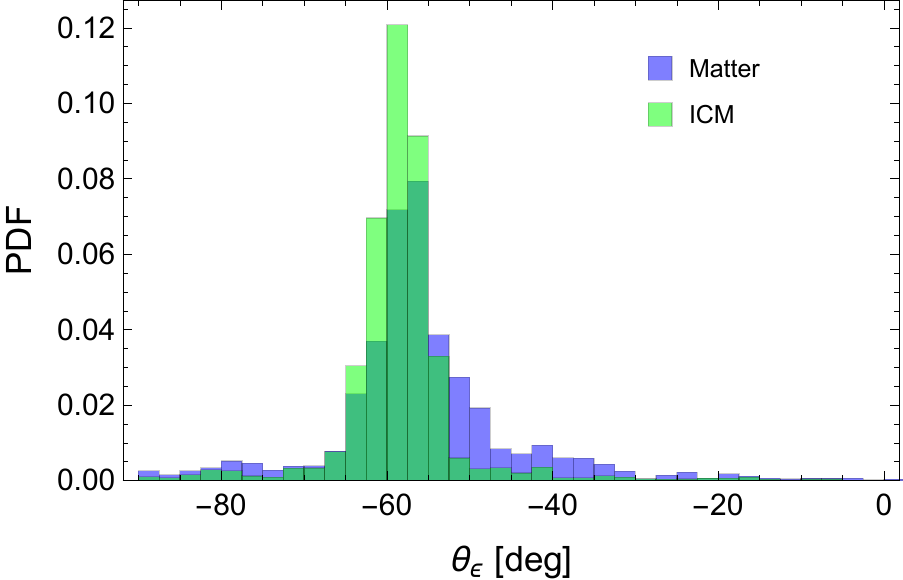}}
       \caption{Marginalised PDFs (plotted as histograms) of the orientation angle in the plane of the sky $\theta_\epsilon$ (measured in degrees North-over-East NE) of the total matter (blue) and of the gas distribution (green). The regression assumed a flat prior for the axis ratio $q_{\text{mat},1}$ of the matter distribution and $q_\text{ICM}\ge q$.}
	\label{fig_M1206_thetaepsilon}
\end{figure}

\begin{figure}
       \resizebox{\hsize}{!}{\includegraphics{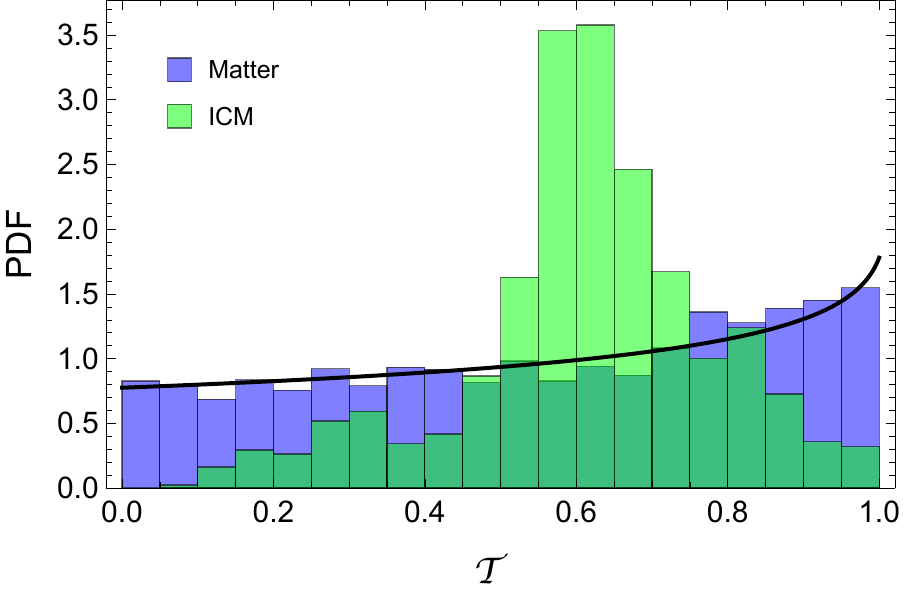}}
       \caption{Marginalised PDFs (plotted as histograms) of the triaxial parameter ${\cal T}$ of the total matter (blue) and of the gas distribution (green). The regression assumed a flat prior for the axis ratio $q_{\text{mat},1}$ of the matter distribution and $q_\text{ICM}\ge q$. The black line shows the prior.}
	\label{fig_M1206_triaxiality}
\end{figure}

A working hypothesis suitable for clusters in near equilibrium is that the triaxial parameters of the gas and of the matter distribution are similar. This is the case for MACS1206. Firstly, the X-ray and the lensing maps share the same orientation in the plane of the sky, see Fig.~\ref{fig_M1206_thetaepsilon}. 

Secondly, our direct measurements of the triaxial parameters are compatible, see Fig.~\ref{fig_M1206_triaxiality}. However, when no strong assumption is made a priori on the matter shape and on the relation between the axial ratios, as in our reference case where we just assumed a flat prior for the axis ratio $q_{\text{mat},1}$ of the matter distribution, the triaxiality parameter of the matter halo cannot be significantly constrained and its final distribution follows the initial prior.

On the contrary, the triaxiality parameter of the gas distribution is strongly constrained by the data.

\subsection{Gas fraction}

\begin{figure}
       \resizebox{\hsize}{!}{\includegraphics{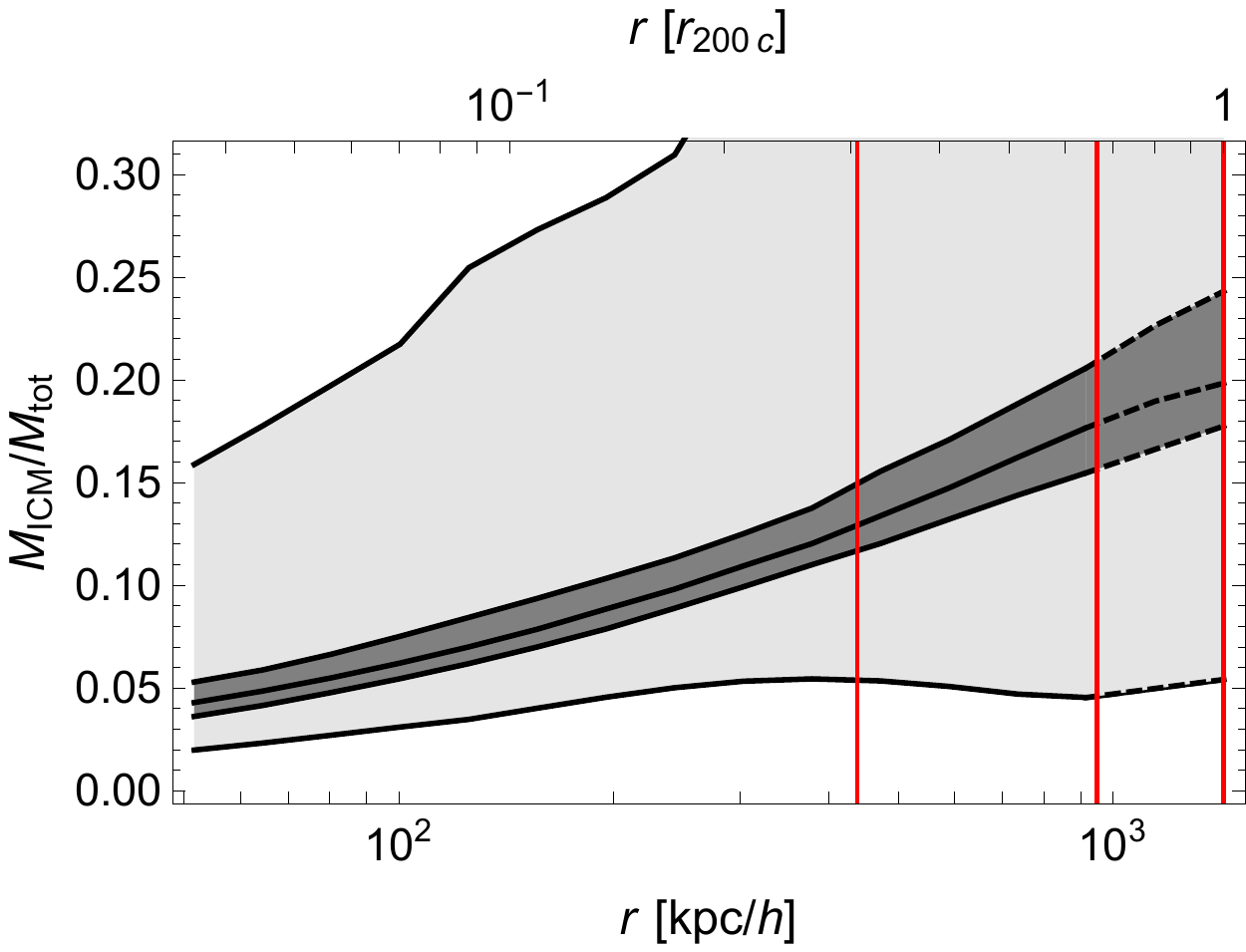}}
       \caption{Ratio of spherically enclosed gas mass to total mass as a function of the spherical radius $r$. The regression assumed a flat prior for the axis ratio $q_{\text{mat},1}$ of the matter distribution and $q_\text{ICM}\ge q$. The middle line tracks the median. The grey shaded regions represent the 68.3 and 95.4 per cent regions around the median, as computed from symmetric quantiles. Dashed lines indicate extrapolations beyond the region covered by X-ray observations. Vertical red lines mark $r_\text{sph,2500}$, $r_\text{sph,500}$, and $r_\text{sph,200}$, as measured in spherical regions.}
	\label{fig_M1206_gasFraction}
\end{figure}

We computed the gas mass fraction in spherically enclosed regions, $f_\text{gas} = M_\text{sph,gas}(<r)/M_\text{sph,tot}(<r)$, using the posterior samples of the ellipsoidal cluster model. The estimate does not rely on the assumption of equilibrium. Despite being computed in spherical regions, this estimate of the cumulative gas mass fraction is free from the assumptions of spherical symmetry. 

The gas fraction profile is shown in Fig.~\ref{fig_M1206_gasFraction}. At the overdensity spherical radius $r_\text{sph,2500}$, we found $f_\text{gas}=0.128\pm0.014$; within $r_\text{sph,500}$, we found $f_\text{gas,500}=0.177\pm0.023$. Based on scaling relations of X-ray luminosity, temperature and gas mass, and employing total masses from weak gravitational lensing measurements, \citet{man+al16} found a constraint on the gas mass fraction of $f_\text{gas,500} = 0.125 \pm 0.005$ in a sample of 40 clusters identified as being dynamically relaxed and hot, consistent with previous measurements using hydrostatic mass estimates for relaxed clusters \citep{man+al16a}. Accounting for the intrinsic scatter affecting gas fraction estimates, our result is consistent with this general trend.

\citet{ume+al12} combined lensing results with {\it Chandra} gas mass measurements and found a cumulative gas mass fraction of $f_\text{gas}=0.137^{+0.045}_{-0.030}$ at $1~\text{Mpc}=0.7~\text{Mpc}h^{-1}$ under the hypothesis of spherical symmetry. Our extrapolated result is $f_\text{gas}(<1~\text{Mpc})=0.157\pm0.019$, in good agreement. 

When compared to the cosmic baryon fraction $f_\text{B}\simeq 0.156$ \citep{planck_2015_XIII}, we find no significant gas depletion, $Y_\text{B}=f_\text{gas,500}/f_\text{B}\sim 1$. Some depletion is predicted by hydrodynamic simulations including radiative cooling, star formation and AGN feedback \citep{bat+al13,pla+al13}. \citet{pla+al13} found that the gas fraction of massive clusters is nearly independent of the physical processes and it is characterised by a negligible redshift evolution, $Y_\text{B}=0.85\pm0.03$ at $r_{500}$. At smaller radii, $Y_\text{B}$ slightly decreases, in agreement with what found in MACS1206, by an amount that depends on the physics included in the simulations. 

MACS1206 may be baryon rich. However, angular variance of the anisotropically distributed gas that originates from the recent formation epoch of clusters and from the strong internal baryon-to-dark-matter density bias, along with density clumpiness, can bias high the measured gas fraction too \citep{bat+al13}.

\subsection{Non-thermal pressure}

\begin{figure}
       \resizebox{\hsize}{!}{\includegraphics{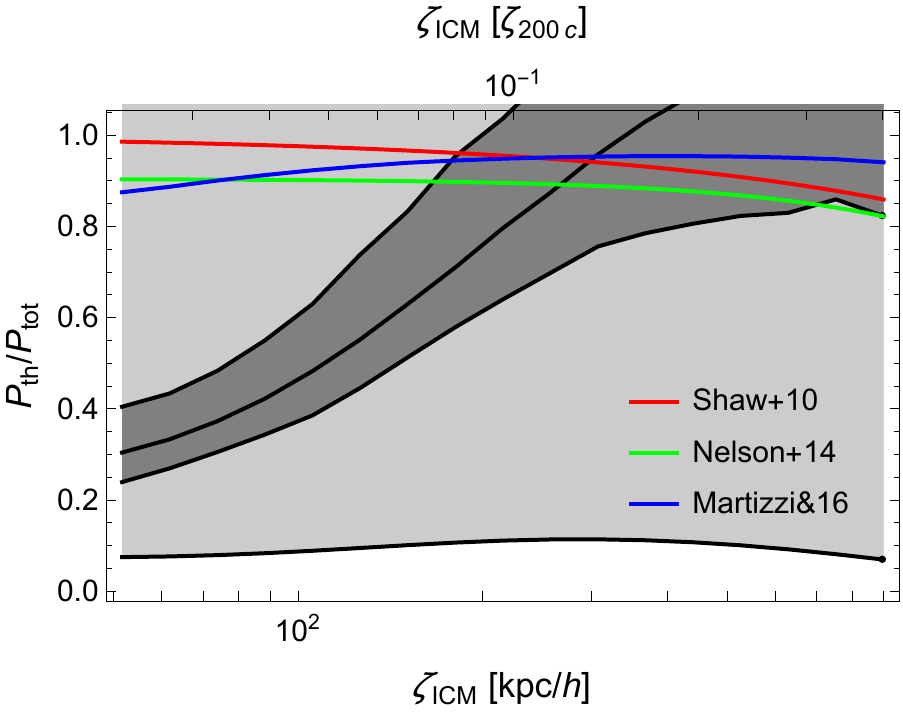}}
       \caption{Ratio of the thermal gas pressure to the total equilibrium pressure as a function of the ellipsoidal radius $\zeta_\text{ICM}$ measured along the major axis of the gas distribution. The regression assumed a flat prior for the axis ratio $q_{\text{mat},1}$ of the matter distribution and $q_\text{ICM}=q_{\Phi}$. The middle line tracks the median. The grey shaded regions show the 68.3 and the 99.7 per cent regions around the median. The red, green and blue lines plot the theoretical predictions from \citet{sha+al10}, \citet{nel+al14}, and \citet{ma+ag16}, respectively.}
	\label{fig_M1206_pressure}
\end{figure}

Non-thermal pressure $P_\mathrm{nth}$ can significantly contribute to the overall balance \citep{ras+al14}. Neglecting the contribution from bulk and/or turbulent motions can systematically bias low the X-ray mass determination \citep{men+al10,ras+al12}. High resolution cosmological simulations showed that the contribution can be significant in the core of relaxed clusters too \citep{lau+al09,mol+al10}. 

Based on ten simulations of massive relaxed clusters, \citet{mol+al10} found that pressure support from subsonic random gas motions can contribute up to 40 per cent in the inner regions and up to 20 per cent within one tenth of the virial radius. The non-thermal contribution is expected to increase with radius in the very outer regions \citep{sha+al10,ma+ag16}. \citet{lau+al09} found a non-thermal pressure contribution of the order of 5-15 per cent at about one tenth of the viral radius, also increasing with radius in the outer regions.

\citet{chi+al12} tested the assumption of strict hydrostatic equilibrium in MS2137.3-2353 and found a significant contribution from non-thermal pressure in the core region, independently of the assumed shape of the cluster.

Since our inference did not rely on the assumption of hydrostatic equilibrium, the equilibrium hypothesis can be used a posteriori to asses the level of non-thermal pressure. The generalised equilibrium condition reads
\beq
\label{eq_he_1}
\nabla P_\mathrm{tot}=-\rho_\text{ICM} \nabla \Phi_\text{mat},
\eeq
where $P_\mathrm{tot}(=P_\mathrm{th}+P_\mathrm{nth})$ is the total pressure, $\rho_\text{ICM}$ the gas density and $\Phi_\text{mat}$ the gravitational potential. When the equilibrium is hydrostatic, the pressure is only thermal, $P_\mathrm{th}= k_\mathrm{B} T n_\text{ICM}$ for an ideal gas.

Due to the shape theorem for gas in equilibrium \citep{bu+ca94}, we considered the posteriori probabilities for the cluster parameters inferred under the prior of flat distribution for the matter shape and $q_\text{ICM}=q_{\Phi}$. The gravitational potential of the ellipsoidal NFW halo was computed using the formulae in \citet{le+su03}. 

The result is presented in Fig.~\ref{fig_M1206_pressure},where the ratio of thermal to equilibrium gas pressure, $P_\mathrm{th}/P_\mathrm{tot}$, is plotted as function of the ellipsoidal radius of the ICM distribution, $\zeta_\text{ICM}$.

We find that the contribution of thermal pressure is dominant. The apparent drop in the inner regions can be an artefact due to the the poor modelling of the temperature profile in the inner $\sim100~\text{kpc}$, see Fig.~\ref{fig_M1206_Te_profile}.

\section{Systematics}
\label{sec_syst}

\begin{table*}
\caption{Results under different methods or data-sets. All regressions assumed a flat prior for the axis ratio $q_{\text{mat},1}$ of the matter distribution and $q_\text{ICM}\ge q$. Differently from the reference analysis (reported in col. 2), we considered the SaWLens convergence maps (col.~3), a 1D analysis of the X-ray surface brightness (col.~4), or a treatment of the SZ effect based on Bolocam data  only (col.~5). Units are as in Table~\ref{tab_parameters}. Typical values and dispersions are computed as bi-weighted estimators.}
\label{tab_systematics}
\begin{tabular}{ l r@{$\,\pm\,$}l r@{$\,\pm\,$}l r@{$\,\pm\,$}l r@{$\,\pm\,$}l}     
\hline
			&		\multicolumn{2}{c}{reference}	&  \multicolumn{2}{c}{SaWLens}&	\multicolumn{2}{c}{1D-X}	&	\multicolumn{2}{c}{no Planck}	\\
	\hline
$M_{200}$                       	&	1.137  	&	0.229 	&	1.115  	&	0.113	&	1.035  	&	0.196	&	1.071  	&	0.187	\\
$c_{200}$                       	&	6.277  	&	1.188 	&	4.400  	&	0.556	&	6.247  	&	1.106	&	5.586  	&	0.879	\\
$q_{\text{mat},1}$              	&	0.466  	&	0.119 	&	0.540  	&	0.145	&	0.434  	&	0.114	&	0.557  	&	0.107	\\
$q_{\text{mat},2}$              	&	0.735  	&	0.176 	&	0.794  	&	0.094	&	0.688  	&	0.158	&	0.749  	&	0.140	\\
$\cos \vartheta$                	&	0.297  	&	0.204 	&	0.302  	&	0.239	&	0.266  	&	0.229	&	0.333  	&	0.209	\\
$\varphi$                       	&	-0.438 	&	1.609 	&	-0.549 	&	1.530	&	0.200  	&	1.306	&	0.096  	&	0.987	\\
$\psi$                          	&	1.027  	&	0.123 	&	0.979  	&	0.149	&	0.936  	&	0.203	&	0.975  	&	0.156	\\
$q_\text{ICM,1}$                	&	0.587  	&	0.109 	&	0.614  	&	0.122	&	0.528  	&	0.124	&	0.684  	&	0.090	\\
$q_\text{ICM,2}$                	&	0.779  	&	0.057 	&	0.778  	&	0.069	&	0.787  	&	0.096	&	0.814  	&	0.067	\\
$n_0$                           	&	0.010  	&	0.001 	&	0.010  	&	0.001	&	0.011  	&	0.001	&	0.011  	&	0.001	\\
$\zeta_\text{c}$                	&	169.780	&	12.02	&	165.950	&	5.880	&	174.490	&	6.728	&	145.060	&	5.839	\\
$\zeta_\text{t}/\zeta_\text{c}$ 	&	8.250  	&	1.635 	&	8.057  	&	1.290	&	7.272  	&	1.293	&	8.041  	&	1.092	\\
$\beta$                         	&	0.600  	&	0.025 	&	0.592  	&	0.027	&	0.613  	&	0.024	&	0.568  	&	0.020	\\
$\eta$                          	&	0.627  	&	0.047 	&	0.643  	&	0.054	&	0.592  	&	0.042	&	0.574  	&	0.046	\\
$\gamma_\text{ICM}$             	&	1.810  	&	0.632 	&	2.336  	&	0.434	&	2.093  	&	0.477	&	2.151  	&	0.527	\\
$T_0$                           	&	24.245 	&	2.646 	&	21.469 	&	2.961	&	21.426 	&	2.885	&	18.384 	&	1.790	\\
$\zeta_{cT}/\zeta_\text{c}$	&	9.069  	&	1.034 	&	8.518  	&	0.882	&	8.666  	&	1.285	&	17.646 	&	2.173	\\
$c_T$                    	&	2.576  	&	0.313 	&	2.790  	&	0.158	&	2.891  	&	0.097	&	2.201  	&	0.482	\\
$T_\text{cc}$                   	&	4.221  	&	1.328 	&	3.654  	&	1.107	&	3.894  	&	0.861	&	4.316  	&	1.258	\\
$\zeta_\text{cc}/\zeta_\text{c}$	&	1.140  	&	0.211 	&	0.986  	&	0.132	&	0.950  	&	0.126	&	1.006  	&	0.133	\\\hline	
\end{tabular}
\end{table*}

The level of systematics errors plaguing an analysis can be checked by comparing results derived with distinct methodologies or exploiting different data-sets. We considered independent treatments of either the lensing, X-ray, or SZ part of our analysis. We only changed one module of the total $\chi^2$ for each test, keeping all the other probes fixed. Results are summarised in Table~\ref{tab_systematics}. Whereas we considered different data-sets or likelihoods, we used the same Bayesian inference and the same treatment of the posterior. As priors, we assumed a flat distribution for the minor-to-major matter axis ratio and $q_\text{ICM}\ge q$.

\subsection{Lensing}

As an alternative to the lensing treatment, we considered the convergence maps obtained with the SaWLens (Strong -and Weak-lensing) method \citep{mer+al15}, which can consistently combine weak and strong lensing with no a priori assumptions about the underlying mass distribution. The convergence map can then be fitted to get unbiased parameters for the lensing model of choice. 

SaWLens performs a reconstruction of the lensing potential on adaptively refined grids. For the inner regions of MACS1206, \citet{mer+al15} used the position and redshift of 33 critical line estimators derived from the 13 multiple-image systems (4 of which spectroscopically confirmed) and measured the shapes of $\la 600$ background galaxies in seven broad-band Advanced Camera for Surveys (ACS) filters. In the outer regions, \citet{mer+al15} exploited the Subaru data for the shear analysis.

For our analysis, we considered two different grid sizes. Even though each grid exploits all data-sets, it is most sensitive to some of them. The low resolution grid, covering $25\arcmin\times25\arcmin$ with a pixel resolution of $50\arcsec$, is well suited for weak lensing on the wide field, and is dominated by the data from the Subaru ground-based telescope. 

The fine grained grid, covering the inner $150\arcsec\times150\arcsec$ with a pixel resolution of $\sim 8\arcsec$, traces strong lensing features near the inner-most core of the cluster and exploits the weak lensing constraints from the {\it HST} on a much smaller field of view but with considerably higher spatial resolution. 

When combining the two maps, we excluded the inner square of side $2\arcmin$ of the low-resolution grid to prevent overlap with the strong lensing region already sampled by the fine-resolution grid.

Statistical uncertainty within SaWLens is based on bootstrap re-samplings and noise realisations. However, the number of resampled maps is not big enough to compute the uncertainty covariance matrix of the 2D-WL map. In these cases, regularisation schemes already employed in lensing analyses \citep{ume+al12,ume+al15a} are not so effective. In fact the covariance matrix is singular not due to intrinsic features of the data sample but due to the number of bootstrap realisations being smaller or comparable to the number of pixels. We then assumed the covariance matrix to be diagonal. This conservative assumption may under-estimate the $\chi^2$, and consequently over-estimate the confidence regions of the parameters but can still recover the main features of mass distribution. 

The pixels of the low resolution map are large enough to contain a significant number of background galaxies ($\sim10$ galaxies per pixel). Even though the reconstruction of the convergence map from the shear field is not local, the main contribution comes from the galaxies in the pixel itself, which reduces the correlations among pixels. 

Since we modelled the whole cluster as one three-dimensional NFW halo and we do not assume any subhalos, our conservative approach can be still effective. However, due to the limited knowledge of the covariance matrix, we used the SaWLens map only for model comparison.

Even though SaWLens fits all data at once, the different grids and the different weights of the lensing data-sets make the low and the fine resolution maps nearly independent. For the inner regions, we considered the azimuthally averaged convergence in 6 equally spaced angular annuli between $5\arcsec$ and $1\arcmin$. To measure the uncertainties, we re-computed the convergence for the noise realisations and we computed the uncertainty covariance matrix from the sample distribution.

As far as data-sets are considered, the SaWLens analysis exploited the shear signal in the outer regions, whereas our reference analysis considered magnification and number counts too.

The results based on the SaWLens maps are in good agreement with our reference model even though the SaWLens-based concentration is smaller and uncertainties on mass and concentration are smaller. As a consequence of the smaller concentration, the shape is a bit rounder too. Even though the shift is compatible with the statistical uncertainties of our reference model, we notice that the NFW profile is not an excellent modelling of the SaWLens convergence profile, which is steep in the inner regions and flat in the outer regions, see Fig.~\ref{fig_M1206_kappa_profile}. In fact, the smaller formal uncertainties are more a signal of problems in modelling than of increased accuracy, which is not reasonable since the data-set is the same apart from the magnification data. However, to fully understand if this deviation is statistical significant, we would need an accurate computation of the SaWLens uncertainty covariance matrix for the WL regime, which at this time is still missing.

\subsection{X-ray surface brightness}

To check the analysis of the X-ray surface brightness, we considered an alternative procedure. The measurement of ellipticity and slope are nearly uncorrelated \citep{def+al05}. The analysis of the 2D surface brightness can then be approximated as a 1D problem. The ellipticity and orientation angle are evaluated in a first step and the surface brightness is measured in elliptical annuli (following the morphology determined before) in a second step \citep{ser+al12a}. The $\chi^2$ function can be written as
\beq
\begin{aligned}
\label{eq_SB6}
\chi^2_\text{SB}  &= \sum_{i=1}^{N_{S}}  \left(\frac{S_{\text{X},i}-\hat{S}_{\text{X},i}}{\delta_{S,i}}\right)^2 \\ 
 &+  \left(\frac{\epsilon_\text{ICM}-\hat{\epsilon}_\text{ICM}}{\delta\epsilon_\text{ICM}}\right)^2
+  \left(\frac{\theta_{\text{ICM},\epsilon}-\hat{\theta}_{\text{ICM},\epsilon}}{\delta\theta_{\text{ICM},\epsilon}}\right)^2,
\end{aligned}
\eeq
with $\hat{S}_\text{X}$, $\hat{\epsilon}_\text{ICM}$, and $\hat{\theta}_{\text{ICM},\epsilon}$ are the model predictions for the corresponding X-ray observables with measurements uncertainties $\delta_{S,i}$, $\delta\epsilon_\text{ICM}$, and $\delta\theta_{\text{ICM},\epsilon}$, respectively. $\epsilon_\text{ICM}$ and $\theta_{\text{ICM},\epsilon}$ are the ellipticity and the orientation angle, respectively, of the X-ray isophotes in the plane of the sky.

The ellipticity and the orientation angle can be determined as the parameters of the ellipse enclosing a fraction of the total cluster light. We considered light thresholds of 50, 60, 70, and 80 per cent and we determined the parameters as the bi-weight estimators. We found $\epsilon_\text{ICM} = 0.23 \pm 0.07$ and orientation angle $\theta_{\mathrm{ICM},\epsilon}= -54.2\pm 2.7\deg$ (measured North over East) in remarkable agreement with the 2D analysis, see Table~\ref{tab_ICM_shape}. These parameters were then used to define the elliptical bins in which the surface brightness profile is resolved.

The agreement with the full 2D analysis is substantial, showing that modelling the gas shape as an ellipsoid with fixed axis ratios and fixed orientation is a very good approximation at the present level of accuracy and precision.

\subsection{SZe}

The integrated Compton parameter was alternatively computed relying on the Bolocam data only. Results are in very good agreement with the reference analysis exploiting {\it Planck} data too. The {\it Planck} data are important to better resolve the elongation of the gas distribution, which favours more triaxial structure. {\it Planck} data also improves the modelling of the gas profile at large radii, favouring a slightly larger core radius and a temperature profile decrement at large radii.

\section{Conclusions}
\label{sec_conc}

MACS1206 is a remarkably regular, massive, face-on cluster. Its concentration is in line with theoretical predictions. The measured triaxial shape is common to clusters simulated in the $\Lambda$CDM concordance cosmology. The gas has settled in the potential well and its distribution traces the iso-potential surfaces. This is evidence under the shape theorem that the cluster is in pressure equilibrium. The level of baryonic depletion is small. The thermal pressure can balance the cluster in hydrostatic equilibrium.

Multi-probe analyses are needed to achieve one of the main goals of precise and accurate cosmology: unbiased cluster mass measurements at the per cent level \citep{ben+al13_spt}. The compatibility of independent observables, from X-ray to lensing to radio observations of the SZe, under the same coherent picture guarantee that the measurement of mass and other intrinsic properties are unbiased and systematic-free. In our modelling, we do not rely on the assumptions of spherical symmetry or hydrostatic equilibrium, which could bias results.

At the same time, the joint exploitation of different data-sets improves the statistical accuracy and enables us to expand the scope of the analysis. We can measure the cluster shape and the non-thermal pressure too.

The multi-wavelength analysis we presented in this paper is a development of the Bayesian method introduced by \citet{ser+al13}. We used the same modelling of the mass and gas distribution and the same inference scheme but: we extended and refined the priors on mass and gas; we fitted the 2D map of the X-ray surface brightness; we used the exact likelihoods for gravitational lensing and X-ray plus SZe rather than approximating them as smooth kernel or multi-variate distributions of the projected parameters. Whereas in the ideal case the two approaches are equivalent, the new approach is more flexible and relies less on geometrical assumptions.

In forthcoming papers, we plan to analyse the full CLASH sample.

\section*{Acknowledgements}
The authors thank M. Bonamigo and M. Limousin for stimulating discussions and C. Grillo per useful comments. SE, MM and MS acknowledge the financial contribution from contracts ASI-INAF I/009/10/0, PRIN-INAF 2012 `A unique dataset to address the most compelling open questions about X-Ray Galaxy Clusters', and PRIN-INAF 2014 1.05.01.94.02 `Glittering Kaleidoscopes in the sky: the multifaceted nature and role of galaxy clusters'. SE acknowledges the financial contribution from contracts NARO15 ASI-INAF I/037/12/0 and ASI 2015-046-R.0. MM acknowledges support from the Italian Ministry of Foreign Affairs and International Cooperation, Directorate General for Country Promotion, and from ASI via contract ASI/INAF/I/023/12/0. JS was supported by NSF/AST-1617022. KU acknowledges support from the Ministry of Science and Technology of Taiwan through grants MOST 103-2112-M-001-030-MY3 and MOST 103-2112-M-001-003-MY3. This research has made use of NASA's Astrophysics Data System (ADS) and of the NASA/IPAC Extragalactic Database (NED), which is operated by the Jet Propulsion Laboratory, California Institute of Technology, under contract with the National Aeronautics and Space Administration.

\bibliographystyle{mn2e_fix_Williams}


\setlength{\bibhang}{2.0em}

\appendix

\section{Projection}
\label{sec_proj}

The problem of finding the volume density distribution of systems whose surface density contours are similar ellipses is a recurrent astronomical problem. Here, we follow the formalism presented by \citet{sta77}, who first discussed the projection of an ideal triaxial galaxy onto the plane of the sky, and further developed by \citet{bin80,bin85}. The formalism was later introduced in the context of gravitational lensing \citep{ogu+al03,co+ki07,ser+al10b,se+um11} and multi-wavelengths analyses of galaxy clusters \citep{def+al05,ser+al06,ser07,bu+hu12,ser+al12a,ser+al13}. Here, we review the main results in terms of the notation used in the present paper. 

The observed system is an ellipsoid whose principal axes define the intrinsic coordinate system. The semi-major axis $l_\text{s}$ is oriented along the third axis of the intrinsic system and the minor-to-major and intermediate-to-major axis ratios are $q_1\le q_2 \le 1$. The orientation of the ellipsoid in the observer system is defined by the Euler angles $\vartheta$, $\varphi$ and $\psi$.

The system appears in the plane of the sky as an ellipse with projected axis ratio
\begin{equation}
\label{eq_tri1} 
q_\perp= \sqrt{ \frac{j+l -\sqrt{(j-l)^2+4 k^2 }}{j+l + \sqrt{(j-l)^2+4 k^2 } } },
\end{equation}
where  $j, k$ and $l$ are defined as
\begin{eqnarray}
j & = &  \left( \frac{\sin\vartheta}{q_1 q_2}\right)^2 + \cos^2\vartheta \left[ \left( \frac{ \cos \varphi}{q_1}\right)^2  +  \left(\frac{\sin\varphi}{q_2}\right)^2 \right] ,  \label{eq_tri2}  \\
k & = &  \left(\frac{1}{q_1^{2}} - \frac{1}{q_2^{2}} \right) \sin \varphi \cos \varphi  \cos \vartheta   , \label{eq_tri3}   \\
l & = &   \left( \frac{\sin\varphi}{q_1}\right)^2 +  \left(\frac{\cos \varphi}{q_2}\right)^2 . \label{eq_tri4} 
\end{eqnarray}
We also use the ellipticity
\beq
\label{eq_tri1a}
\epsilon_\perp = 1-q_\perp
\eeq

The (tangent of the) orientation angle in the plane of the sky of the projected ellipse (measured North over East as usual in astronomy) is
\beq
\label{eq_tri5} 
\tan \theta_\epsilon =  \tan [ \Delta\theta_{\psi=0}+\frac{\pi}{2} H(\Delta_{\pi/2}) -\psi  ], 
\eeq
where $H$ is the Heaviside function and $\Delta\theta_{\psi=0}$ is the angle between the projection of the major axis of the ellipsoid and the principal axes of the ellipse,
\beq
\label{eq_tri6} 
\Delta\theta_{\psi=0} = \frac{1}{2} \arctan \left[\frac{2 k}{j-l} \right].
\end{equation}
The apparent principal axis that lies furthest from the projection of the 3D major-axis onto the plane of the sky is the apparent major axis if \citep{bin85}
\begin{equation}
\label{eq_tri7} 
\Delta_{\pi/2}=(j-l)\cos (2 \Delta\theta_{\psi=0})  +2k \sin (2\Delta\theta_{\psi=0})  \leq 0.
\eeq

The semi-major axis $l_\perp$ of the projected ellipse in the plane of the sky, i.e. perpendicularly to the line of sight, is
\beq
l_\perp = \frac{l_\text{s}}{e_\parallel \sqrt{f}}
\eeq
where
\begin{equation}
\label{eq_tri8} 
f =   \sin^2\vartheta \left[ \left(\frac{  \sin \varphi}{q_1}\right)^2+   \left(\frac{\cos \varphi}{q_2}\right)^2\right]  + \cos^2 \vartheta  ,
\end{equation}
and
\beq
\label{eq_tri9} 
e_\parallel  = \frac{q_\perp}{q_1q_2 f^{3/4}} .
\eeq

The half-size $l_\parallel$ along the line of sight of the ellipsoid projected perpendicularly to the line of sight, i.e. as seen from above, is
\beq
\label{eq_tri10} 
l_\parallel =  \frac{l_\text{s}}{\sqrt{f}} .
\eeq
The parameter $e_\parallel$ quantifies the extent of the cluster along the line of sight. It can be expressed as the ratio of the size of the cluster along the line of sight to the size in the plane of the sky,
\beq
e_\parallel =\frac{l_\parallel}{l_\perp}. 
\eeq
The larger $e_\parallel$, the more the orientation bias toward the observer. If $e_\parallel >1$, the cluster is more elongated along the line of sight than wide in the plane of the sky.

A general form for the volume density is $\rho_\text{3D}  = \rho_\text{s} f_\rho(x)$, where $\rho_\text{s}$ sets the density scale and $x$ is an dimensionless variable. The functional $f_\rho$ describes the density profile and be characterised by a number of parameters, e.g. the concentration, the outer slope, the truncation radius. 

The projected surface distribution is obtained as
\beq
\Sigma_\text{2D}  = \int_\parallel \rho_\text{3D}  dl,
\eeq
where the subscript $_\parallel$ denotes integration along the line of sight.

If the volume density is constant on similar ellipsoids, the functional $f_\rho$ can be expressed in terms of $x=\zeta/l_\text{s}$, where $\zeta$ is the ellipsoidal radius. The projected isocontours are elliptical,
\beq
\Sigma_\text{2D} = \Sigma_\text{s} f_\Sigma (\xi/l_\perp),
\eeq
where $\xi$ is the observed elliptical radius, the density scale is given by
\beq
\Sigma_\text{s} = \rho_\text{s} l_\parallel,
\eeq
and the functional form is obtained as
\beq
f_\Sigma (x) = 2 \int_x^\infty \frac{f_\rho(x') x'}{\sqrt{x'^{2}-x^2}}dx',
\eeq
analogously to the spherical case.

\section{Potential shape}
\label{sec_pote_shap}

\begin{figure}
       \resizebox{\hsize}{!}{\includegraphics{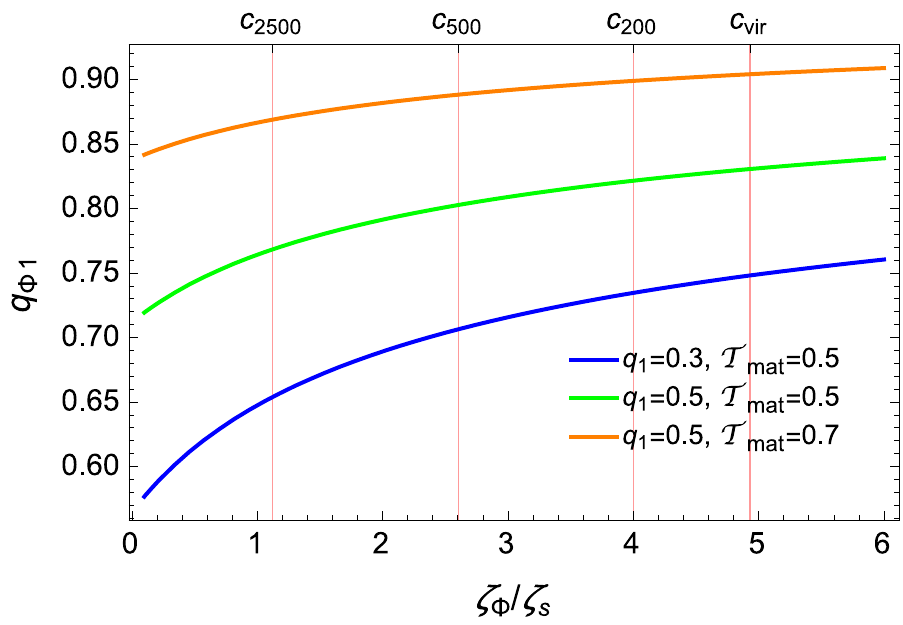}}
       \caption{Minor to major axis ratio of the isopotential surfaces of a NFW halo with $c_{200}=4$ at $z=0.3$ as a function of the  ellipsoidal radius.}
	\label{fig_potential_axial_ratio}
\end{figure}

\begin{figure}
       \resizebox{\hsize}{!}{\includegraphics{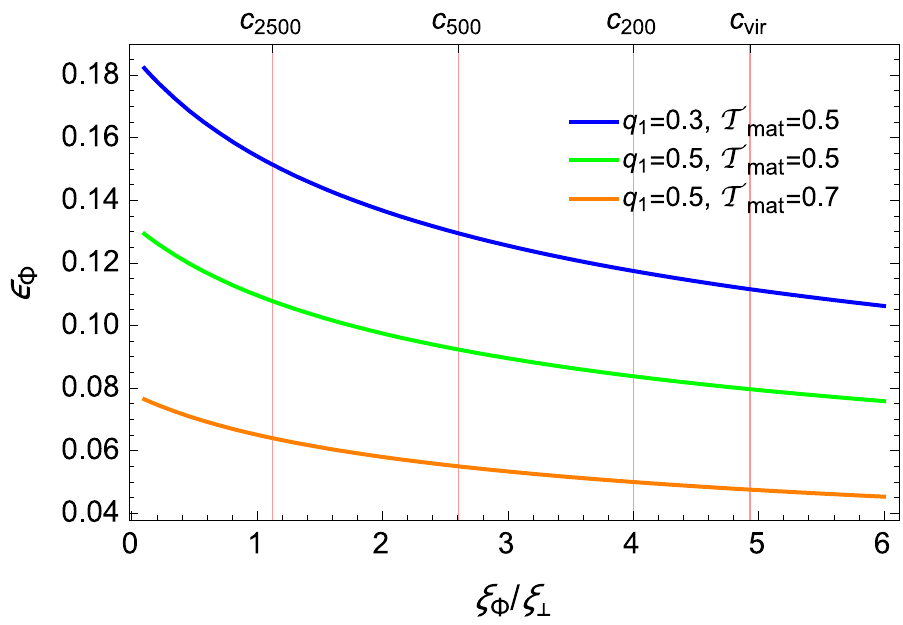}}
       \caption{Projected ellipticity of the isopotential surfaces of a NFW halo with $c_{200}=4$ as a function of the elliptical radius. The orientation of the halo is fixed by $\cos \vartheta =0.5$, and $\phi=\pi/3$.}
	\label{fig_potential_epsilon}
\end{figure}

\begin{figure}
       \resizebox{\hsize}{!}{\includegraphics{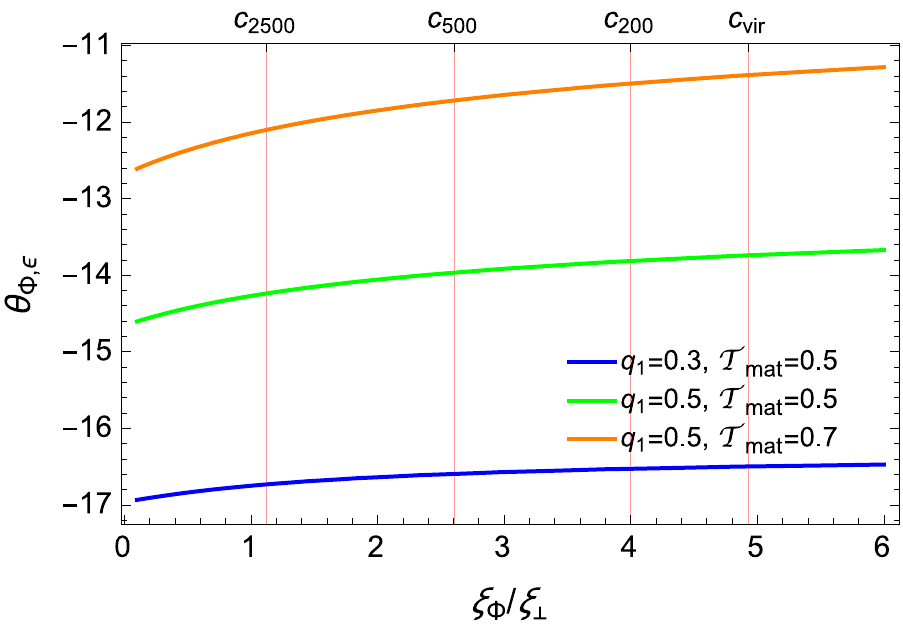}}
       \caption{Orientation angle of the isopotential surfaces of a NFW halo with $c_{200}=4$ as a function of the elliptical radius. The orientation of the halo is fixed by $\cos \vartheta =0.5$, $\phi=\pi/3$, and $\psi=0$.}
	\label{fig_potential_theta_epsilon}
\end{figure}

Even though the isopotential surfaces of ellipsoidal dark halos are not exact ellipsoids, the isopotential surfaces are still well approximated as ellipsoids whose shape slightly varies with the radius \citep{le+su03}. Here, we follow \citet{kaw10} and derive the axis ratios of the potential by numerical integration. We considered the gravitational potential $\Phi$ of a triaxial NFW halo, but results are nearly independent of the peculiar matter density profile.

The isopotential surface (in the intrinsic system aligned with the axis ratios) defined by $\Phi (x,y,z)=\text{const.}=\Phi (0,0,c_\Phi)$ is approximated by the triaxial ellipsoid with minor axis $a_\Phi$, intermediate axis $b_\Phi$, and major axis $c_\Phi$, such that $\Phi (a_\Phi,0,0) =\Phi (0,b_\Phi,0)=\Phi (0,0,c_\Phi)$; the axis ratios are then $q_{\Phi,1}=a_\Phi/c_\Phi$ and $q_{\Phi,2}=b_\Phi/c_\Phi$.

The shape variation is small between $\zeta_{500}$ and $\zeta_\text{vir}$, the scales most interested by weak lensing observations. In Figs.~\ref{fig_potential_axial_ratio}, \ref{fig_potential_epsilon} and \ref{fig_potential_theta_epsilon}, we plot the intrinsic minor to major axis ratio of the potential, the projected ellipticity and the orientation angle of the projected potential, respectively, as a function of the radius. Radial variations are small even for very elongated clusters ($q_{\text{mat},1} \la 0.3$), and far below the usual observational accuracy between $\zeta_{500}$ and $\zeta_\text{vir}$. The potential and the matter distribution have nearly the same triaxiality parameter and $e_{\Phi,i}/e_{\text{mat},i}\sim 0.7$.

\end{document}